\documentclass[%
reprint,
twocolumn,
superscriptaddress,
nofootinbib,
amsmath,amssymb,
aps,
]{revtex4-2}
\usepackage{natbib}
\usepackage{url}
\usepackage{orcidlink}
\usepackage{float}
\usepackage{graphicx}% Include figure files
\usepackage{dcolumn}% Align table columns on decimal point
\usepackage{bm}% bold math
\begin{document}
	
	\def\aap{AA}
	\def\prl{Phys. Rev. Lett.}
	\def\aapr{AA Rev}
	\def\apjl{ApJL}
	\def\na{NewA}
	\def\apss{APSS}
	\def\aaps{AAPS}
	\def\mnras{MNRAS}
	\def\apj{ApJ}
	\def\apjs{ApJS}
	\def\aj{AJ}
	\def\pasp{PASP}
	\def\pasj{PASJ}
	\def\nat{Nat}
	\def\memsai{MmSAI}
	\def\prd{PhysRevD}
	\def\jcap{JCAP}
	\def\physrep{Phys. Rept.}
	\def\araa{Annual Review of Astronomy and Astrophysics}
	\def\ssr{Space Science Reviews}
	
	\newcommand{{\qsub}}{\bf{q}}
	\newcommand{{\rsub}}{{\bf{r}}}
	\newcommand{{\xlight}}{{\bf{n}}_{\rm{light}}}
	\newcommand{{\xlens}}{{\bf{n}}_{\rm{mac}}}
	\newcommand{{\data}}{\bf{D}}
	\newcommand{{\datan}}{{\bf{d}}_{\rm{n}}}
	\newcommand{{\datanprime}}{{{\bf{d}}_{\rm{n}}^{\prime}}}
	\newcommand{{\dimg}}{{\bf{d}}_{\mathcal{I}}}
	\newcommand{{\dimgprime}}{{\bf{d}}_{\mathcal{I}}^{\prime}}
	\newcommand{{\dptsrc}}{{\bf{d}}_{\rm{ptsrc}}}
	\newcommand{{\dptsrcprime}}{{\bf{d}}_{\rm{ptsrc}}^{\prime}}
	\newcommand{{\dfr}}{{\bf{d}}_{\rm{fr}}}
	\newcommand{{\dfrprime}}{{\bf{d}}_{\rm{fr}}^{\prime}}
	\newcommand{{\zlens}}{{z_{\rm{d}}}}
	\newcommand{{\zsrc}}{{z_{\rm{s}}}}
	
	\preprint{APS/123-QED}
	
	\title{JWST lensed quasar dark matter survey IV: Stringent warm dark matter constraints from the joint reconstruction of extended lensed arcs and quasar flux ratios}
	
	\author{D.~Gilman\orcidlink{0000-0002-5116-7287}}
	\thanks{Brinson Prize Fellow}
	\email{gilmanda@uchicago.edu}
	\affiliation{Department of Astronomy \& Astrophysics, University of Chicago, Chicago, IL 60637, USA}
	
	\author{A.~M.~Nierenberg\orcidlink{0000-0001-6809-2536}}
	\affiliation{University of California, Merced, 5200 N Lake Road, Merced, CA 95341, USA}
	
	\author{T.~Treu\orcidlink{0000-0002-8460-0390}}
	\affiliation{Department of Physics and Astronomy, University of California, Los Angeles, CA,  90095, USA}
	
	\author{C.~Gannon\orcidlink{0009-0009-0443-3181}}
	\affiliation{University of California, Merced, 5200 N Lake Road, Merced, CA 95341, USA}
	
	\author{X.~Du\orcidlink{0000-0003-0728-2533}}
	\affiliation{Department of Physics and Astronomy, University of California, Los Angeles, CA,  90095, USA}
	
	\author{H.~Paugnat\orcidlink{0000-0002-2603-6031}}
	\affiliation{Department of Physics and Astronomy, University of California, Los Angeles, CA,  90095, USA}
	
	\author{S.~Birrer\orcidlink{0000-0003-3195-5507}}
	\affiliation{Department of Physics and Astronomy, Stony Brook University, Stony Brook, NY 11794, USA}
	
	\author{A.~J.~Benson\orcidlink{0000-0001-5501-6008}}
	\affiliation{Carnegie Institution for Science, Pasadena CA 91101, USA }
	
	\author{P.~Mozumdar\orcidlink{0000-0002-8593-7243}}
	\affiliation{Department of Physics and Astronomy, University of California, Los Angeles, CA,  90095, USA}
	\affiliation{Department of Physics and Astronomy, University of California, Davis, 1 Shields Ave., Davis, CA 95616, USA}
	
	\author{K.~C.~Wong\orcidlink{0000-0002-8459-7793}}
	\affiliation{Research Center for the Early Universe, Graduate School of Science, The University of Tokyo, 7-3-1 Hongo, Bunkyo-ku, Tokyo 113-0033, Japan}
	
	\author{D.~Williams\orcidlink{0000-0002-8386-0051}}
	\affiliation{Department of Physics and Astronomy, University of California, Los Angeles, CA,  90095, USA}
	
	\author{R.~E.~Keeley\orcidlink{0000-0002-0862-8789}}
	\affiliation{University of California, Merced, 5200 N Lake Road, Merced, CA 95341, USA}
	
	\author{K.~N.~Abazajian\orcidlink{0000-0001-9919-6362}}
	\affiliation{Department of Physics and Astronomy, University of California, Irvine, CA 92697-4575, USA}
	
	\author{T.~Anguita\orcidlink{0000-0003-0930-5815}}
	\affiliation{Instituto de Astrofisica, Departamento de Fisicas y Astronomia, Universidad Andres Bello, Santiago, Chile}

	\author{V.~N.~Bennert\orcidlink{0000-0003-2064-0518}}
	\affiliation{Physics Department, California Polytechnic State University, San Luis Obispo, CA 93407, USA }
	
	\author{S.~G.~Djorgovski\orcidlink{0000-0002-0603-3087}}
	\affiliation{California Institute of Technology, Pasadena CA 91125, USA }
	
	\author{S.~F.~Hoenig\orcidlink{0000-0002-6353-1111}}
	\affiliation{School of Physics and Astronomy, University of Southampton, Southampton SO17 1BJ, United Kingdom }
	
	\author{A.~Kusenko\orcidlink{0000-0002-8619-1260}}
	\affiliation{Department of Physics and Astronomy, University of California, Los Angeles, CA,  90095, USA}
	\affiliation{Kavli Institute for the Physics and Mathematics of the Universe (WPI), UTIAS, The University of Tokyo, Kashiwa, Chiba 277-8583, Japan}
	
	% \author{C.~Lemon
		% \affiliation{Oskar Klein Centre, Department of Physics, Stockholm University, SE-106 91 Stockholm, Sweden }
		
		\author{M.~Malkan\orcidlink{0000-0001-6919-1237}}
		\affiliation{Department of Physics and Astronomy, University of California, Los Angeles, CA,  90095, USA}
		
		\author{T.~Morishita\orcidlink{0000-0002-8512-1404}}
		\affiliation{IPAC, California Institute of Technology, MC 314-6, 1200 E. California Boulevard, Pasadena, CA 91125, USA}
		
		\author{V.~Motta\orcidlink{0000-0003-4446-7465}}
		\affiliation{Instituto de F\'{\i}sica y Astronom\'{\i}a, Universidad de Valpara\'{\i}so, Avda. Gran Breta\~na 1111, Valpara\'{\i}so, Chile}
		
		\author{L.~A.~Moustakas\orcidlink{0000-0003-3030-2360}}
		\affiliation{Jet Propulsion Laboratory, California Institute of Technology, 4800 Oak Grove Dr, Pasadena, CA 91109}
		
		\author{W.~Sheu\orcidlink{0000-0003-1889-0227}}
		\affiliation{Department of Physics and Astronomy, University of California, Los Angeles, CA,  90095, USA}
		
		\author{D.~Sluse\orcidlink{0000-0001-6116-2095}}
		\affiliation{STAR Institute, University of Li{\`e}ge, Quartier Agora, All\'ee du six Ao\^ut 19c, 4000 Li\`ege, Belgium}
		
		\author{D.~Stern\orcidlink{0000-0003-2686-9241}}
		\affiliation{Jet Propulsion Laboratory, California Institute of Technology, 4800 Oak Grove Dr, Pasadena, CA 91109}
		
		\author{M.~Stiavelli\orcidlink{0000-0002-8512-1404}}
		\affiliation{Space Telescope Science Institute, 3700 San Martin Drive, Baltimore, MD 21218, USA}
		
		%     \author{R.~H.~Wechsler\orcidlink{0000-0003-2229-011X}}
		% \affiliation{Kavli Institute for Particle Astrophysics \& Cosmology, P.O. Box 2450, Stanford University, Stanford, CA 94305, USA }
		% \affiliation{Department of Physics, Stanford University, 382 Via Pueblo Mall, Stanford, CA 94305, USA }
		% \affiliation{SLAC National Accelerator Laboratory, Menlo Park, CA 94025, USA }
		
		\date{\today}
		
		\begin{abstract}
			We present a measurement of the free-streaming length of dark matter (DM) and subhalo abundance around 28 quadruple image strong lenses using observations from JWST MIRI presented in Paper III of this series. We improve on previous inferences on DM properties from lensed quasars by simultaneously reconstructing extended lensed arcs with image positions and relative magnifications (flux ratios). Our forward modeling framework generates full populations of subhalos, line-of-sight halos, and globular clusters, uses an accurate model for subhalo tidal evolution, and accounts for free-streaming effects on halo abundance and concentration. Modeling lensed arcs leads to more-precise model-predicted flux ratios, breaking covariance between subhalo abundance and the free-streaming scale parameterized by the half-mode mass $m_{\rm{hm}}$. Assuming subhalo abundance predicted by the semi-analytic model {\tt{galacticus}} ($N$-body simulations), we infer (Bayes factor of 10:1) $m_{\rm{hm}} < 10^{7.4} \mathrm{M}_{\odot}$ ($m_{\rm{hm}} < 10^{7.2} \mathrm{M}_{\odot}$), a 0.4 dex improvement relative to omitting lensed arcs. These bounds correspond to lower limits on thermal relic DM particle masses of $6.5$ and $7.4$ keV, respectively. Conversely, assuming DM is cold, we infer a projected mass in subhalos ($10^6 < m/M_{\odot}<10^{10.7}$) of $1.7_{-1.2}^{+2.6} \times 10^7 \ \mathrm{M}_{\odot} \ \rm{kpc^{-2}}$ at $95 \%$ confidence. This is consistent with {\tt{galacticus}} predictions ($0.9 \times 10^7 \mathrm{M}_{\odot} \ \rm{kpc^{-2}}$), but in mild tension with recent $N$-body simulations ($0.6 \times 10^7 \mathrm{M}_{\odot} \ \rm{kpc^{-2}}$). Our results are among the strongest bounds on WDM, and the most precise measurement of subhalo abundance around strong lenses. Further improvements will follow from the large sample of lenses to be discovered by Euclid, Rubin, and Roman.    
		\end{abstract}
		\maketitle
		
		%\tableofcontents
		
		\section{Introduction}
		The concordance cosmological model of Cold Dark Matter plus a cosmological constant, $\Lambda$ ($\Lambda$CDM), includes particle dark matter as the dominant form of  mass in the Universe \citep{Planck++20}. However, dark matter continues to evade direct detection, and we lack a detailed understanding of its particle nature. Gravity remains the only known interaction between dark and baryonic matter, and as such, cosmic probes of dark matter take center stage. 
		
		Many cosmic probes focus on the abundance and internal structure of halos, gravitationally bound concentrations of dark matter that permeate the cosmos and envelop galaxies \citep[for a review, see][]{DrlicaWagner22}. CDM makes clear predictions for the properties of these objects. First, simulations predict that dark matter halos should exhibit a universal density profile \citep[][hereafter the NFW profile]{Navarro++97}. This prediction follows from the collisionless nature of dark matter in CDM. Second, CDM predicts that the cosmic dark matter structure is approximately scale free across many orders of magnitude in halo mass \citep{Wang++20b}. This second prediction is associated with the free-streaming length, a scale determined by the formation mechanism and mass of dark matter particle, which becomes imprinted in cosmic structure \citep{Bond++83}. In CDM, the free-streaming scale is so small that it becomes irrelevant for galactic structure formation; in some theories, the minimum halo mass is comparable to that of a planet. Other classes of dark matter models predict free-streaming lengths of order kpc, which leads to a suppression of structure on halo mass scales $\sim 10^7 - 10^{10} \mathrm{M}_{\odot}$ \citep{Colin++00,AvilaReese++01,Bode++01,Schneider++13}. 
		%These models fall under the category of 
		The canonical such model is warm dark matter (WDM), which
		will be the subject of this analysis. 
		
		Strong gravitational lensing enables inferences of the abundance and internal structure of low-mass halos, distinguishing halo properties predicted by CDM from alternative theories \citep[for a review, see][]{Vegetti++24}. Lensing provides the unique capability to infer halo properties across cosmological distances, including objects in the field, irrespective of their baryonic mass. Alongside complementary probes from dwarf galaxies \citep[e.g.][]{Nadler++21,Correa21,Dekker++22} and stellar streams \citep{Bonaca++19,Banik++21,Nibauer++25}, strong lensing has pushed the observational frontier of halo properties to scales below $10^8 \mathrm{M}_{\odot}$. 
		
		Quadruply imaged quasars, the particular class of lens system considered in this work, can push measurements of halo structure to mass scales below the threshold of galaxy formation $\sim10^7 \mathrm{M}_{\odot}$. In quadruple-image systems, a massive foreground galaxy produces four unresolved images of a background quasar. For a single-plane lens system, such as a foreground galaxy, host dark matter halo, and its associated subhalo population, the relative magnifications (flux ratios) among the images depends on second derivatives of the gravitational potential projected onto the plane of the lens. Through the Poisson equation, the image magnifications therefore depend on the projected mass surface density near an image. Any object that dominates the local projected mass will significantly impact the flux ratio, even if the perturber is orders of magnitude less massive than the main deflector \citep{MaoSchneider98,Nierenberg++14,Nierenberg++17}. In addition to flux-ratio studies, distortions caused by dark matter substructure can be detected in extended lensed arcs. With certain exceptions \citep[e.g.][]{Birrer++17,He++22,Powell++22,Wagner-Carena++24,Riordan25}, these analyses measure the mass and position of individual perturbers, and provide a complementary means to probe dark matter substructure gravitationally \citep{Vegetti++12,Vegetti++14,Hezaveh++16,Minor++21,He++22,Despali++22,Sengul++22,He++22,Minor25,Lange++25,Enzi++25,He++25,Powell++25}. 
		
		Analyses of quadruply imaged quasars characterize properties of dark matter substructure at the population level---the domain where dark matter models predict halo properties. The current state-of-the-art involves using multi-plane ray tracing to compute perturbations to the image flux ratios and image positions from subhalos around the main deflector and halos along the entire line of sight, generating millions of model-predicted datasets per lens \citep{Gilman++19}. These methods can constrain any theory that predicts the abundance and density profile of low-mass halos, including CDM \citep{Dalal++02}, WDM \citep{Hsueh++20,Gilman++19,Gilman++20,Keeley++23,Keeley++24}, self-interacting dark matter \citep{Gilman++21,Gilman++23}, primordial black hole dark matter \citep{Dike23}, fuzzy dark matter \citep{Chan++20,Laroche++22}, and the primordial matter power spectrum \citep{ZentnerBullock03,Gilman++22}. 
		
		Through a Cycle 1 JWST program \citep[GO-2046; PI Nierenberg][]{Nierenberg++21,Nierenberg++24}, we have measured the flux ratios in a sample of 31 strongly lensed quasars, tripling the existing sample size of lens systems suitable for a dark matter inference. The JWST survey remedies a historical limitation of flux-ratio analyses of dark matter substructure associated with stellar microlensing. For sources more compact than a microarcsecond, microlensing by stars can dominate millilensing perturbations by halos~\citep{Sluse++13}. By measuring emission from the warm dust region surrounding the background quasar, which has a typical size of 1--10 pc \citep{Sluse++13}, JWST targets a region large enough to avoid microlensing, but compact enough to experience millilensing. Nuclear narrow-line emission from the background quasar \citep{MoustakasMetcalf03,Nierenberg++20} and radio-loud quasars with extended radio jets \citep{Koopmans++03,McKean++07} also meet the criteria to avoid stellar microlensing. However, radio-loud systems are fairly uncommon, and the size of the nuclear narrow-line region, roughly ten times larger than the warm dust region, makes it a less sensitive probe of low-mass halos. While the nuclear narrow-line region does enable flux-ratio investigations of dark matter substructure \citep[e.g.][]{Nierenberg++14,Nierenberg++17,Gilman++22,Gilman++23}, the new data from JWST push the sensitivity to lower halo mass scales. For example, a single $10^6 \mathrm{M}_{\odot}$ halo, if it is located in projection within $\sim 10 \ $m.a.s. of a lensed image, can cause a change in the image magnification at the level of $\sim 5\%$ for a source size of 5 pc---compared to a perturbation of $< 1 \%$ for a source size of 80 pc, typical of the nuclear narrow-line region \citep{Nierenberg++24}. 
		
		To exploit the JWST warm dust flux ratios to their full potential, we address a limitation of previous dark matter studies with quadruply imaged quasars associated with the lens modeling. Previous analyses used only the image positions and flux ratios to constrain the lens model and infer substructure properties. These data have a limited ability to constrain the mass profile on the main deflector on angular scales comparable to the image separation. As a result, residual lens modeling uncertainties constitute the largest source of statistical uncertainty when predicting image flux ratios, which leads to weaker constraints on substructure properties. In addition, recent analyses have drawn attention to the role of angular structure in the mass profile of the main deflector, such as boxyness or diskyness. If not accounted for in the lens model, these additional degrees of freedom can masquerade as flux-ratio perturbation by halos  \citep{Gilman++17,Hsueh++18,Cohen++24,Gilman++24,Paugnat++25}. 
		
		To address limitations associated with the lens modeling, we analyze the full sample of lenses observed through our JWST program while incorporating constraints from image positions, flux ratios, and extended lensed arcs. Jointly modeling these data tightens the constraints on the mass profile of the main deflector, thus improving the precision of model-predicted flux ratios and in turn enabling more robust inferences of substructure properties. While the modeling of extended lensed arcs has become routine in current strong lensing studies, this work addresses the specific challenge of modeling these data simultaneously with image positions and flux ratios in the presence of potentially tens of thousands of low-mass perturbers. For this work, we compile observations of extended lensed arcs using archival imaging data from the Hubble Space Telescope (HST), as well as new observations of extended lensed arcs from JWST-NIRCam and JWST-MIRI. We use flux ratios and astrometric measurements obtained through JWST GO-2046 \citep{Nierenberg++24,Keeley++24,Keeley++25}, as well as Keck adaptive optics \citep{Nierenberg++14} and HST grism measurements of narrow-line flux ratios \citep{Nierenberg++20}. Regarding the modeling of small-scale structure in the lens system, we include additional sources of flux-ratio perturbation from globular clusters, and angular structure in the main deflector mass profile in the form of elliptical multipoles \citep{PaugnatGilman25}. This analysis closely follows the methodology presented by \citet{Gilman++24}, who validated the methods on simulated datasets. 
		
		This is the fourth publication in a series of papers \citep{Nierenberg++24,Keeley++24} that present results from the JWST lensed quasar dark matter survey. This work appears at the same time as \citet{Keeley++25}, which presents the flux-ratio measurements for the full sample, and a WDM interpretation using inference methodology presented by \citet{Gilman++19,Gilman++20} that models only the image positions and flux ratios. In this paper, we consider the same models for dark matter substructure, globular clusters, and the main deflector mass profile as \citet{Keeley++25}, but incorporate the additional observational constraints from the extended lensed arcs to demonstrate the improved constraining power they provide on dark matter properties. In future papers in this series, we will extend the analysis presented in this work to other dark matter and early Universe models that alter halo abundance and internal structure. 
		
		This paper is organized as follows. In Section~\ref{sec:inference}, we discuss the Bayesian inference framework used to infer substructure properties from a sample of lensed quasars. In Section~\ref{sec:data} we discuss the datasets used in this analysis. Section~\ref{sec:dmmodel} details the modeling of dark matter substructure, including free-streaming effects in WDM and the tidal evolution of subhalos. In Section~\ref{sec:lensmodeling}, we describe aspects of the lens modeling procedure, including the general approach and specific considerations for individual systems. Section \ref{sec:results} presents the results of the analysis, including constraints on the free-streaming length of dark matter and a measurement of the amplitude of the subhalo mass function in CDM. We summarize our findings and provide concluding remarks in Section \ref{sec:discussion}. 
		
		Throughout this work, we assume cosmological parameters as measured by \citep{Planck++20}, although we note that our dark matter inference is insensitive to small changes in cosmological parameters. %, with $\Omega_{M}=0.31$, $H_0 = 67.5 \ \rm{km}\  \rm{s^{-1}} \ \rm{Mpc^{-1}}$, $\sigma_8 = 0.81$, and $n_s = 0.965$%.
		We perform gravitational lensing calculations using {\tt{lenstronomy}}\footnote{\url{https://github.com/lenstronomy/lenstronomy}}\citep{Birrer++18,Birrer++21}. We render (sub)halo and globular cluster populations with {\tt{pyHalo}}\footnote{\url{https://github.com/dangilman/pyHalo}}\citep{Gilman++20}. We use {\tt{samana}}, an open-source code that wraps {\tt{pyHalo}} and {\tt{lenstronomy}}, to perform our statistical inference of halo properties. We make analysis scripts to reproduce the results of this analysis, together with the posterior distributions\footnote{\url{https://github.com/dangilman/samana/tree/main/notebooks/jwst_DM_survey_IV}} and notebooks that present lens models for 24 quadruply imaged quasars\footnote{\url{https://github.com/dangilman/samana/tree/main/notebooks/baseline_lensmodels}} publicly available. 
		
		\section{Bayesian inference framework}
		\label{sec:inference}
		The goal of this analysis is to calculate the posterior probability distribution of a set of hyper-parameters that describe dark matter substructure, $\qsub$, given a sample of quadruply imaged quasars. As hyper-parameters, these quantities do not have one-to-one connections to observables; instead, they specify probability distributions for parameters that describe full populations, or \textit{realizations}, of dark matter halos and subhalos. For the analysis in this paper, $\qsub$ specifies the free-streaming length of dark matter, among other quantities, but we can apply the methodology presented in this section to consider any other structure formation or dark matter model. We will therefore present the methodology in fully general terms, and discuss the particular structure formation model used in this work in Section \ref{sec:dmmodel}. The techniques described in this section closely follow previous analyses, which involved extensive tests on simulated datasets in a variety of dark matter theories \citep[e.g.][]{Gilman++19,Gilman++20,Gilman++21,Gilman++24}. 
		
		The Bayes theorem for the hyper-parameters is
		\begin{equation}
			\label{eqn:bayestheorem}
			p \left(\qsub | \data \right) \propto \pi \left(\qsub \right) \prod_{n=1}^{N} \mathcal{L}\left(\datan | \qsub \right).
		\end{equation}
		Here, $\data$ represents the statistically-independent observations of a sample of $N$ strongly-lensed quasars. For each individual lens, the data vector $\datan$ consists of the following observations:
		\begin{itemize}
			\item $\dptsrc$ represents the relative image positions between the four images of the lensed quasar. 
			\item $\dfr$ represents the three flux ratios (or relative magnifications) among the four lensed images. 
			\item $\dimg$ represents the imaging data, or lensed arc, which partially encircles the main deflector.
		\end{itemize}
		Previous studies of dark matter using image flux ratios used only the point source positions $\dptsrc$ and flux ratios $\dfr$. In this work, we extend this methodology to jointly model the image positions and flux ratios with extended arcs, closely follow the methodology presented by \citet{Gilman++24}.  
		
		In the context of evaluating the likelihood function for each lens in Equation \ref{eqn:likelihood}, we begin by introducing notation: 
		\begin{itemize}
			\item $\qsub$: A set of hyper-parameters that describe a structure formation or dark matter model, such as the normalization of the halo mass function or the free-streaming length of dark matter. 
			\item $\rsub$: A realization of dark matter halos, subhalos, and other small-scale structures, such as globular clusters. These parameters specify the mass, redshift, angular position, and density profiles of a static population of perturbers. 
			\item $\xlight$: Parameters that describe the lens and source light profiles. 
			\item $\xlens$: Parameters that describe the lens macromodel, or the combined baryonic and dark matter density profile of the main deflector, as well as the combined luminous and dark matter from satellite galaxies detected near main deflector.  
		\end{itemize}
		The likelihood function for an individual system involves an integral over the nuisance parameters $\xlight$ and $\xlens$, and all possible configurations $\rsub$ corresponding to the model $\qsub$:
		\begin{widetext}
			\begin{eqnarray}
				\label{eqn:likelihoodfull}
				\nonumber  \mathcal{L}\left(\datan | \qsub \right) =  \int p\left(\datan | \rsub, \xlight, \xlens \right)  p\left(\rsub | \qsub \right) p\left(\xlight\right)  p\left(\xlens\right) d \xlight d \xlens d \rsub.\\
			\end{eqnarray}
		\end{widetext}
		The quantities $p\left(\xlight\right)$ and $p\left(\xlens \right)$ are priors on the lens and source light profiles, and the macromodel parameters, respectively. The term $p\left(\rsub | \qsub \right)$ represents the probability of having a population of halos $\rsub$ in a universe described by $\qsub$. We account for this term by using {\tt{pyHalo}} to generate realizations $\rsub$ according to the predictions of the dark matter model specified by $\qsub$. 
		
		Equation \ref{eqn:likelihood} requires that we reconstruct the image positions, flux ratios, and imaging data simultaneously, in the presence of a particular substructure realization $\rsub$. This leads to concerns related to systematic biases associated with modeling imaging data with dark matter substructure, a known challenge in strong lens modeling \citep[e.g.][]{Nightingale++18,Galan++24,Gilman++24,Ephremidze++25}. To avoid potential bias, we will instead propagate constraints from imaging data onto the dark matter inference through importance sampling weights on the macromodel parameters. With this approximation the likelihood function becomes
		\begin{widetext}
			\begin{eqnarray}
				\nonumber  \mathcal{L}\left(\datan | \qsub \right)= \int w_{\rm{img}}\left(\xlens | \dptsrc, \dimg \right) p\left(\dptsrc, \dfr | \rsub,\xlens \right)  p\left(\rsub | \qsub \right) p\left(\xlight\right)  p\left(\xlens\right) d \xlight d \xlens d \bf{{\rm{r}}}\\
				\label{eqn:likelihood}
			\end{eqnarray}
		\end{widetext}
		where $w_{\rm{img}}$ are the importance sampling weights on the macromodel parameters $\xlens$, which depend on $\dptsrc$ and $\dimg$, but not on $\dfr$. In this framework, we use the lensed arcs to constrain the mass profile of the main deflector. This leads to more-precise model-predicted flux ratios, which enables stronger constraints on the nature of dark matter. 
		
		To evaluate Equation \ref{eqn:likelihood}, we use an end-to-end forward modeling framework that accounts for the relevant physical processes: strong lensing by the main deflector and nearby satellite galaxies, small-scale perturbations by dark matter halos and globular clusters, small-scale deviations from ellipticity in the main deflector mass profile, and statistical measurement uncertainties in the observations. In these calculations, we forward model the data with potentially tens-of-thousands of halos in the lens model, both at the redshift of the main deflector and along the line of sight.
		
		Equation \ref{eqn:likelihoodfull} includes an integral over all possible main deflector and substructure models. Some configurations of the main deflector mass profile and (sub)halo populations produce nearly identical image positions and flux ratios, and constitute a set of lensing degeneracies. Some well known examples of lensing degeneracies, such as the mass sheet transformation, modify the underlying mass profile while leaving some observables, including the flux ratios, unchanged. To break these degeneracies, one can introduce additional data, or impose additional assumptions on the lens models used in the analysis. In this work, we use both approaches. As discussed in Sections \ref{sec:lensmodeling} and \ref{sec:dmmodel}, our analysis considers a subset of all possible lens models that are consistent with observed properties of massive elliptical galaxies and the (sub)halo density profiles predicted by cold and warm dark matter. Further, as discussed in Section \ref{ssec:casestudies}, the simultaneous modeling of lensed arcs alongside image positions and flux ratios breaks some degeneracies between the main deflector mass profile and the (sub)halo population. As prescribed by Equation \ref{eqn:likelihood}, we simultaneously vary the parameters of these components to marginalize over any remaining covariance between the dark matter parameters and the nuisance parameters, within the family of lens models we consider physically plausible.
		
		The following subsections review the methods we use to evaluate Equation \ref{eqn:likelihood}. We begin in Section \ref{ssec:lensing} with a discussion of the gravitational lensing formalism and multi-plane ray-tracing we use to compute observables. Section \ref{ssec:likelihood} describes how we evaluate the likelihood function by forward modeling the image positions, flux ratios, and how we compute the importance sampling weights $w_{\rm{img}}$. Section \ref{ssec:imgdatareconstruction} provides details about how we reconstruct the lensed arcs during lens modeling calculations. 
		
		\subsection{Gravitational lensing formalism}
		\label{ssec:lensing}
		We perform lensing calculations and backwards ray tracing using the multi-plane lens equation \citep{BlandfordNarayan86}
		\begin{equation}
			\label{eqn:lenseqn}
			{\boldsymbol{\theta}}_{\rm{n+1}}= \boldsymbol{\theta} - \frac{1}{D_{\rm{s}}} \sum_{i=1}^{n} D_{\rm{is}}{\boldsymbol{\alpha}_{\rm{i}}} \left(D_{\rm{i}} \boldsymbol{\theta}_{\rm{i}}\right),
		\end{equation}
		where $\boldsymbol{{\theta}}$ represents an angular coordinate from the perspective of an observer, $\boldsymbol{\theta}_{\rm{n}}$ represents the angular position of a light ray at the $n$th lens plane, $\boldsymbol{\alpha}_{\rm{i}}$ is a deflection field at the $i$th lens plane, $D_{\rm{s}}$ is the angular diameter distance to the source plane and $D_{ij}$ is the angular diameter distance between lens planes $i$ and $j$. 
		
		Throughout this paper we will show convergence\footnote{Convergence is the projected mass in units of the critical density for lensing \citep{SEF92}.} maps of dark matter substructure. We define the convergence, $\kappa$, for a multi-plane lens system $\kappa \equiv \frac{1}{2}\bf{\nabla} \cdot \boldsymbol{\alpha}$, which is equivalent to the Poisson equation $2 \kappa = \nabla^2 \Psi$ for a single-plane lens system. We compute the substructure convergence, $\kappa_{\rm{DM}}$, by subtracting the contribution from the main deflector mass model
		\begin{equation}
			\label{eqn:effectivekappa}
			\kappa_{\rm{DM}} = \frac{1}{2}\bf{\nabla} \cdot \boldsymbol{\alpha} - \kappa_{\rm{macro}}.
		\end{equation}
		The convergence maps are computed relative to the mean dark matter density, so some regions will appear under-dense relative to the average.
		
		Exact ray-tracing calculations with Equation \ref{eqn:lenseqn} become prohibitively expensive when including tens-of-thousands of line-of-sight halos in the lens model. The problem stems from the recursive nature of Equation \ref{eqn:lenseqn} and the presence of background halos, objects behind the main deflector but in front of the source, which produce a deflection field $\boldsymbol{\alpha_{\beta}}$. In contrast to the population of foreground halos and subhalos at the lens redshift, for which we only need to calculate deflections once for a given $\boldsymbol{\theta}$, the recursive nature of Equation \ref{eqn:lenseqn} requires that we continuously reevaluate $\boldsymbol{\alpha_{\beta}}$ for any new proposal of the macromodel parameters. The computation time for a lens mass and source light reconstruction therefore scales with the number of background halos, which can be of order $10^4$. 
		
		To make the problem tractable, we use the decoupled multi-plane formalism, an approximation for full multi-plane ray tracing introduced by \citet{Gilman++24}. Using an initial proposal for the deflection field produced by the main deflector, $\boldsymbol{\hat{\alpha}}_{\rm{m}}$, we ray-trace through the lens system to the source plane using Equation \ref{eqn:lenseqn}. For the initial proposal of $\boldsymbol{\hat{\alpha}}_{\rm{m}}$, we choose a deflection field that satisfies the lens equation for the four image positions without substructure in the lens model. From this initial ray-tracing calculation, we compute the deflection field caused by halos behind the main deflector, $\boldsymbol{\hat{\alpha}_{\beta}}\equiv\boldsymbol{\alpha_{\beta}}\left(\boldsymbol{\theta},\boldsymbol{\hat{\alpha}_{\rm{m}}} \right)$, which map to source plane coordinates $\hat{\boldsymbol{\beta}}$. 
		
		The decoupled multi-plane approximation involves the substitution of  
		$\boldsymbol{\hat{\alpha}_{\beta}}$ in place of $\boldsymbol{\alpha_{\beta}}$ for any subsequent lensing calculation involving the same population of halos, breaking the recursive nature of Equation \ref{eqn:lenseqn}. The deflection field that maps from the main lens plane to the source plane becomes $\boldsymbol{\alpha}_{\rm{m}}+\boldsymbol{\hat{\alpha}_{\beta}}$, where $\boldsymbol{\alpha_{\rm{m}}}$ is a deflection field produced by the macromodel, and the lens equation can be written
		\begin{equation}
			\label{eqn:lenseqndecoupled}
			\boldsymbol{\beta}= \boldsymbol{\hat{\beta}} + \frac{T_{\rm{ds}}}{T_{\rm{s}}} \left(\boldsymbol{\hat{\alpha}_{\rm{m}}} - \boldsymbol{\alpha_{\rm{m}}}\right),
		\end{equation}
		where $T_{\rm{ds}}$ is the transverse comoving distance between the main deflector lens plane and the source plane, $T_{\rm{s}}$ is the transverse comoving distance to the source plane, and $\boldsymbol{\beta}$ are coordinates on the source plane. Because Equation \ref{eqn:lenseqndecoupled} depends linearly on $\boldsymbol{\alpha_{\rm{m}}}$, and does not involve recursive calculations with potentially thousands of line-of-sight halos, it accelerates lensing calculations by factors of several thousand relative to full ray tracing. However, since we compute $\boldsymbol{\hat{\alpha}_{\beta}}$ with exact ray tracing, the approximation preserves the non-linear features of multi-plane lensing that could affect the flux ratios, and reduces to exact ray tracing when $\boldsymbol{\hat{\alpha}_{\rm{m}}}=\boldsymbol{{\alpha}_{\rm{m}}}$. For additional discussion on the validity and accuracy of this approximation we refer to \citet{Gilman++24}. 
		\begin{figure*}
			\centering
			\includegraphics[trim=2cm 0.4cm 1cm
			0cm,width=0.9\textwidth]{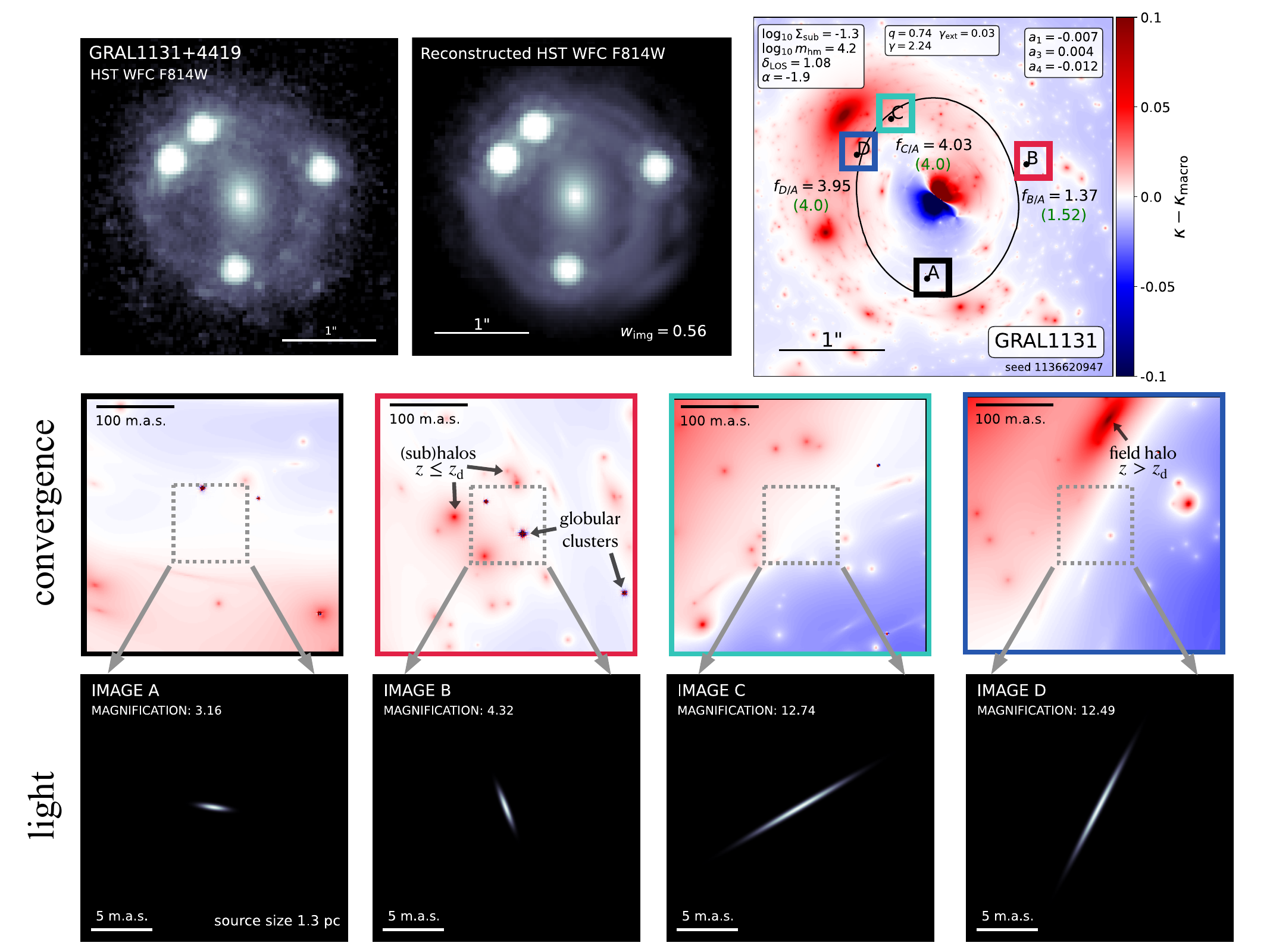}
			\caption{\label{fig:1131zooms} Illustrations of the various steps in the calculation of the likelihood function, as detailed in Section \ref{ssec:likelihood}. This example shows one lens model constructed for GRAL1131+4419. The calculations depicted here are performed millions of times per lens to compute the likelihood function (Equation \ref{eqn:likelihood}). {\bf{Top row:}} Observed (left) and reconstructed (center) image of GRAL1131-4419. The imaging data importance weight (Equation \ref{eqn:imageweights}) is given in the center panel. On the right we show effective convergence (Equation \ref{eqn:effectivekappa}) for this lens model. The dark matter hyper-parameters $\qsub$ (see Section \ref{sec:dmmodel} for details on the dark matter model) and the strength of the multipole perturbations to the macromodel (see Section \ref{ssec:macromodel} for discussion on the macromodel) are quoted in the upper left and right corners, respectively. The black line is the critical curve. {\bf{Center row:}} Zoomed in regions around each lensed image showing the effective convergence near the image. Subhalos and field halos at $z \leq z_{\rm{d}}$ appear round in the convergence maps, while halos behind the main deflector, whose lensing effects are amplified by the macromodel, appear elongated along the tangential direction of the critical curve. {\bf{Bottom row:}} Model-predicted quasar images obtained by ray tracing through the lens system to a background source with a size 1.3 pc, characteristic of the compact warm dust region from which we measure flux ratios. We compute image magnifications (quoted in the top left) by computing the total integrated flux on a high resolution grid (416 by 416 pixels, resolution 0.00011 arcsec/pixel). The panels show how a lensed image would appear, as seen through a telescope with exquisite angular resolution and deconvolution with a perfectly-known PSF. }
		\end{figure*}
		\subsection{Calculation of the likelihood function}
		\label{ssec:likelihood}
		Using the decoupled multi-plane formalism allows us to predict observables with ray-tracing calculations involving full populations of subhalos and line-of-sight halos. In the remainder of this section, we describe our approach for using this machinery to compute the likelihood function in Equation \ref{eqn:likelihood}. 
		
		We will evaluate the integral in Equation \ref{eqn:likelihood} by drawing samples of $\qsub$ from $\pi_{S}\left(\qsub\right)$, or the \textit{sampling distribution}, on $\qsub$, and then weighting these samples based on the probability of measuring the data $\datan$, given the dark matter model corresponding to the $\qsub$ sample. The sampling distribution is always implemented as a uniform (or log-uniform) distribution $\pi_{S}\left(\qsub\right) = \mathcal{U}\left(\qsub_{\rm{min}},\qsub_{\rm{max}}\right)$ for each hyper-parameter. For each draw of $\qsub$ from $\pi_{S}\left(\qsub\right)$, we generate a realization of halos and other small scale perturbers, $\rsub$, using {\tt{pyHalo}}. We then create model-predicted datasets in the presence of the static population of objects specified by $\rsub$. Writing these datasets $\datanprime = \datanprime \left(\rsub,\xlens,\xlight\right)$, we can write the likelihoods as $p\left(\datan | \rsub, \xlight, \xlens \right) = p\left(\datan | \datanprime \right)$. Provided we can calculate $p\left(\datan | \datanprime\right)$, we can evaluate the integral for each lens by generating model-predicted datasets for many realizations. The number of simulations required for each lens depends on the dimension of the $\qsub$ parameter space, with what frequency model-predicted datasets approximately match the observations, and the complexity of the probability landscape. In this work, we generate between 0.3 and 21 million realizations per lens to evaluate the integral. Appendix \ref{app:convergence} discusses convergence tests that show we have simulated long enough to adequately explore the parameter space.
		
		We will now describe in additional detail how we evaluate the likelihood of the observed data given of the model parameters $\rsub$, $\xlens$, and $\xlight$, using our forward model for $\datanprime$. Once we have calculated the likelihood following the approach outlined in the following sections, we use a Gaussian kernel density estimator to obtain a continuous approximation for each $\mathcal{L}\left(\datan | \qsub\right)$ before taking the product, following Equation \ref{eqn:bayestheorem}. We discuss some technical aspects of the kernel density estimation in Appendix~\ref{app:kde}. 
		
		Throughout the following sections we will make frequent references to Figure \ref{fig:1131zooms}, which illustrates various steps in the forward modeling procedure for one lens model reconstruction of the system GRAL1131. We will discuss what appears in each panel and how it relates to the forward modeling procedure as we describe the calculation of astrometric likelihood $p\left(\dptsrc | \dptsrcprime \right)$, flux ratio likelihood $p\left(\dfr | \dfrprime \right)$, and imaging data likelihoods $p\left(\dimg | \dimgprime \right)$. As the measurements are statistically independent we will sometimes use the shorthand notation $p\left(\dptsrc, \dfr, \dimg | \dptsrcprime, \dfrprime, \dimgprime \right) = p\left(\dptsrc | \dptsrcprime\right) \times p\left(\dfr | \dfrprime \right) \times p\left(\dimg | \dimgprime \right)$.
		
		\subsubsection{The astrometric likelihood}
		\label{sssec:astrometriclike}
		The astrometric likelihood, $p\left(\dptsrc | \dptsrcprime \right)$, depends on the model-predicted image positions, which in turn depend on the substructure realization $\rsub$ and the lens macromodel $\xlens$, i.e. $\dptsrcprime \left(\rsub,\xlens\right)$. The relative image positions are measured with a precision of $\sim 5$ milliarcseconds for the systems in our sample \citep{Schmidt++23,Keeley++24,Keeley++25}. This immediately poses a computational challenge in the forward model, as most random configurations of the macromodel and the halo population will not match the observations to this level of precision. 
		
		To handle the astrometric likelihood, we solve for a subset of macromodel parameters $\xlens$ such that the deflection field satisfies the lens equation for the observed image positions in the presence of the full population of halos. For this task, we use the COBYQA\footnote{https://www.cobyqa.com/stable/} optimization routine \citep{rago_thesis,razh_cobyqa}. As we will discuss further in Section \ref{ssec:macromodel}, a subset of $\xlens$ parameters are held fixed during this process, and a subset are allowed to vary in order to satisfy the lens equation\footnote{The logarithmic profile slope, projected mass axis ratio, the orientation and strength of multipole perturbations to an elliptical power-law profile, and the mass and positions of satellite or companion galaxies are held fixed to values sampled from a macromodel prior $p\left(\xlens\right)$ during this calculation, while the remaining parameters in $\xlens$ are then solved for to satisfy the lens equation.}. We handle statistical measurement uncertainties in the image positions by adding random astrometric perturbations to each image position, and require that the macromodel satisfy the lens equation for these perturbed coordinates. With this approach, all of the lens models generated in our forward model satisfy the lens equation to high precision for the measured image positions. Typically, the non-linear solver will find a set of macromodel parameters that solves the lens equation for a given population of halos. However, in models with very high (sub)halo abundance, the solver will occasionally not find a solution. In these cases, we reject the proposed lens model, generate a new realization, and restart the procedure.  
		
		At this stage, we have used only astrometric information $\dptsrc$ and we have solved for a set of macromodel parameters that satisfy the lens equation for the observed image positions, plus astrometric perturbations based on statistical measurement uncertainties of 5 milliarcseconds. Our model reproduces the relative point-source locations of a given lens, one example of which is shown in the top-left panel of Figure \ref{fig:1131zooms} for the system GRAL1131. For this lens model, we can also generate the full effective convergence map $\kappa_{\rm{DM}}$ (see Equation \ref{eqn:effectivekappa}) in dark matter substructure for this realization, one example of which is shown in the top right panel of Figure \ref{fig:1131zooms}. The dark matter and macromodel parameters (see Sections \ref{sec:dmmodel} and \ref{ssec:macromodel}, respectively, for details) used to create this lens model and population of halos are quoted in the white boxes in the panel showing the effective convergence map. In the middle row, we zoom in around each lensed image to show the dark matter subhalos, line-of-sight halos, and globular clusters that appear, in projection, near each lensed image. We now proceed to compute the image magnifications, accounting for these various sources of small-scale perturbation. 
		
		\subsubsection{The flux ratio likelihood}
		\label{ssec:frlike}
		We calculate magnifications of a compact, but finite-size, area around the background quasar. As discussed in Section \ref{sssec:sourcelight}, the finite-size of the lensed background source washes out contamination from stellar microlensing, but also affects the strength of millilensing perturbations by halos. We model the emission region around the quasar as a circular Gaussian light profile with a full-width at half-maximum that depends on the size of the emission region, as discussed further in Section \ref{ssec:lmspecific}. To compute the image magnifications, we ray trace on a high-resolution grid around each observed image coordinate. The panels in the bottom row of Figure~\ref{fig:1131zooms} show how each lensed image appears when we calculate its magnification. The small arcs show how the lensed quasar images would appear if they were observed by a telescope with 0.00011 arcsec-per-pixel spatial resolution, after deconvolution with a perfectly-known point spread function. The image magnifications are listed in the top-left of each panel. 
		
		For each lens system, the measurement uncertainties on the flux ratios are correlated. As discussed by \citet{Keeley++25}, the uncertainties are approximately Gaussian, so we compute the flux-ratio likelihood as 
		\begin{equation}
			\label{eqn:frlike}
			p\left(\dfr | \dfrprime \right) = \frac{\exp\left[ \left(\dfr -  \dfrprime \right)^{\rm{T}} \bf{\Sigma_{\rm{fr}}}^{-1} \left(\dfr -  \dfrprime \right)\right]}{\sqrt{\left(2 \pi\right)^3| \bf{\Sigma_{\rm{fr}}}|}}.
		\end{equation}
		where $\bf{\Sigma_{\rm{fr}}}$ is the covariance matrix of the measured flux ratios.
		
		The calculation of the flux ratio likelihood differs from previous inferences, which used Approximate Bayesian Computing (ABC) to approximate a flux ratio likelihood. Previously, we used ABC to deal with narrow-line flux measurements, which have non-Gaussian uncertainties when flux uncertainties are propagated onto the flux ratios. The Gaussian covariance matrix for the flux ratios measured by JWST obviates the need for using ABC to handle the error propagation. In the ABC approach, we add perturbations to the model-predicted fluxes or flux ratios, and then compute a summary statistic using the perturbed model-predicted flux ratios 
		\begin{equation}
			S = || \dfr -  \dfrprime||,
		\end{equation}
		which is simply the metric distance between the model-predicted and measured data. One then rejects samples of $\qsub$ if $S >  \epsilon$, where $\epsilon$ is a tolerance threshold. 
		
		We apply the ABC approach to one system in our sample, J0924, because the flux ratios deviate strongly from the predictions of a smooth lens model. As a result, the flux ratio likelihood heavily downweights the majority of lens models. To generate enough samples to fully explore the parameter space we use the ABC approximation of the exact flux ratio likelihood. With ABC, we can guarantee that we have enough samples to explore the parameter space. We accept the 6,000 samples corresponding to the smallest $S$ out of $\sim 6,000,000$ total lens models. This amounts to an implicit choice of $\epsilon$.  
		
		\subsubsection{Imaging data importance sampling}
		\label{ssec:imglike}
		In this work, we use the imaging data to constrain the macromodel and the mass profile of the main deflector across angular scales significantly larger than the size of a pixel. Using lensed arcs to detect individual (sub)halos typically involves pixel-based source reconstructions, computations of sensitivity functions across the image plane -- essentially a calibration of the likelihood function given choices made in the modeling -- supersampling of the imaging data, and in some cases, modeling of perturber light profiles \citep{Vegetti++12,Hezaveh++16,Springola++18,Despali++22,Nightingale++24,Powell++22,Ballard++24,He++25,Minor25,McKean++25}. Our treatment of the imaging data is not intended to provide meaningful insights into the small-scale structure of particular lens systems, so we do not take the approach of reconstructing the lensed arcs for each system generated in the forward model.  
		
		As discussed at the beginning of this section, we use importance sampling weights on the macromodel parameters $\xlens$ to propagate constraints from imaging data onto the dark matter inference. We perform the calculation of the importance weights, $w_{\rm{img}}$, separately from the simulations used to evaluate the flux ratio likelihood, and then re-weight the astrometric and flux ratio likelihood, $p\left(\dptsrc, \dfr | \rsub, \xlens \right)$, by $w_{\rm{img}}$. In terms of model-predicted datasets $\dimgprime\left(\xlens, \xlight, \rsub \right)$ and $\dptsrcprime \left(\xlens, \rsub \right)$, we define the importance weights as
		\begin{widetext}
			\begin{equation}
				\label{eqn:imageweights}
				w_{\rm{img}}\left(\xlens | \dptsrc, \dimg \right) = \int \left[\frac{p\left(\xlens, \xlight, \rsub | \dimg, \dptsrc \right)}{p\left(\xlens, \xlight, \rsub | \dptsrc \right)}\right] p\left(\rsub | \qsub \right)\pi_{S} \left(\qsub\right) p\left(\xlens \right) p\left(\xlight \right) d \rsub \ d \qsub. 
			\end{equation}
		\end{widetext}
		The term in brackets represents the additional information provided by reconstructing the lensed arcs. It is the likelihood of the macromodel parameters obtained from imaging data plus astrometry, relative to only astrometry. Without imaging data, this term evaluates to one, the integral evaluates to a constant, and the likelihood function (Equation \ref{eqn:likelihood}) reduces to what we would obtain with only knowledge of image positions and flux ratios. 
		\begin{figure*}
			\includegraphics[trim=1cm 0.5cm 0cm
			1cm,width=0.49\textwidth]{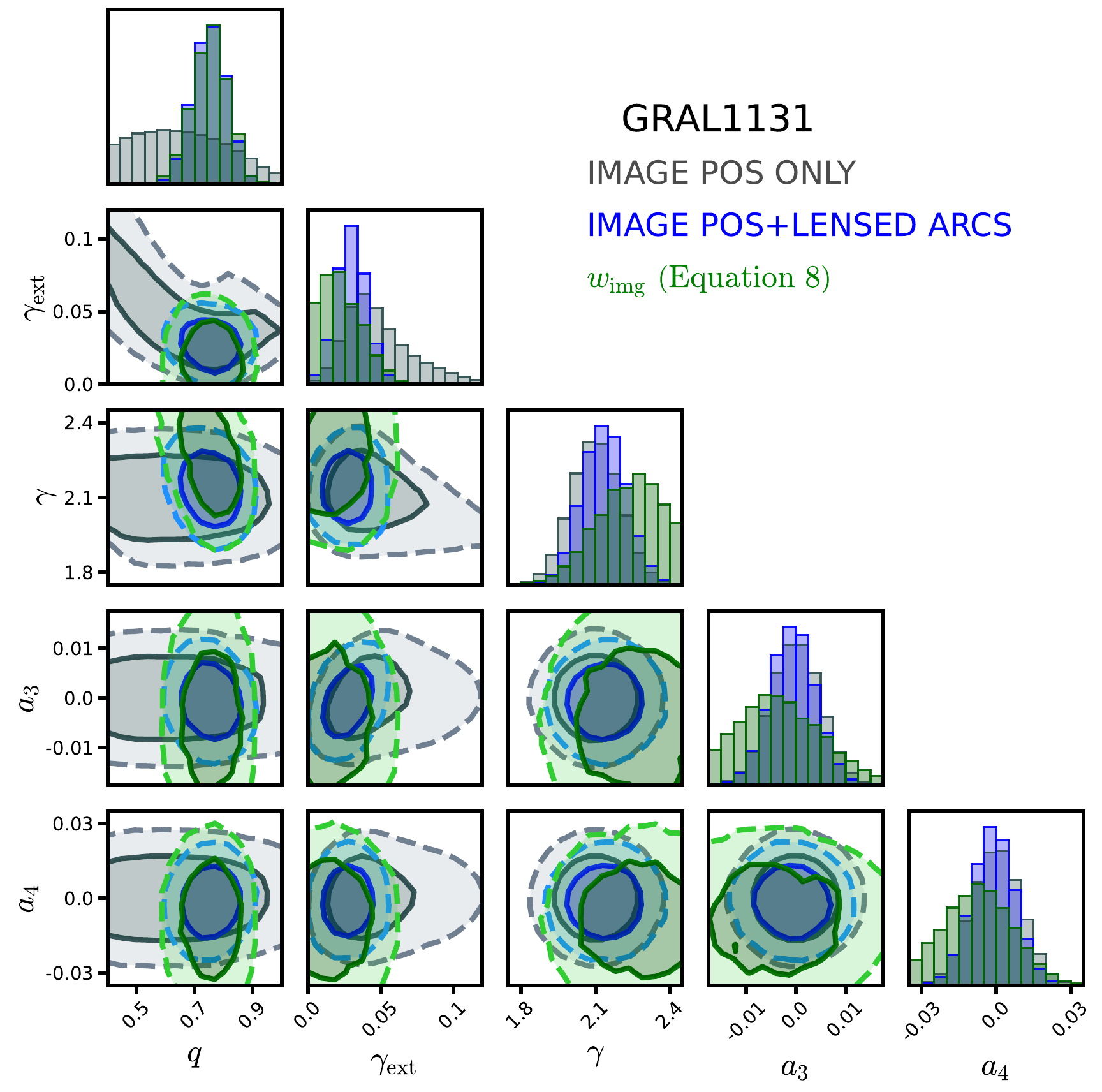}
			\includegraphics[trim=0.5cm 0.5cm 1cm
			0cm,width=0.46\textwidth]{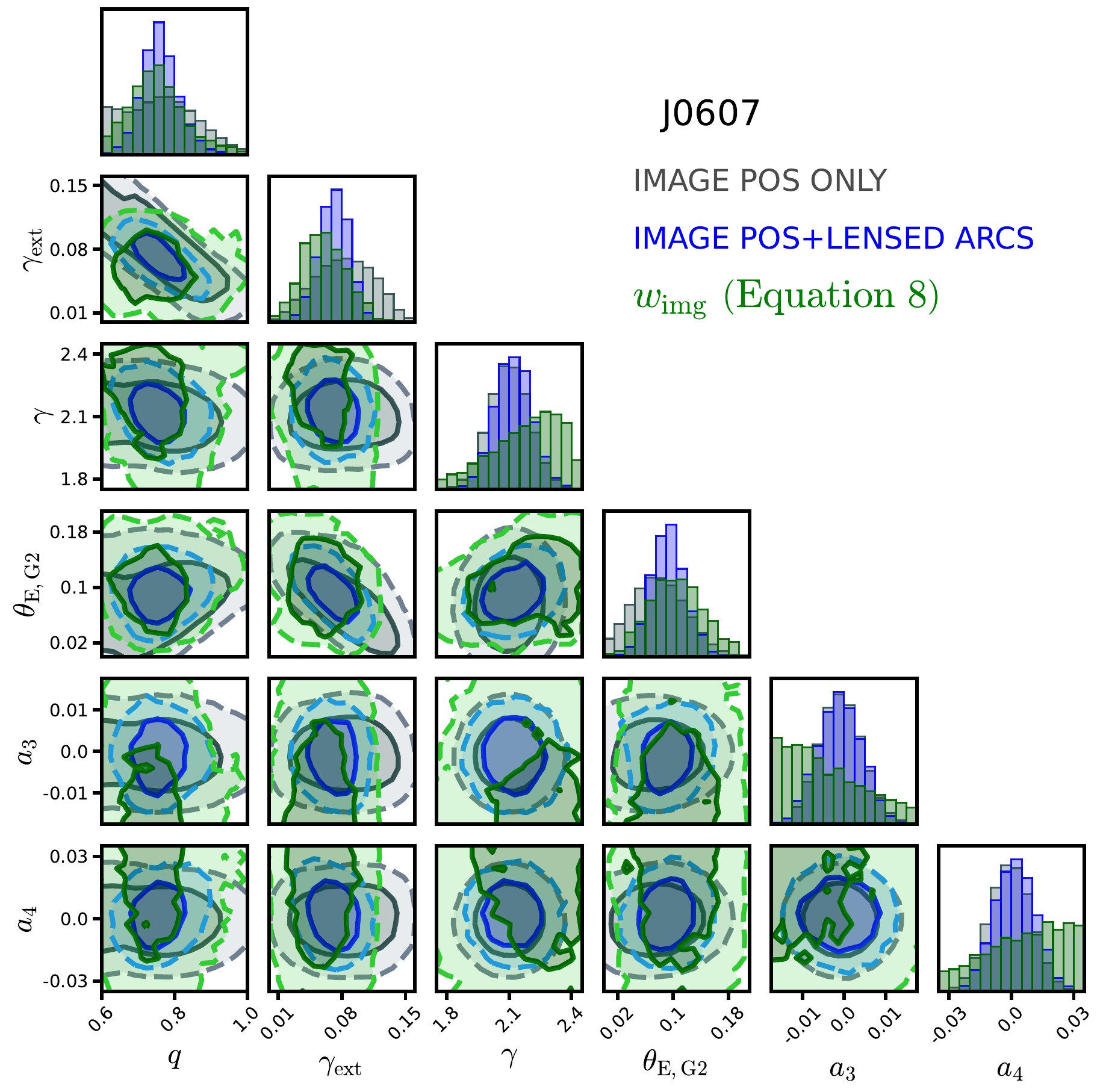}
			\caption{\label{fig:wimgpdfs} Distributions of macromodel parameters that illustrate the role of the importance sampling weights (Equation \ref{eqn:imageweights}) used to incorporate constraints from lensed arcs. Solid (dashed) contours are $68 \%$ ($95 \%$) confidence intervals. {\bf{Left:}} Parameters inferred for GRAL1131 (see Figure \ref{fig:1131zooms}) include the main deflector axis ratio $q$, external shear strength $\gamma_{\rm{ext}}$, the logarithmic profile slope of the main deflector mass profile $\gamma$, the strength of $m=3$ and $m=4$ multipole perturbations, and the mass of satellite galaxy near J0607, $\theta_{\rm{E},G2}$. The gray distribution shows the prior $p\left(\xlens\right)$ updated with constraints from the astrometric likelihood. The blue distributions shows the inference from image positions and lensed arcs. The green distribution represents the importance weights $w_{\rm{img}}$ defined in Equation \ref{eqn:imageweights}, marginalized over $\xlight$, that encode new information from the imaging data (gray $\times$ green = blue).  {\bf{Right:}} The same as the left panel, but showing parameters for J0607 (see second row of Figure \ref{fig:mosaic}), including the mass of its satellite galaxy $\theta_{\rm{E},G2}$.}
		\end{figure*}
		\begin{figure*}
			\includegraphics[trim=1cm 0.5cm 0cm
			1.2cm,width=0.46\textwidth]{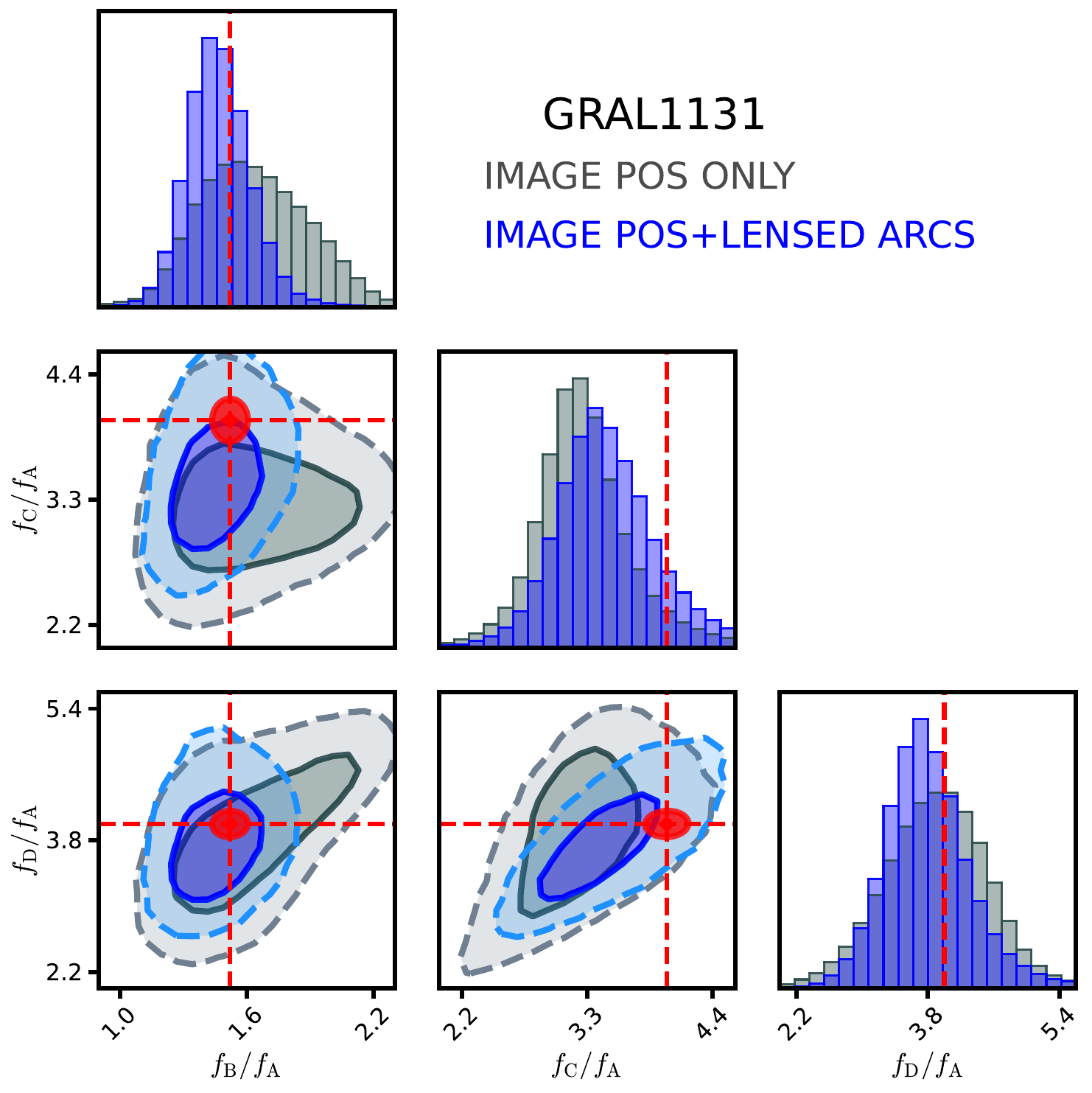}
			\includegraphics[trim=0.5cm 0.5cm 1cm
			1.5cm,width=0.46\textwidth]{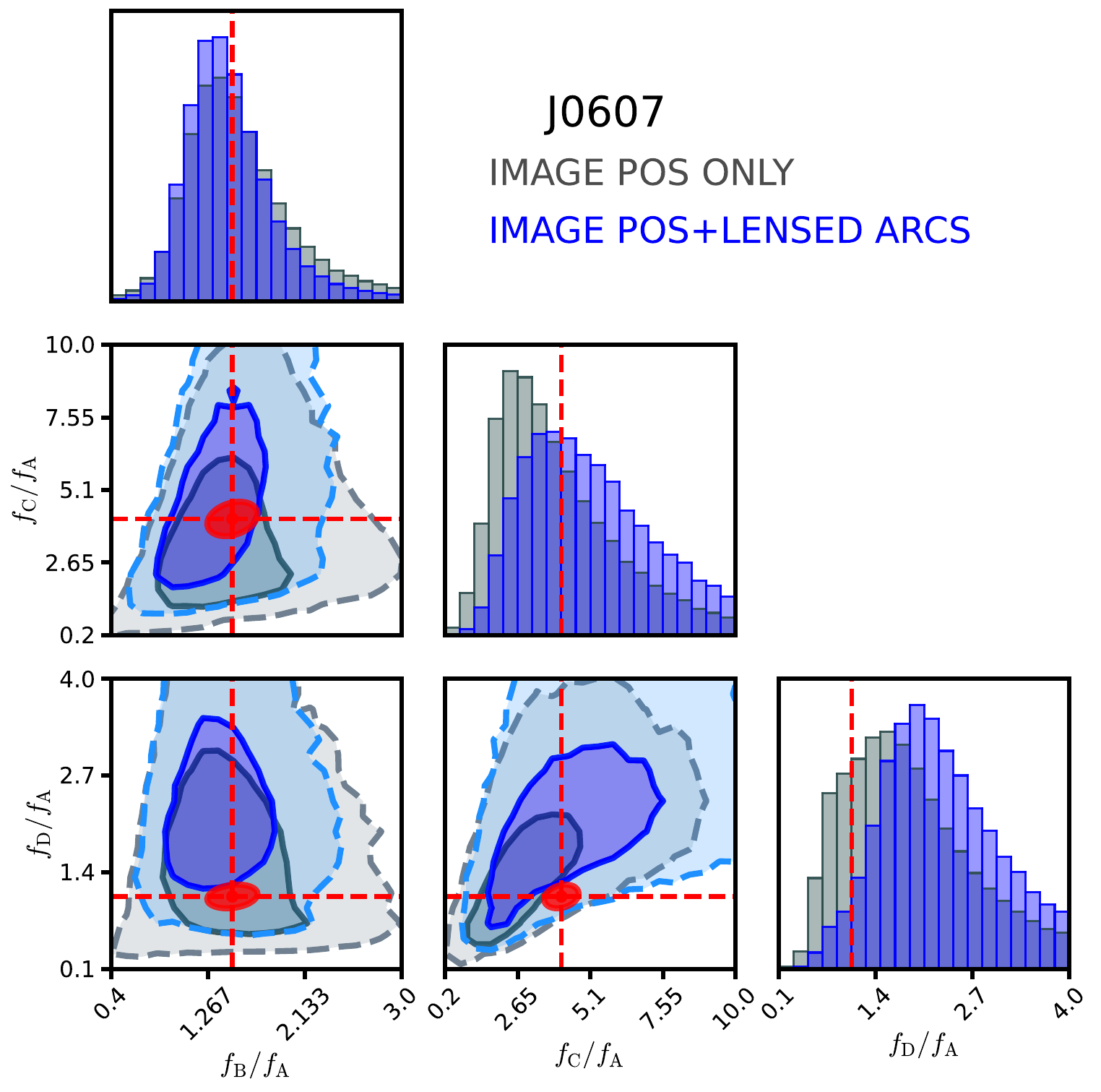}
			\caption{\label{fig:frwimg} The model-predicted flux ratios for GRAL1131 (left) and J0607 (right) obtained by propagating the image data importance weights $w_{\rm{img}}$ through the inference pipeline. The gray distribution corresponds to the gray distribution in Figure \ref{fig:wimgpdfs}, and shows model-predicted flux ratios given the prior $p\left(\xlens\right)$ and the astrometric likelihood. The blue distributions, which correspond to the blue distributions in Figure \ref{fig:wimgpdfs}, show the flux ratios after weighting by $w_{\rm{img}}$. The improved precision in the model predicted flux ratios enables stronger constraints on dark matter properties. Red crosshairs are measured values.}
		\end{figure*}
		
		We include dark matter substructure in the calculation of $w_{\rm{img}}$ because the macromodel parameters $\xlens$ will absorb some large-scale features of the deflection field produced by a particular substructure realization $\rsub$. Marginalizing over $\rsub$ effectively increases the uncertainties of the $\xlens$ parameters to account for their correlations with $\rsub$. To give a specific example, if we place several $10^{10} M_{\odot}$ halos inside the Einstein radius, their lensing effects would be absorbed by the normalization of the main deflector mass profile, $\theta_{\rm{E}}$. 
		
		It is important to distinguish correlations between $\xlens$ and $\rsub$, which vary on a realization-by-realization basis, from correlations between $\xlens$ and the hyper-parameters $\qsub$. Averaged over many realizations, correlations between $\xlens$ and $\rsub$ will manifest as increased uncertainties on the $\xlens$ parameters, relative to what we would infer without substructure in the lens model. On the other hand, a requirement for our approach to yield unbiased results is that $\xlens$ should not be correlated or otherwise strongly dependent on $\qsub$ when incorporating constraints from the imaging data. To understand this requirement, consider the example mentioned in the previous paragraph related to the normalization of the main deflector mass profile $\theta_{\rm{E}}$. The total mass rendered in halos varies from realization to realization with Poisson scatter, and these fluctuations around the average dark matter density are absorbed into $\theta_{E}$. In other words, the standard deviation of $\theta_{E}$ is strongly correlated with $\qsub$, with dark matter models that have more substructure corresponding to a broader distribution of $\theta_{\rm{E}}$. Importance sampling weights derived without substructure would select $\theta_{\rm{E}}$ with a typical precision of $<1 \%$, penalizing dark matter models that include additional substructure based solely on the diameter of the lens. This would effectively transform the inference pipeline into an Einstein radius detector, and bias inferences towards lens models with less substructure. 
		
		To avoid biasing our inferences through the importance weights, we apply them to a subset of macromodel parameters that are not correlated or strongly dependent on $\bf{\qsub}$: the main deflector axis ratio $q$, the external shear strength $\gamma_{\rm{ext}}$, the logarithmic power law slope of the main deflector mass profile $\gamma$, the strengths of the $m=3$ and $m=4$ multipole terms, and the mass of the nearest satellite or companion galaxy, $\theta_{\rm{E},G2}$, if one is present\footnote{In principle, we could also include the $m=1$ multipole term. However, the $m=1$ term does not appear to be constrained by imaging data}. We refer to Section \ref{ssec:macromodel} for additional discussion regarding the macromodel parameters and the modeling of the main deflector mass profile. 
		
		To verify that the importance sampling weights do not bias our inferences, we can check that our posterior distribution has the property
		\begin{equation}
			\label{eqn:requirement}
			p\left(\qsub | \dptsrc, \dimg \right) = p\left(\qsub | \dptsrc \right).
		\end{equation}
		In order words, without flux ratio information, the imaging data constraints implemented through $w_{\rm{img}}$ should not affect our inferences regarding the nature of dark matter. We note that if we can solve for a set of macromodel parameters that fit the observed image positions for any realization $\rsub$, Equation \ref{eqn:requirement} reduces to $p\left(\qsub | \dptsrc, \dimg \right) = \pi \left(\qsub \right)$, where $\pi\left(\qsub \right)$ is the prior probability. We verify that Equation \ref{eqn:requirement} is satisfied by propagating $w_{\rm{img}}$ through the inference pipeline with the flux-ratio uncertainties increased by a large $\sim 10^{10}$ factor. 
		
		It is easiest to interpret $w_{\rm{img}}$ as a probability distribution that updates our inference on $\xlens$ with information from imaging data. To illustrate, Figure \ref{fig:wimgpdfs} shows examples of the inferred macromodel parameters for two systems, GRAL1131 (see Figure \ref{fig:1131zooms}) and J0607 (far left in second row of Figure \ref{fig:mosaic}). In the left panel we show the inferred $q$, $\gamma_{\rm{ext}}$, $\gamma$, and the $a_3$ and $a_4$ multipole terms for GRAL1131. In gray we show the inferred parameters using only the prior $p\left(\xlens\right)$ (see Section \ref{ssec:macromodel}) and astrometric constraints. In blue, we show the inferred parameter values after modeling the imaging data. The green distribution shows the importance sampling distribution that corresponds to $w_{\rm{img}}$ -- evaluating the green probability distribution at any coordinate is equivalent to evaluating the importance weight $w_{\rm{img}}$ at this coordinate. Multiplying the gray distribution (our inferences on the macromodel informed by astrometry, given the prior on the macromodel parameters) with the green distribution (the additional information from imaging data) yields blue distribution (our combined inference on $\xlens$ from imaging data and astrometry). The right panel shows the same calculation for J0607, which has a nearby satellite galaxy whose mass is constrained by the imaging data. 
		
		Figure \ref{fig:frwimg} shows the model-predicted flux ratios for GRAL1131 and J0607 after incorporating the imaging data information through $w_{\rm{img}}$ (blue), relative to the model-predicted flux ratios obtained without imaging data (gray). These flux-ratio distributions include substructure and globular clusters (see Section \ref{sec:dmmodel}), and are marginalized over the main deflector mass profile (see Section \ref{sec:lensmodeling}). The red cross hairs show the measured flux ratios for these systems. As is evident from Figure \ref{fig:frwimg}, the additional constraints on the main deflector mass profile encoded by $w_{\rm{img}}$ affect the model-predicted flux ratios. 
		
		\subsection{Imaging data reconstruction}
		\label{ssec:imgdatareconstruction}
		We calculate $w_{\rm{img}}$ by reconstructing the imaging data, while requiring that the macromodel satisfy the lens equation for the observed image positions, with substructure included in the lens model. To evaluate Equation \ref{eqn:imageweights} we generate $N\sim 200,000$ lens models. For each model, we evaluate $p\left(\xlens, \xlight, \rsub | \dimg, \dptsrc \right)$ by, first, obtaining a lens model that satisfies the lens equation for the observed image positions, as discussed in Section \ref{sssec:astrometriclike}. Next, we simultaneously reconstruct the main deflector lens light with the extended lensed arcs. The specific modeling choices we implement in this analysis for the background source light profile and the lens light profile are discussed in Section \ref{sec:lensmodeling}. We connect the source-plane surface brightness to the image plane surface brightness by ray tracing backwards, and then convolve the lens plane surface brightness with a model for the point-spread function (PSF)\footnote{We do not super-sample for the source reconstruction because we are using the imaging data to constrain the large-scale mass profile of the main deflector, not to detect small-scale structure in the lens. For the latter, super-sampling is necessary to capture the small-scale lensing effects of a perturber}. We discuss the model for the PSF in Section \ref{sssec:psf}. 
		
		As we have already constructed a lens model that satisfies the lens equation for the image positions, we perform the reconstruction of the lens light and the lensed arcs for a fixed lens model. To focus computational time in regions of parameter space that fit the imaging data, we use a particle swarm optimization (PSO) with 10 particles and 80 iterations to move the $\xlight$ parameters toward high-probability regions of parameter space. The top-center panel of Figure \ref{fig:1131zooms} shows one example outcome of this calculation for GRAL1131, for one population of halos. 
		
		Once we have run the PSO, we compute the probability
		\begin{equation}
			p\left(\dimg | \dimgprime\right) = \frac{\exp\left[ \left(\dimg -  \dimgprime\right)^{\rm{T}} \bf{\Sigma_{\rm{\mathcal{I}}}}^{-1} \left(\dimg -  \dimgprime\right)\right]}{\sqrt{\left(2 \pi\right)^{n_{\rm{pix}}}| \bf{\Sigma}_{\rm{\mathcal{I}}}|}},
		\end{equation}
		where $n_{\rm{pix}}$ is the number of pixels in the image plane, and $\bf{\Sigma_{\rm{\mathcal{I}}}}^{-1}$ is a diagonal covariance with entries $\Sigma_{\mathcal{I},ii} = \sigma_{ii,\rm{(Poisson)}}^2 + \sigma_{ii,(\rm{rms})}^2 + \delta \sigma_{ii,(\rm{noise})}^2$. The first and second terms represent Poisson fluctuations and Gaussian background fluctuations, respectively, which we estimate directly from the observations. 
		
		The third term, $\delta \sigma_{\rm{noise}}^2$, dilutes the information content of the imaging data by down-weighting residuals in surface brightness of lensed arcs and lens light. We include $\delta \sigma_{\rm{noise}}$ to make the problem computationally tractable. To provide a specific example of the importance of this weighting, we first define an effective sample size, $N_{\rm{eff}}$, in terms of the probabilities $p\left(\dimg | \dimgprime \right)$ across an set of $N$ lens models
		\begin{eqnarray}
			\label{eqn:neff}
			\nonumber	\bf{p_{\rm{img}}} \equiv \frac{p\left(\dimg | \dimgprime\right)}{\max | p\left(\dimg | \dimgprime \right)|}\\
			N_{\rm{eff}}\equiv \sum_{i=1}^{N} p_{\rm{img(i)}}.
		\end{eqnarray}
		With HST-quality imaging data and thousands of degrees of freedom, a small change to the lens or source light model can result in a relative likelihood $\Delta \log p_{\rm{img}} \sim 100$ between the most probable and second-most probable lens model, while the reduced $\chi^2$ and goodness of fit remains approximately the same. Relative to the most probable solution, all other lens models would be down-weighted $p_{\rm{img(i)}} < e^{-\Delta \log w} = e^{-100}$. This would result in $N_{\rm{eff}} \approx 1$ (unless we generate of order $e^{100}$ lens models in the forward model), meaning only one realization contributes to the importance sampling weights in Equation \ref{eqn:imageweights}. Including $\sigma_{\rm{noise}}$, the difference between log-likelihoods becomes $\Delta \log p / \sigma_{(\rm{noise})}$, and we can incorporate the imaging data constraints through Equation \ref{eqn:imageweights} in a way that is computationally tractable. 
		
		There is not a unique way to choose $\sigma_{\rm{noise}}$. We have experimented with defining this term in a way that connects directly to the uncertainties in the imaging data on a pixel level. For example, we can generate 10,000 background noise realizations (using $\sigma_{ii,(\rm{rms})}$ and $\sigma_{ii,\rm{(Poisson)}}$), render them on top of a best-fit lens model for each system, and set $\sigma_{ii,(\rm{noise})}$ equal to the standard deviations of the resulting distribution of log-likelihoods. However, strategies for defining $\sigma_{(\rm{noise})}$ in a way that connects to the background and Poison noise in the image tends to over-correct, in the sense that we discard more information from the imaging data than is necessary to make the problem computationally tractable. Instead, since the primary motivation for including $\sigma_{(\rm{noise})}$ is related to available CPU hours, $n_{\rm{cpu-hour}}$, we define $\sigma_{(\rm{noise})}$ in such a way that $\sigma_{(\rm{noise})} \rightarrow 0$ (meaning we do not down-weight constraints from lensed arcs) as $n_{\rm{cpu-hour}}\rightarrow \infty$. To compute $w_{\rm{img}}$, we generate $N\sim 200,000$ lens models per system, on average, and set $\sigma_{(\rm{noise})}$ such that $N / N_{\rm{eff}} = 2000$. The threshold $N / N_{\rm{eff}} = 2000$ represents a compromise between extracting more information from the lensed arcs, and having enough sampling of the parameter space to suppress shot-noise in the $w_{\rm{img}}$ distributions. This approach effectively bootstraps uncertainties in the imaging data through $\sigma_{\rm{noise}}$ until the top $\sim 1 \%$ of lens models have comparable log-likelihoods. 
		
		The time-per-lens-reconstruction, including generating a population of (sub)halos, ray tracing through the realization, constructing a lens model that solves the lens equation, and reconstructing the lensed arcs, varies widely among the systems in our sample. It depends most strongly on the complexity of the quasar host galaxy, with additional complexity in the quasar host galaxy requiring more detailed models of the source morphology. The time-per-lens-reconstruction ranges from $\sim 1$ minute per lens model for systems like H1413, for which we model the background source with a single elliptical S{\'e}rsic profile (as discussed in Section \ref{sssec:sourcelight}), to $\sim 30$ minutes per lens model for RXJ1131, which exhibits complex morphology in the lensed arcs. Following our approach of propagating constraints from lensed arcs to the dark matter inference through $w_{\rm{img}}$, we do not need to reconstruct the lensed arcs for the lens models used in the dark matter inference, and the time-per-lens-reconstruction for the dark matter inference is $\mathcal{O}\left(1 \ \rm{min}\right)$ for every system.  
		
		\section{Datasets}
		\label{sec:data}
		We apply the analysis methods outlined in the previous sections to a sample of 28 quadruply imaged quasars. These systems, together with their lens and source redshifts and the telescope used to collect the imaging data, are listed in the first three columns of Table \ref{tab:tablenmax}. 
		
		For 26 of the 28 systems, we use flux ratios from warm dust surrounding the background quasar obtained through the JWST lensed quasar dark matter survey. Lenses in this sample were selected to ensure that the unresolved image fluxes be bright enough in WISE W4 to ensure high signal to noise measurements with MIRI, and that the main deflector be an elliptical galaxy. The latter requirement ensured that systems in the sample show no significant complexity in the form of stellar disks, spirals arms, or bars, structures that should be explicitly modeled, if present \citep{Gilman++17,Hsueh++18}. The sample we analyze includes deflectors with morphological properties consistent with those of massive elliptical galaxies, the kinds of systems that tend to dominate the strong lensing cross section \citep{Auger++10}. As discussed by \citet{Keeley++25}, we omit three systems from the dark matter analysis because they exhibit morphological properties inconsistent with massive ellipticals, are triple-image systems that our inference pipeline does not currently accommodate, or have blended images when computing the image magnifications\footnote{For an example of non-blended quasar images, see the examples shown in the bottom row of Figure~\ref{fig:1131zooms}.}.
		
		We use multi-band spectral energy distribution fitting of the MIRI data in order to isolate the more spatially extended warm dust emission from potential microlensing contamination from blueward features, such as light from the dust sublimation zone and the quasar accretion disk. These measurements, and the techniques used to perform the SED fitting to isolate emission from the warm dust torus, are summarized in the first \citet{Nierenberg++24}, second \citet{Keeley++24} and third \citet{Keeley++25} papers in this series. \citet{Keeley++25} also validates the robustness of the measurements to microlensing with regards to the dark matter measurements. 
		
		We included 2  additional systems which did not meet the flux requirements for inclusion in the JWST sample. For these lenses, we use measurements from the nuclear narrow-line region surrounding the background quasar presented by \citet{Nierenberg++14} and \citet{Nierenberg++20}. The JWST and HST data differ due the physical size of the emission region, with emission from the compact warm dust region experiencing stronger perturbations by dark matter halos than emission from the more extended narrow-line region. Both the warm dust and nuclear narrow-line emission are large enough to avoid microlensing by stars \citep{MoustakasMetcalf03,Sluse++13}, and the light-crossing time of these regions should wash out variability in the quasar light curve. The size of the emission regions and the priors on the source sizes are discussed further in Section \ref{sssec:quasarlight}.
		
		For systems with mid-IR flux ratios from JWST, we use the image positions measured during the SED fitting, as discussed by \citet{Nierenberg++24}, \citet{Keeley++24} and \citet{Keeley++25}. We assume typical astrometric uncertainties of 5 milli-arcseconds for each system. For the two narrow-line systems, B1422 and RXJ0911, we use image positions as reported by \citet{Nierenberg++14} and \citet{Nierenberg++20}, respectively, again measured to 5 milliarcsecond precision. 
		
\begin{figure*}
			\centering
			\includegraphics[trim=0cm 0cm 0cm
			0cm,width=0.3\textwidth]{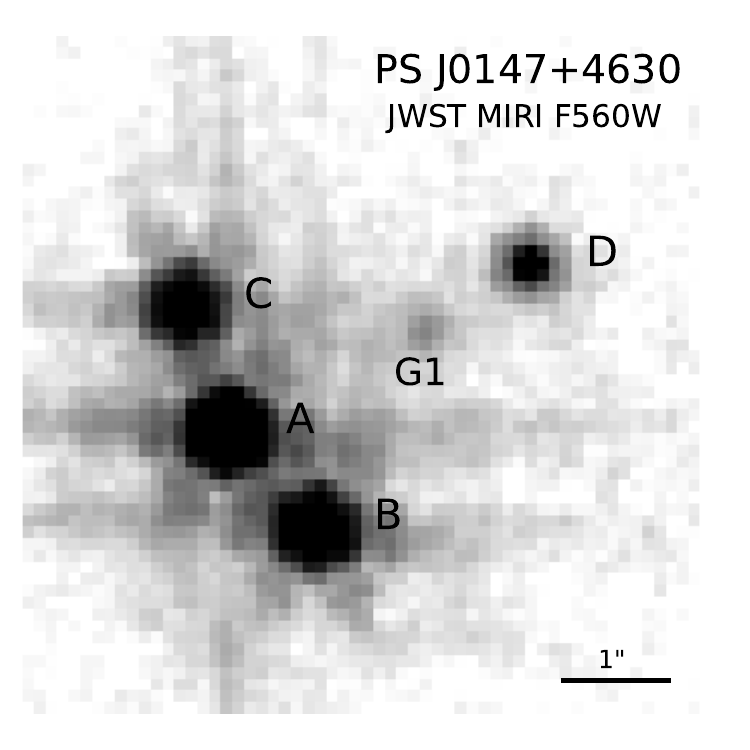}
			\includegraphics[trim=0cm 3cm 0cm
			2cm,width=0.48\textwidth]{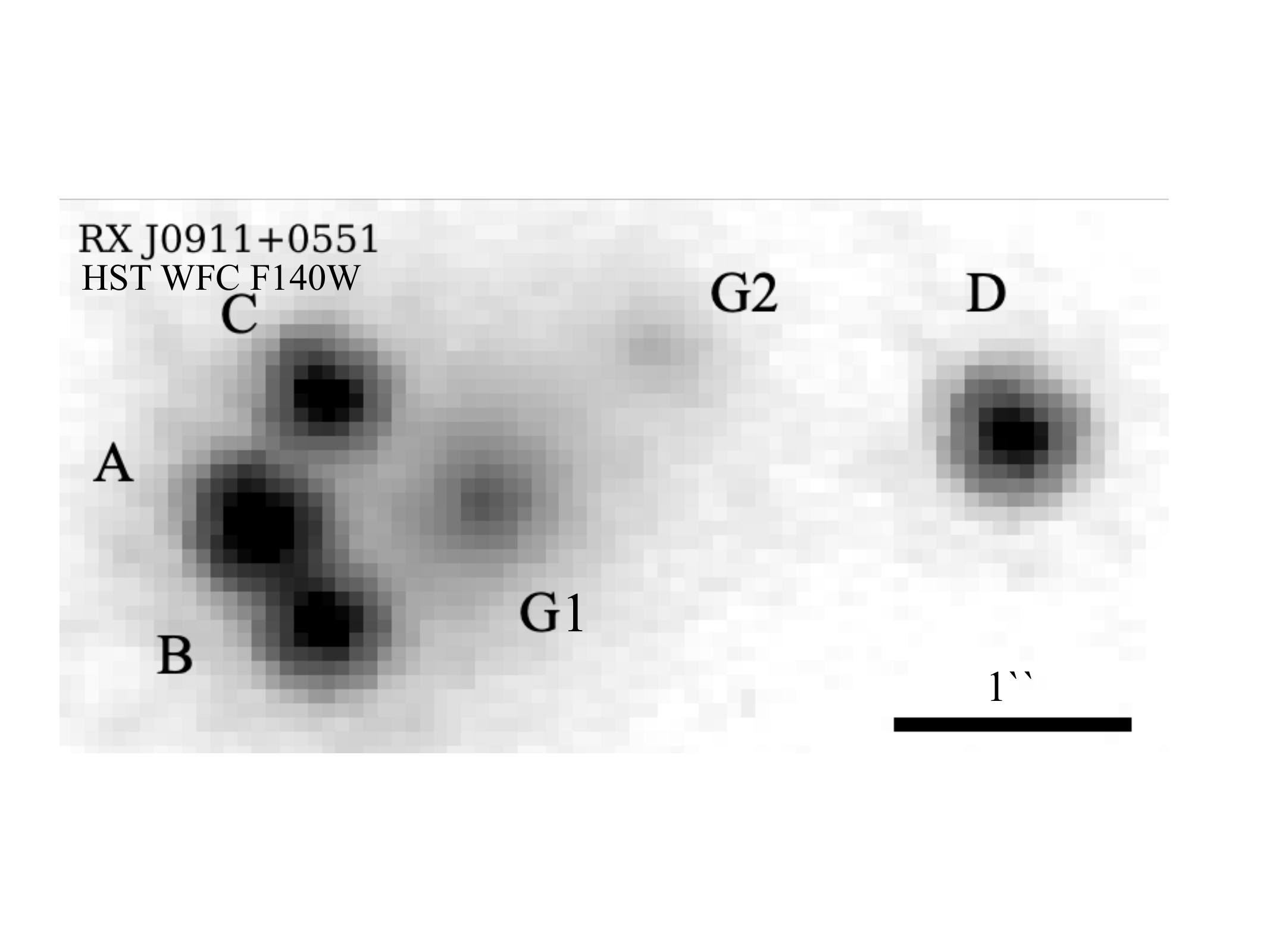}
			\includegraphics[trim=0cm 3cm 0cm
			2cm,width=0.45\textwidth]{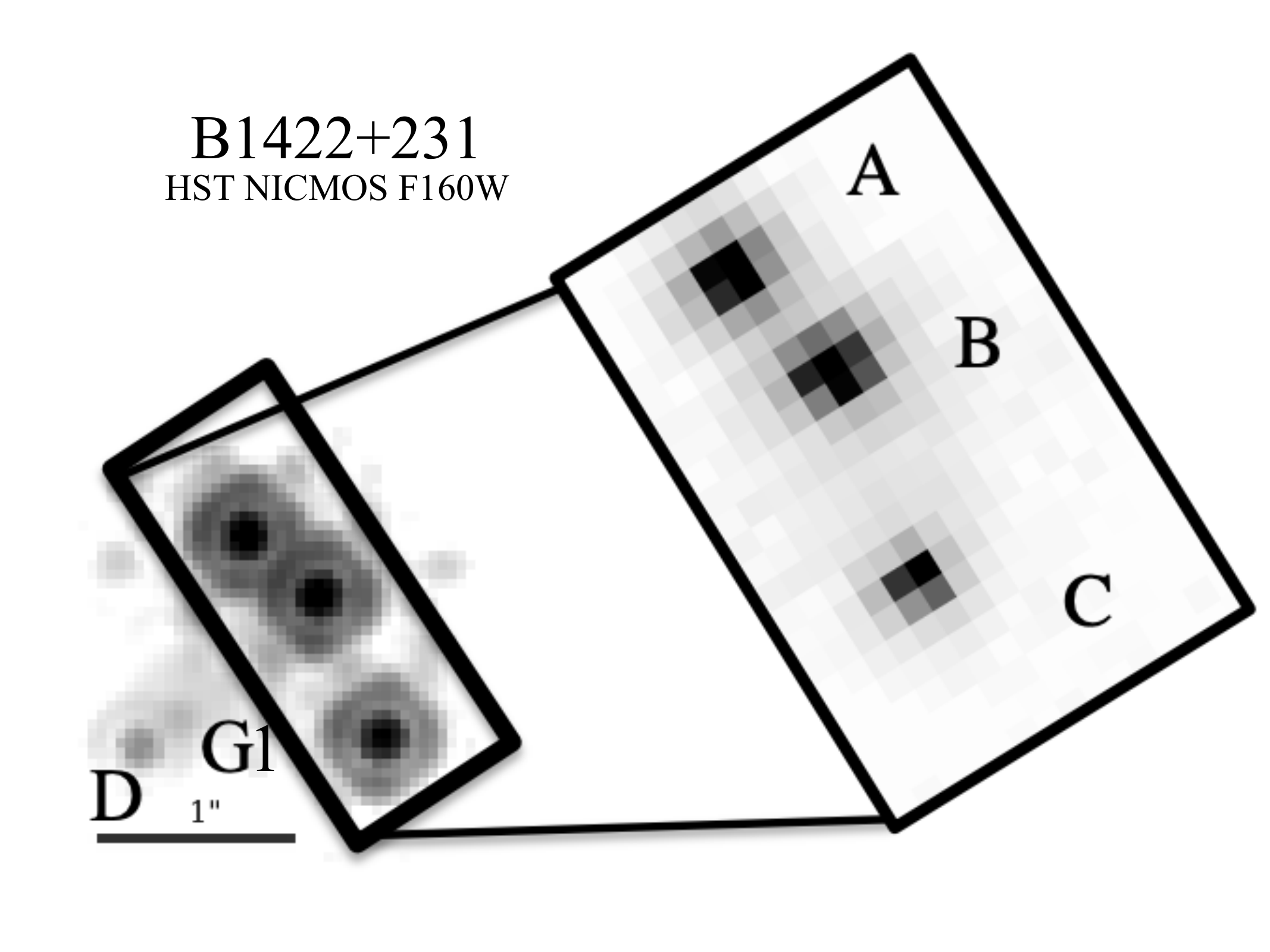}
			\includegraphics[trim=0.5cm 3cm 3cm
			2cm,width=0.45\textwidth]{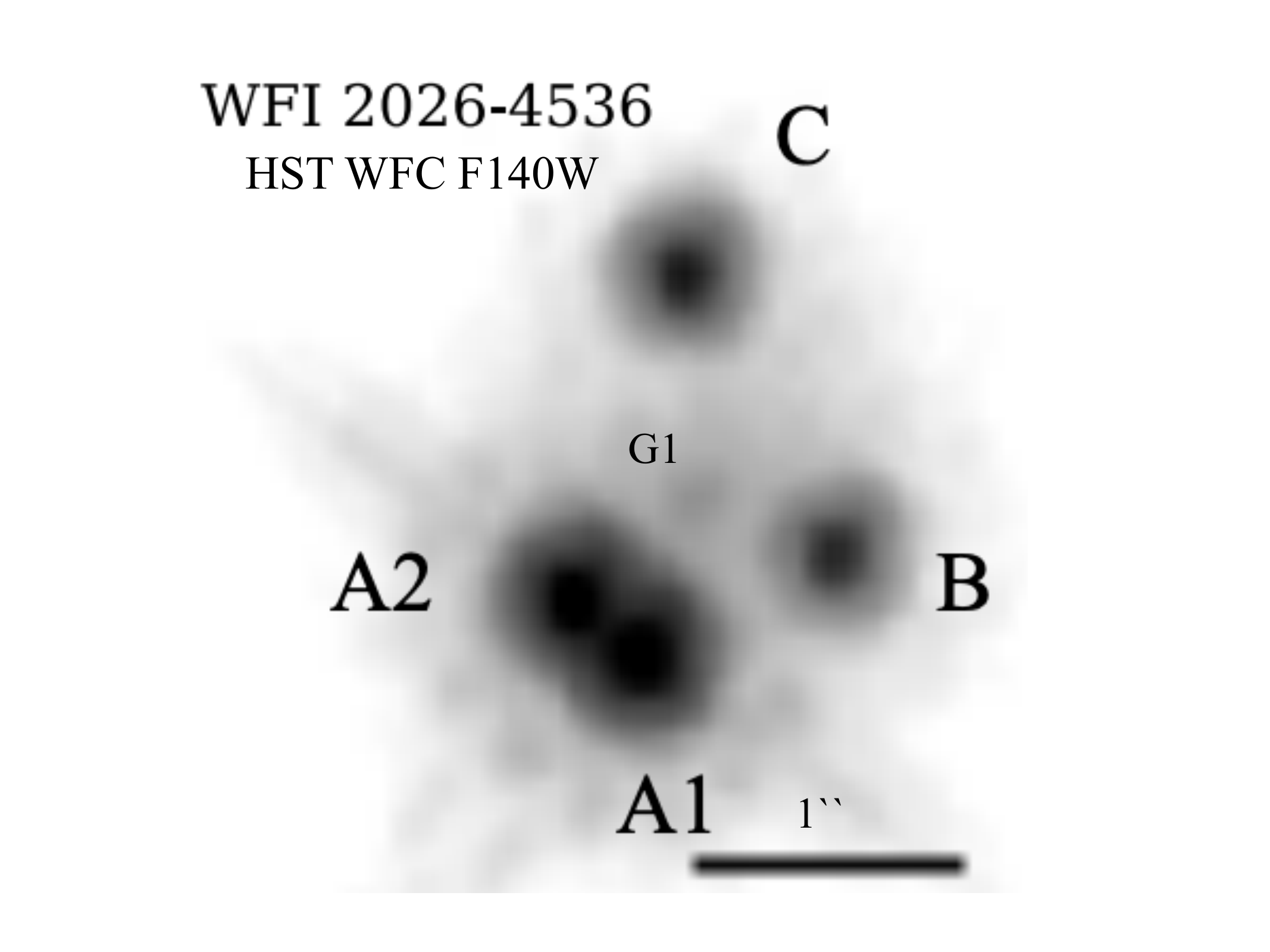}
			\caption{\label{fig:mosaicnoimg} The 4 systems for which we only model image positions and flux ratios, for reasons discussed in Section \ref{ssec:lmspecific}. The images of RXJ0911+0551 and WFI2026-4536 are adapted from \citep{Nierenberg++20}, and the image of B1422+231 is adapted from \citep{Nierenberg++14}.}
		\end{figure*}

		We model the extended lensed arcs and compute the imaging data likelihood for 24 out of the 28 systems. These systems are shown in Figure \ref{fig:mosaic}. As discussed in Section \ref{ssec:lmspecific}, we do not model the lensed arcs in 4 cases, shown in Figure \ref{fig:mosaicnoimg}, either because there are no lensed arcs in the imaging data, as in the case of RXJ0911, B1422, WFI2026, or because PSF cruciform artifact dominates the surface brightness of the lensed arcs, in the case of PSJ0147. For these systems, the dark matter constraints are based only on point source position and flux ratios.
		
		For 3 of the 24 for which we model lensed arcs, we use NIRCam imaging from JWST, which provides exquisite angular resolution and high signal-to-noise. These data, for systems WFI2033, HE0435, and PG1115, were acquired through the program JWST GTO-1198 \citep{StiavelliMorishita17,Williams++25}. For 10 systems, we use HST imaging data in either F160W or F814W. These observations were obtained through HST-GO-15320 and HST-GO-15652 (PI:Treu). Some lens models using HST data for these systems were presented by \citet{Shajib++19} and \citet{Schmidt++23} using an automated modeling pipeline. For the remaining 11 systems we use imaging data from JWST-MIRI F560W. These observations were obtained through JWST GO-2046 (PI:Nierenberg). F560W is also the bluest filter used during the SED fitting in the flux ratio measurements. Relative to HST or NICam images, the MIRI data has degraded angular resolution by about a factor of 2, and relatively short exposures. Despite these limitations, in some cases the extended lensed arcs are more prominent in MIRI F560W than in HST data, and they are bright enough to enable some constraining power on the lens macromodel (for example, see the discussion in Section \ref{ssec:casestudies} regarding MG0414). Section \ref{ssec:lmspecific} discusses for which systems we choose to model the MIRI data over the HST data, and presents further considerations regarding the lens modeling of each individual lens system. The second column of Table~\ref{tab:tablenmax} lists the deflector and source redshifts. For systems without a spectroscopic or otherwise reliable redshift measurement, we place them at $z_{\rm{d}} = 0.5$. These systems are marked $0.5^{\star}$. 
		
		\begin{table*}
			\setlength{\tabcolsep}{12pt}
			\caption{\label{tab:tableqsub} Description of the dark matter hyper-parameters introduced in Section \ref{sec:dmmodel}. The predictions for $\delta_{\rm{LOS}}$, $\alpha$ and $\Sigma_{\rm{sub}}$ remain unchanged in WDM, as these parameters determine the form of the subhalo mass function anchored at high masses.}
			\begin{tabular}{cccc}
				\hline
				Hyper-parameter & Description & Sampling distribution &  Remarks\\	
				\hline
				$\delta_{\rm{LOS}}$ & rescales the amplitude of the& $\mathcal{U}\left(0.9, 1.1\right)$ & $\delta_{\rm{LOS}}=1$ corresponds to the \\
				& field halo mass function  & & Sheth--Tormen prediction\\ \\ 
				$\alpha$ & logarithmic slope of the & $\mathcal{U}\left(-1.95, -1.85\right)$ & CDM predicts $\alpha \sim -1.9$\\
				& subhalo mass function at infall & & \\ \\ 
				$\Sigma_{\rm{sub}} \ \left[\rm{kpc^{-2}}\right]$ & amplitude of the differential & $\log_{10} \mathcal{U}\left(-2.2, 0.2\right)$ & $N$-body predicts $\sim 0.1 \ \rm{kpc^{-2}}$\\
				& subhalo mass function at infall & & SAM predicts $\sim 0.15 \ \rm{kpc^{-2}}$ \\
				&  & & (see Section \ref{ssec:priors}) \\ \\ 
				$m_{\rm{hm}} \left[\mathrm{M}_{\odot}\right]$ & half-mode mass scale & $\log_{10} \mathcal{U}\left(4.0, 10\right)$ & mass function $\&$ concentrations\\
				& & & suppressed for $ m \lesssim m_{\rm{hm}}$ \\
				\hline
			\end{tabular}
		\end{table*}
		
		\section{Modeling of dark matter substructure and globular clusters}
		\label{sec:dmmodel}
		This section describes the modeling of small-scale structure in the lens system, including dark matter halos and globular clusters. We begin in Section \ref{ssec:mfuncs} by describing the models for dark matter halos and subhalos in CDM, including their mass function, concentration--mass relation, and density profiles. In Section \ref{ssec:wdm}, we describe how we modify the CDM relations to account for free-streaming in WDM. Section \ref{ssec:priors} summarizes the sampling distributions and Bayesian priors assigned to each hyper-parameter introduced in Sections \ref{ssec:mfuncs} and \ref{ssec:wdm}. Section \ref{ssec:globularclusters} describes the modeling of globular clusters. 
		
		The parameters introduced throughout this section comprise the hyper-parameters $\qsub$ that we will infer from the data. We summarize these parameters, together with their physical interpretation, in Table~\ref{tab:tableqsub}. We generate substructure and globular cluster populations using the open-source software {\tt{pyHalo}}\footnote{https://github.com/dangilman/pyHalo} \citep{Gilman++20}. 
		
		\subsection{Halo and subhalo mass functions in CDM}
		\label{ssec:mfuncs}
		In this section we discuss the modeling of the (sub)halo mass function, halo density profiles, and the concentration--mass relation. In Section \ref{sssec:field} we begin by describing the modeling of field halos along the line of sight. Section \ref{sssec:subahlos} describes how we model subhalos, including considerations related to their tidal evolution. 
		
		Throughout this section, we define halo masses in terms of $m_{200}$ calculated with respect to critical density of the Universe at the halo redshift, or---in the case of subhalos---at their infall redshift. We generate masses (infall masses) for field halos (subhalos) in the range $10^6 \mathrm{M}_{\odot}$--$10^{10.7} \mathrm{M}_{\odot}$. Halos less massive than $10^6 \mathrm{M}_{\odot}$ are too small to affect the flux ratios given the physical size of the emission region around the background quasar (see Section~\ref{sssec:quasarlight}). Halos more massive than $10^{10.7} \mathrm{M}_{\odot}$ are quite rare and, if they were present, would very likely host enough luminous material to be detectable, and thus included explicitly in the lens model (see Section~\ref{ssec:luminoussats}). 
		
		\subsubsection{Field halos} 
		\label{sssec:field}
		We render halos along the line of sight in a double-cone geometry that opens towards the lens and closes at the source position. The opening angle of the cone is taken to be six times the Einstein radius for each lens, which has been shown through extensive testing to be sufficient for our intended precision. We divide the line of sight into circular planes with a uniform spacing in redshift of $\Delta z = 0.02$. At each lens plane along the line of sight, halos are distributed following a uniform spatial distribution\footnote{CDM predicts the positions of halos to be correlated, but this correlation is negligible on the halo mass and distance scales relevant for substructure lensing.}. 
		
		CDM and WDM predict the total number and masses of field halos, and of subhalos at infall. For field halos, we draw halo masses according to 
		\begin{equation}
			\label{eqn:mfunclos}
			\frac{\mathrm{d}^2 N_{\rm{CDM}}}{\mathrm{d}m \mathrm{d}V} = \delta_{\rm{LOS}} \left[1+\xi\left(m_{\rm{host}},z_{\rm{d}}\right)\right] \frac{\mathrm{d}^2 N_{\rm{ST}}}{\mathrm{d}m \mathrm{d}V},
		\end{equation}
		where $\frac{\mathrm{d}^2 N_{\rm ST}}{\mathrm{d}m \mathrm{d}V}$ represents the Sheth--Tormen mass function model \citep{ST01}. The term in brackets in Equation~\ref{eqn:mfunclos},  $1+\xi\left(m_{\rm{host}},z_{\rm{d}}\right)$, which includes the 2-halo term $\xi\left(m_{\rm{host}},z_{\rm{d}}\right)$, boosts the number of field halos in the region $z_{\rm{d}} \pm \Delta z$ due to the presence of a group-scale host $m_{\rm{host}} \sim 10^{13} \mathrm{M}_{\odot}$. We evaluate both the Sheth--Tormen mass function and the 2-halo term using {\tt{colossus}} \citep{Diemer18}, and include the modifications to the two-halo boost term suggested by \citet{Lazar++21}. This leads to a $\sim 20 \%$ increase in the number of halos near the main deflector, and we place these additional objects at the main deflector redshift. By integrating Equation~\ref{eqn:mfunclos}, we compute the mean number $\langle N \rangle$ of halos predicted by the halo mass function, and then generate $N$ halos from a Poisson distribution with a mean $\langle N \rangle$. 
		
		The factor $\delta_{\rm{LOS}}$ in Equation~\ref{eqn:mfunclos} scales the total number of field halos, and is intended to capture systematic uncertainties associated with cosmological parameters and the halo mass function model. We use this term to allow for up to $20\%$ differences in the average line-of-sight towards strong lenses, relative to a random line of sight. Although  strong lenses are typically massive elliptical galaxies and thus preferentially reside in dense environments \citep{Treu++09}, these considerations apply to the immediate vicinity of the main deflector on scales of Mpc, and this enhancement of structure near the lens redshift is captured through $\xi\left(m_{\rm{host}},z\right)$. The parameter $\delta_{\rm{LOS}}$ applies to the entire line of sight (Gpc scales), for which the bias from selection effects and correlated structure, relative to a random line of sight, is expected to be at the percent level \citep{Tang++25}. 
		
		We describe the density profile of each halo as a truncated NFW profile \citep{Baltz++09}
		\begin{equation}
			\rho_{\rm{NFW}}\left(r,r_\mathrm{s},r_\mathrm{t},\rho_\mathrm{s}\right) = \frac{\rho_\mathrm{s}}{x\left(1+x\right)^2} \frac{\tau^2}{\tau^2+x^2},
		\end{equation}
		with $x\equiv r/r_\mathrm{s}$ and $\tau\equiv r_\mathrm{t}/r_\mathrm{s}$. We compute $\rho_\mathrm{s}$ and $r_\mathrm{s}$ for a given halo mass and concentration using the concentration--mass relation, $c\left(m,z\right)$, presented by \citet{DiemerJoyce19} with a scatter of 0.2 dex \citep{Wang++20}. We calculate $\rho_\mathrm{s}$ and $r_\mathrm{s}$ using the halo mass definition of $m_{200}$ with respect to $\rho_{\rm{crit}}\left(z\right)$, or the critical density of the Universe at the halo redshift. We truncate field halos at $r_\mathrm{t} = r_{200}$. For both the mass function and concentration--mass relation, we use {\tt{colossus}} to perform the cosmological calculations, such as evaluating the Sheth--Tormen mass function and \citet{DiemerJoyce19} concentration--mass relation, and then {\tt{pyHalo}} populates the lensing volume with halos, while accounting for the modifications $\delta_{\rm{LOS}}$ and $\xi\left(m_{\rm{host}},z\right)$. 
		
		At each lens plane along the line of sight, we add a sheet of constant negative convergence $\kappa_{\langle m \rangle} = \langle m \rangle / \left(\Sigma_{\rm{crit}} A\right)$, where $\langle m \rangle$ is the average mass added in halos obtained by integrating Equation~\ref{eqn:mfunclos}, $\Sigma_{\rm{crit}}$ is the critical density for lensing at the redshift of the lens plane, and $A$ is the area of the lens plane\footnote{When considering WDM, we account for the suppression of the halo mass function, discussed in Section~\ref{ssec:wdm}, when calculating $\langle m \rangle$.}. The inclusion of these convergence sheets preserves the correct physical property of multi-plane lensing in a spatially flat Universe: halos add small perturbations on top of a smooth background, but light rays travel on straight lines, on average. Without including the mass sheet correction, we would bias our lens models by introducing an overall bending towards the center of the mass distribution that becomes more severe in models with more substructure.   
		
		\subsubsection{Subhalos} 
		\label{sssec:subahlos}
		When a field halo crosses the virial radius of the main deflector it becomes a subhalo, stops smoothly accreting material, begins losing mass to tidal stripping, and tends to sink towards the center of the main halo (host) as a result of dynamical friction. These processes can affect the density profiles and spatial distribution of subhalos, which in turn affect the lensing signal. 
		
		Regarding the spatial distribution of subhalos, because the typical Einstein radius is much smaller than the host halo scale radius $\theta_{\rm{E}} / r_\mathrm{s} \ll 1$, the spatial distribution of subhalos in projection appears approximately uniform \citep{Xu++15,Gannon++25}. We therefore place subhalos uniformly in the main lens plane out to a maximum radius $3 \theta_{\rm{E}}$, where $\theta_{\rm{E}}$ is the Einstein radius of the system. 
		
		Our strategy for modeling subhalo density profiles uses the tidal evolution model presented by \citet{Du++25}. This model is designed specifically for the analysis of strong gravitational lens systems, and predicts the density profile of subhalos that appear in projection inside $\sim 3 \theta_{\rm{E}}$, or approximately $\sim 30 \ \rm{kpc}$ for a typical system, based on the subhalo properties at infall, given the infall subhalo mass function. 
		
		We begin by defining the differential subhalo mass function per unit area at infall:
		\begin{equation}
			\label{eqn:subhalomfunc}
			\frac{\mathrm{d}^2 N}{\mathrm{d}m \mathrm{d}A} = \frac{\Sigma_{\rm{sub}}}{m_0}\left(\frac{m}{m_0}\right)^{-\alpha} \mathcal{F}\left(m_{\rm{host}},z\right),
		\end{equation}
		where $m$ is the mass at infall, $m_0=10^8 \mathrm{M}_{\odot}$, $\alpha$ is the logarithmic slope, and $\mathcal{F}\left(m_{\rm{host}},z\right)$ accounts for projection effects in the subhalo abundance that scale with the host halo mass and redshift \citep{Gilman++20}. The inclusion of $\mathcal{F}\left(m,z\right)$ allows us to interpret $\Sigma_{\rm{sub}}$ (units $\rm{kpc^{-2}}$) as a hierarchical parameter---i.e. common to all lenses---that determines the number of objects that accrete onto the host halo and eventually appear near the Einstein radius in projection. We calibrate $\mathcal{F}\left(m,z\right)$ using the semi-analytic model {\tt{galacticus}}\footnote{https://github.com/galacticusorg/galacticus/wiki} \citep{Benson12}. As shown by \citet{Gannon++25}, the {\tt{galacticus}} prediction can be written 
		\begin{equation}
			\label{eqn:shmfscaling}
			\mathcal{F}\left(m_{\rm{host}},z\right) = \left(m_{\rm{host}}/10^{13} \mathrm{M}_{\odot}\right)^{k_1}\left(z+0.5\right)^{k_2},
		\end{equation}
		with $k_1=0.55$ and $k_2=0.37$ \citep{Gannon++25}. Integrating the infall mass function over subhalo mass over the mass range $10^6 M_{\odot} - 10^{10.7} M_{\odot}$ determines the expected number of subhalos $\langle N\rangle$. We generate $N$ subhalos sampling from a Poisson distribution with mean $\langle N \rangle$.
		
		Equation \ref{eqn:subhalomfunc} specifies the number density of halos with a mass $m$ at infall that eventually appear within a $30 \ \rm{kpc}$ aperture of the host halo center (the region relevant for strong lensing). After generating this infall population, we use the tidal stripping framework presented by \citet{Du++25} to predict each subhalo's density profile. The calculation begins with assigning infall redshifts to each subhalo using the probability distribution $p\left(z_{\rm{infall}}|m, m_{\rm{host}}\right)$, where the infall redshift depends on $m$ and the host-halo mass $m_{\rm{host}}$. From $z_{\rm{infall}}$ and $m$ we evaluate the concentration of the subhalo at infall $c=c\left(m,z_{\rm{infall}}\right)$. Next, given the host halo concentration $c_{\rm{host}}=c_{\rm{host}}\left(m_{\rm{host}}, z_{\rm{d}}\right)$, which we calculate using the \citet{DiemerJoyce19} concentration--mass relation, we sample the bound mass fraction, $f_{\rm{bound}}\equiv m_{\rm{bound}}/m$, from the probability distribution $p\left(f_{\rm{bound}}| c, z_{\rm{infall}}, c_{\rm{host}}\right)$. \citet{Du++25} derived empirical representations of this distribution by evolving subhalos in a growing host potential with {\tt{galacticus}}. As discussed by \citet{Du++25}, a given set of subhalo infall time, concentration, and host halo concentration specifies a distribution of $f_{\rm{bound}}$ due to differences in subhalo orbits and mass loss histories. Given $f_{\rm{bound}}$, we use tidal tracks \citep[e.g.][]{Errani++21,Du++24} to calculate the truncation radius $r_\mathrm{t}$, and a factor $f_{\rm{t}}$ that changes the normalization of the NFW profile $\rho_\mathrm{s} \rightarrow f_{\rm{t}} \rho_\mathrm{s}$. This model predicts that, at the redshift of the lens, most subhalos have lost over $90 \%$ of the mass they had at infall. The model implemented in $\tt{pyHalo}$ shows excellent agreement with cosmological dark matter only simulations of subhalo evolution performed with {\tt{galacticus}} for $\Sigma_{\rm{sub}} \approx 0.17 \ \rm{kpc^{-2}}$ \citep{Du++25}. In Section \ref{ssec:priors}, we discuss theoretical priors on $\Sigma_{\rm{sub}}$ based on both $\tt{galacticus}$ and the Symphony $N$-body simulations. 
		
		Throughout this and the preceding section, we have referred to (sub)halo abundance and concentration in the context of CDM. In the next section, we explain how we modify the mass function and concentration--mass relation to account for the effects of free-streaming in WDM. We will also comment on how the tidal stripping model discussed in this section predicts an additional suppression of the bound subhalo mass function in WDM due to the lower halo concentrations predicted by WDM, and the explicit dependence of subhalo tidal evolution on infall concentration. 
		\begin{figure*}
			\includegraphics[trim=1cm 0.75cm 0cm
			1cm,width=0.48\textwidth]{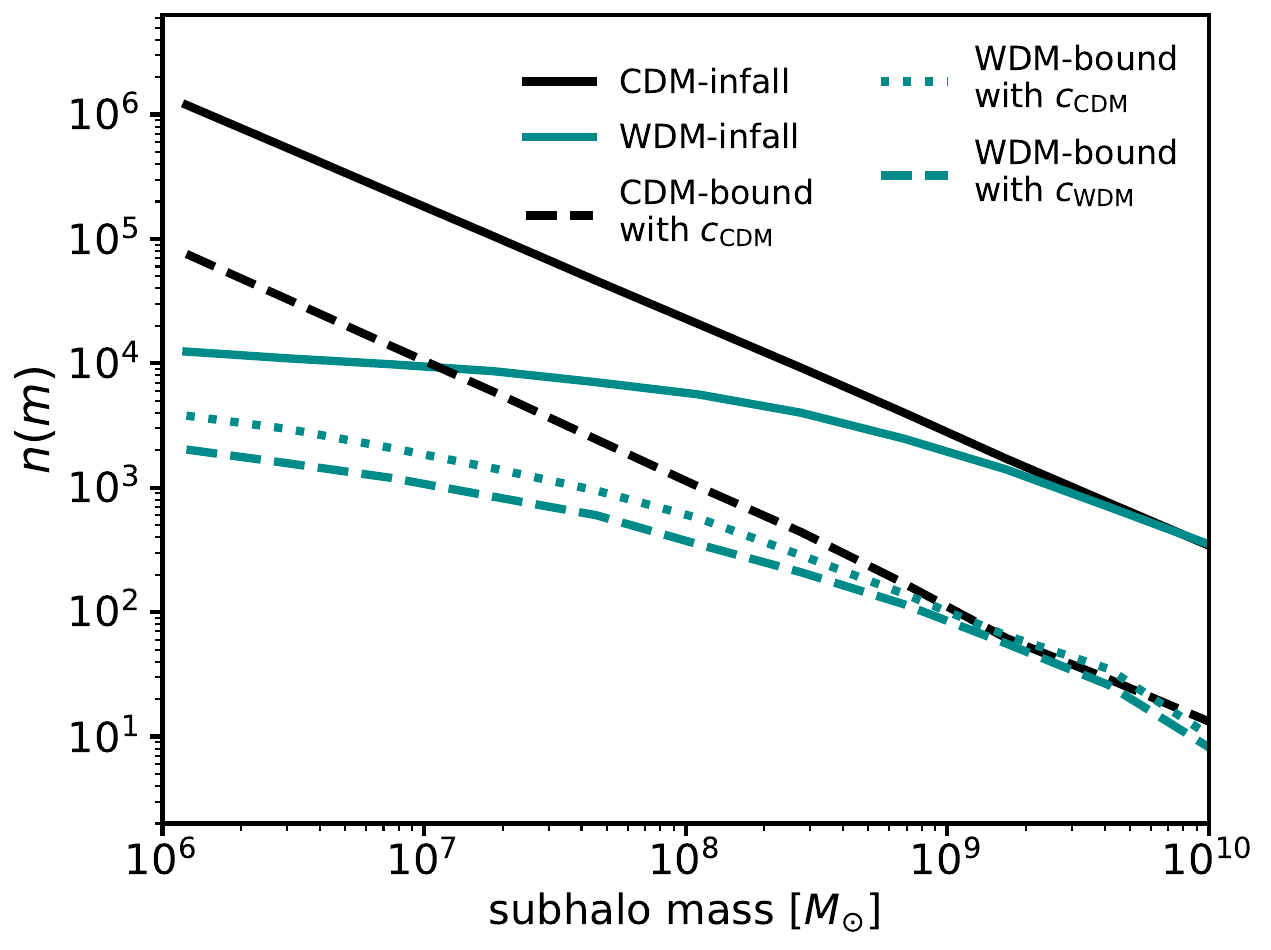}
			\includegraphics[trim=1cm 0.75cm 1cm
			0cm,width=0.48\textwidth]{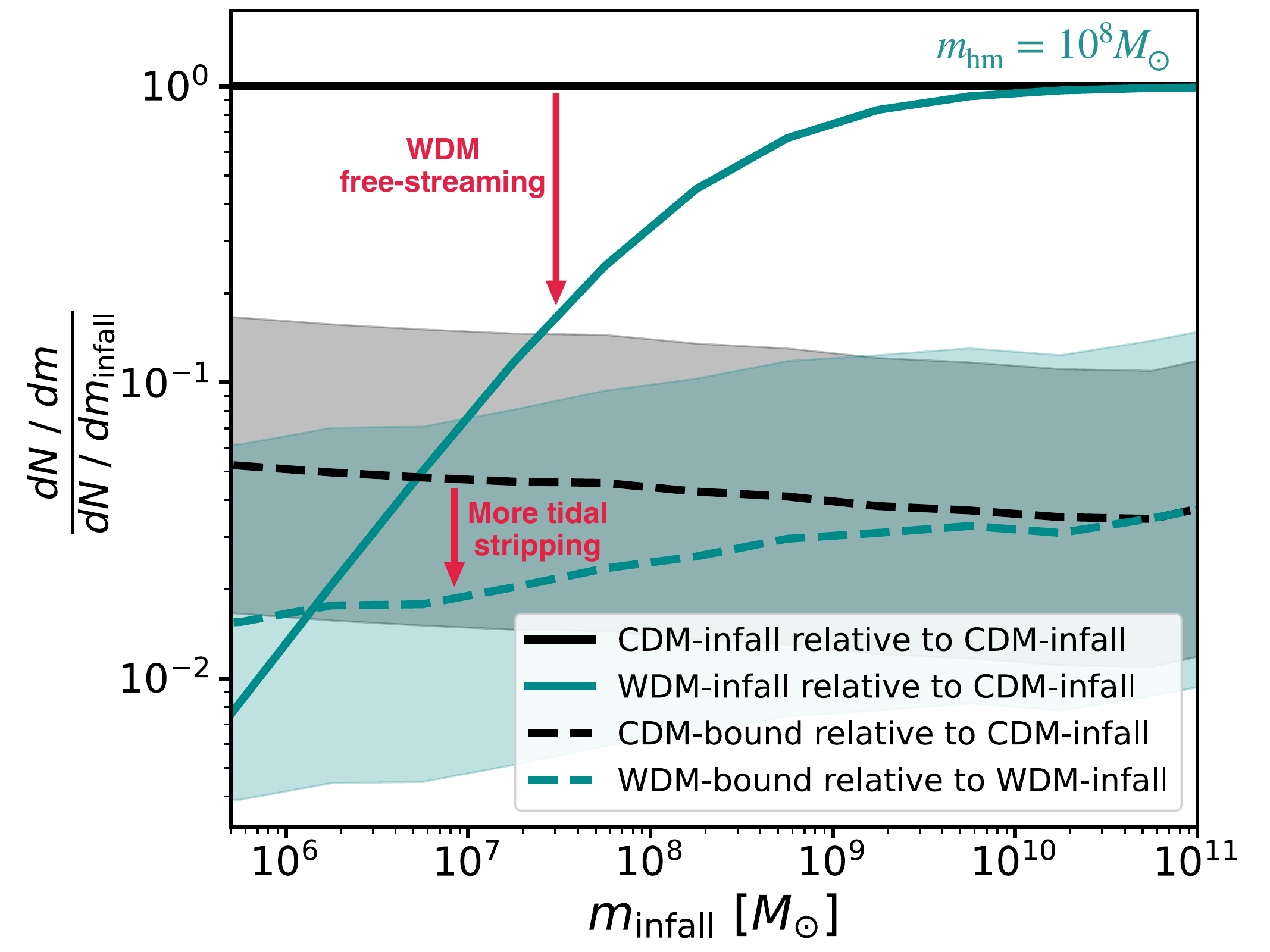}
			\caption{\label{fig:diffmfunc} {\bf{Left:}} The infall (solid lines) and bound (dashed and dotted lines) subhalo mass functions predicted by our tidal evolution model for $m_{\rm{hm}}=10^8 \mathrm{M}_{\odot}$. To suppress sample variance we show the median across 100 realizations with $\Sigma_{\rm{sub}} = 0.1 \ \rm{kpc^{-2}}$ in a circular aperture with an area $1086 \ \rm{kpc^2}$. The cyan (black) curves correspond to WDM (CDM). Linestyles correspond to bound mass functions with different assumed concentration--mass relations: The dashed black curve uses a CDM-like concentration--mass relation, the dashed cyan curve uses a WDM-like concentration--mass relation (Equation \ref{eqn:wdmmc}), and the dotted cyan curve assumes a CDM-like concentration--mass relation. {\bf{Right:}} Suppression of the subhalo bound mass function as a function of infall mass assuming $m_{\rm{hm}}=10^8 \mathrm{M}_{\odot}$. Solid curves show the amplitude of the infall mass functions relative to the CDM infall mass function. The dashed black (cyan) curve shows the bound mass function in CDM (WDM), relative to the infall mass function in CDM (WDM).  Shaded regions encompass $1 \sigma$ scatter. The extra suppression on small scales in WDM occurs due to the dependence of tidal stripping on infall concentration: because WDM halos have lower concentrations, they lose more mass.}
		\end{figure*}
		
		\subsection{Extension to warm dark matter}
		\label{ssec:wdm}
		WDM comprises a class of dark matter theory with a suppression of the linear matter power spectrum on scales $k > 1 \ \rm{Mpc^{-1}}$. A transfer function $T\left(k\right)$ characterizes this small-scale cutoff. By convention \citep{Viel++05}, we can model this transfer function
		\begin{equation}
			\label{eqn:transferfunc}
			\sqrt{\frac{P_{\rm{wdm}}\left(k\right)}{P_{\rm{cdm}}\left(k\right)}} \equiv T\left(k\right) = \left(1+\left( \alpha k \right)^{\beta}\right)^{-\gamma}.
		\end{equation}
		Following common practice, we define the half-mode scale $k_{\rm{hm}}$ as the wave number where $T\left(k_{1/2}\right)=1/2$, or $k_{\rm{hm}} = \alpha^{-1}\left(2^{1/\gamma}-1\right)^{1/\beta}$. The scale $k_{\rm{hm}}$ corresponds to a halo mass $m_{\rm{hm}} = \left(4 \pi /3 \right) \Omega_{\rm m} \rho_{\rm{crit}} \left(\pi / k_{\rm{hm}}\right)^3$, where $\rho_{\rm{crit}}$ is the critical density of the Universe at $z=0$ and $\Omega_m$ is the fractional contribution of dark matter to the critical density. 
		
		The values of $k_{\rm{hm}}$, $\beta$, and $\gamma$ depend on the mass and formation mechanism of the dark matter particle.  For example, a sterile neutrino of the same mass can be warm or cold, depending on the production mechanism~\cite{Kusenko:2009up,Abazajian:2019ejt,Zelko++22}. To enable comparison with existing WDM constraints, for this initial analysis of our full JWST dataset we assume the same thermal relic WDM model considered in many previous works. For a spin--1/2 thermal relic WDM particle\footnote{For thermal relic WDM we have $\beta \sim 2$ and can take $\gamma=5$ in Equation \ref{eqn:transferfunc} \citep{Stucker++22}.}, we can relate the half-mode mass to the particle mass
		\begin{equation}
			\label{eqn:mhm}
			m_{\rm{hm}} = M_0 \left(\frac{m_{\rm{therm}}}{3 \,\rm{keV}}\right)^{\zeta} \mathrm{M}_{\odot},
		\end{equation}
		with $\left(M_0,\zeta,\nu\right) = \left(4.0 \times 10^8 M_{\odot},-3.564,1.049\right)$ in our adopted cosmology \citep{Planck++20}. This conversion between $m_{\rm{hm}}$ and $m_{\rm{therm}}$ uses recent calculations of the linear theory transfer function presented by \citet{Vogel++23}. 
		
		Free-streaming suppresses both the number and concentration of halos less massive than $m_{\rm{hm}}$. We model the first effect by modifying the halo mass function predicted by CDM as
		\begin{equation}
			\label{eqn:wdmmfunc}
			\frac{\mathrm{d}^2N_{\rm{wdm}}}{\mathrm{d}m \mathrm{d}V} = \frac{\mathrm{d}^2N_{\rm{cdm}}}{\mathrm{d}m \mathrm{d}V} f\left(m_{\rm{hm}} / m\right)
		\end{equation}
		where
		\begin{equation}
			\label{eqn:suppression}
			f\left(x\right) = \left(1 + ax^b\right)^c
		\end{equation}
		and $a=1.95$, $b=0.8$, and $c=-1.0$ \citep{Lovell++20}. Given the concentration--mass relation in CDM, $c_{\rm{cdm}}\left(m,z\right)$, we compute the halo concentration in WDM as
		\begin{equation}
			\label{eqn:wdmmc}
			c_{\rm{wdm}}\left(m,z,m_{\rm{hm}}\right) = c_{\rm{cdm}}\left(m,z\right)f\left(m_{\rm{hm}}/m\right)
		\end{equation}
		with coefficients $a=60$, $b=1$, $c=-0.17$ for the suppression given by Equation \ref{eqn:suppression} \citep{Bose++16}. When accounting for scatter in the concentration--mass relation, we first generate halo concentrations in CDM with 0.2 dex scatter, and then then scale the resulting values by $f\left(m_{\rm{hm}}/m\right)$. 
		
		To model subhalo populations in WDM, we modify the CDM infall subhalo mass function using the same functional form as in Equation~\ref{eqn:wdmmfunc} for the WDM suppression of the field halo mass function
		\begin{equation}
			\label{eqn:wdmshmf}
			\frac{\mathrm{d}^2N_{\rm{wdm}}}{\mathrm{d}m \ \mathrm{d}A} = \frac{\mathrm{d}^2N_{\rm{cdm}}}{\mathrm{d}m \ \mathrm{d}A} f\left(m_{\rm{hm}} / m\right),
		\end{equation} 
		with $a=1.95$, $b=0.8$, and $c=-1.0$. Given the infall mass, we use the tidal evolution model by \citet{Du++25} to predict the bound mass, and calculate the density profile using the tidal tracks, as described in Section~\ref{sssec:subahlos}.
		
		\begin{figure}
			%\includegraphics[trim=3cm 0cm 3cm
			%0.cm,width=0.48\textwidth]{./figures/halo_profiles_cdm_wdm.pdf}
			\includegraphics[trim=1cm 0.3cm 0.5cm
			0.cm,width=0.45\textwidth]{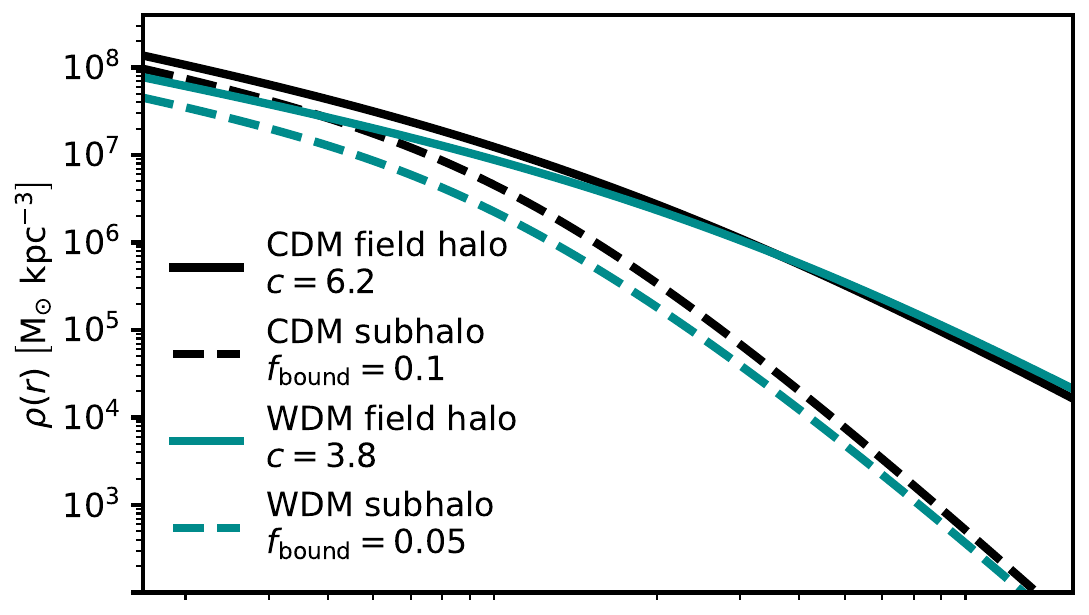}
			\includegraphics[trim=1.2cm 0.15cm 0.55cm
			0.cm,width=0.45\textwidth]{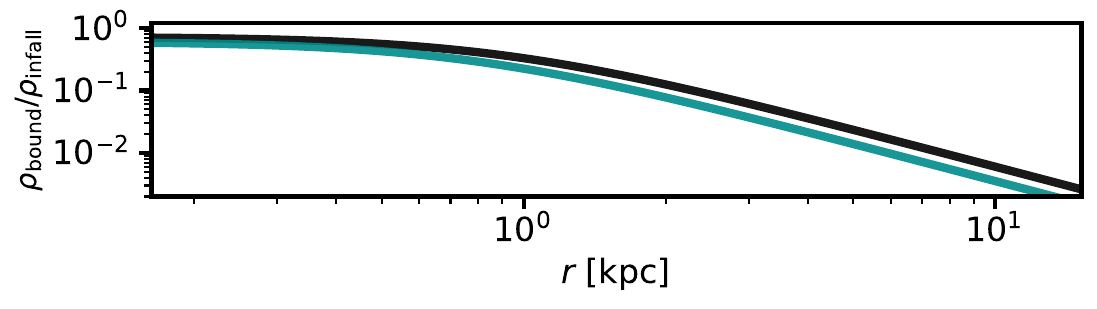}
			\includegraphics[trim=1.3cm 0.7cm 0.5cm
			0.cm,width=0.45\textwidth]{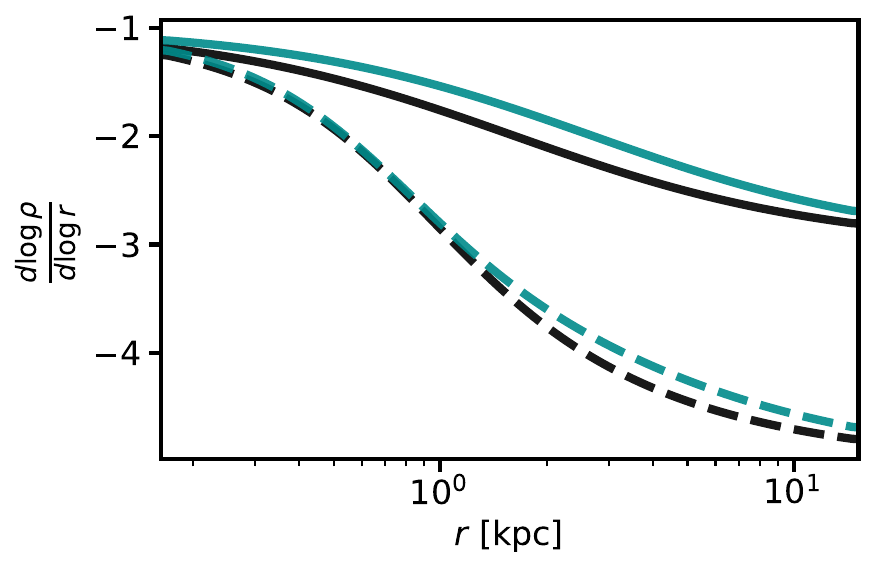}
			\caption{\label{fig:diffprofile} {\bf{Top:}} The density profiles of halos and subhalos in CDM (black), and in a WDM (cyan) model with $m_{\rm{hm}} = 10^{8.5} \mathrm{M}_{\odot}$. The field halos, with density profiles constructed at $z=2$ (solid lines), have a total mass $10^9 \mathrm{M}_{\odot}$. The scale radius of the CDM field halo is $r_\mathrm{s} = 1.6 \ \rm{kpc}$, and the CDM subhalo has this same scale radius at infall. Due to free-streaming effects, the WDM halo and subhalo have a lower concentration, with $r_\mathrm{s} = 2.7 \ \rm{kpc}$. The dashed curves show density profiles of subhalos with $m_{\rm{infall}}=10^9 \mathrm{M}_{\odot}$ obtained with our tidal stripping model \citep{Du++25}, which predicts that this WDM subhalo loses twice as much mass by $z=0.5$. The panel under the x-axis shows the density profile of the tidally-evolved subhalos, relative to the density profiles in the field or at infall, and clearly shows the effects of additional mass loss due to more efficient tidal stripping. {\bf{Bottom:}} The logarithmic profile slope of the field halos and subhalos shown in the top panel. Our tidal stripping model predicts most subhalos have logarithmic profile slopes $|d \log \rho / d \log r| \sim 2-4$ at $r_\mathrm{s}$.}
		\end{figure}
		
		As both WDM and CDM subhalos are predicted to have NFW profiles at infall, the density profiles in each case are fully specified by the tidal tracks, which in turn depend on the bound mass fraction. In WDM, halos have lower concentrations, and traverse the tidal track more quickly. A population of halos with systematically lower concentrations will have, on average, lower bound mass fractions at any given time. 
		
		The explicit dependence of the bound mass on infall concentration leads to more mass loss in WDM compared to CDM, providing an additional means to distinguish the two models. To illustrate, Figure \ref{fig:diffmfunc} shows the infall and bound mass functions in CDM (black) and WDM (cyan). WDM predicts a turnover in the infall mass function (solid cyan curve) according to Equation \ref{eqn:suppression}. Tidal stripping further suppresses the amplitude of the bound mass function in CDM (dashed black) and WDM (dashed cyan). In CDM we have $f_{\rm{bound}}\sim 0.05$, on average, for all subhalos\footnote{The bound mass fraction is only slightly higher at lower masses because lower-mass halos tend to have higher concentrations in CDM.}. In contrast, for WDM we have $f_{\rm{bound}} \sim 0.05$ only on scales $m > m_{\rm{hm}}$, and more suppression on smaller scales due to the lower infall concentrations predicted in WDM. We can see the concentration dependence of the bound mass function more clearly by examining the right panel of Figure \ref{fig:diffmfunc}, which shows the suppression at each infall mass scale. On scales $m \lesssim m_{\rm{hm}}$ the WDM bound mass function is more suppressed relative to the WDM infall mass function than the CDM bound mass function relative to the CDM infall mass function, a consequence of WDM subhalos being more susceptible to tidal stripping due to their lower concentrations. 
		
		In Figure \ref{fig:diffprofile}, we show the halo density profiles for $10^9 \mathrm{M}_{\odot}$ field halos (solid curves), and the subhalos these objects could become if they were absorbed by a $10^{13} \mathrm{M}_{\odot}$ host halo and lose mass to tidal stripping (dashed curves). The density profiles and tidal effects depicted in the figure are the predictions of our tidal evolution model \citep{Du++25}, and follow from moving the NFW profiles along their tidal tracks. These examples are representative of the kinds of perturbers that appear in our forward modeling pipeline. As shown under the $x$-axis of the top panel, this particular WDM halo loses twice as much material as its CDM counterpart, a consequence of its lower concentration at infall. The additional mass loss manifests as a tidal truncation at a smaller radius, and an overall rescaling of the density profile \citep{Du++24}. 
		
		The bottom panel of Figure \ref{fig:diffprofile} shows the logarithmic profile slope as a function of radius for the subhalos and field halos, using the same color scheme as the top panel. Due to their tidal evolution, most subhalos in our simulations have logarithmic profile slopes between -4 and -2 at the halo scale radius. We note that if one such tidally-stripped halo were to appear near an extended lensed arc and modeled as an NFW profile, the perturbation could be interpreted as coming from a halo with an anomalously high concentration \citep[e.g.][]{Vegetti++12,Minor++21,Enzi++25,He++25}, even though the existence of such objects is expected from the tidal evolution of CDM subhalos. While the discussion regarding Figure \ref{fig:diffprofile} considers  a $10^9 \mathrm{M}_{\odot}$ halo, the effects we discuss are qualitatively unchanged for other halo mass scales.
		
		\subsection{Sampling distributions and priors for dark matter hyper-parameters}
		\label{ssec:priors}
		We use a uniform sampling distribution on $\log_{10} m_{\rm{hm}}/ \mathrm{M}_{\odot}$ in the range $\left[4, 10\right]$, and quote our constraints on this parameter with the same Bayesian prior. While CDM encompasses models with halos less massive than $10^4 M_{\odot}$, and therefore $m_{\rm{hm}} =10^4 \mathrm{M}_{\odot}$ is not formally consistent with the theory, for $m_{\rm{hm}} = 10^{4} \mathrm{M}_{\odot}$ the abundance and density profiles of halos deviate by less than  $10 \%$ at $10^6 \mathrm{M}_{\odot}$, the minimum halo mass rendered in our simulations. Therefore, we can identify regions of parameter space with $m_{\rm{hm}} \lesssim \mathcal{O}\left(10^{5} \mathrm{M}_{\odot}\right)$ as effectively indistinguishable from CDM. 
		
		We draw $\delta_{\rm{LOS}}$ (Equation \ref{eqn:mfunclos}) from a sampling distribution that is uniform in $\left[0.9, 1.1\right]$, allowing for $10 \%$ difference in the amplitude of the line-of-sight halo mass function, relative to the Sheth-Tormen model prediction. We draw $\alpha$ from a uniform sampling distribution in $\left[-1.95, -1.85\right]$, around the predictions from $N$-body simulation and semi-analytic models $\alpha \sim -1.9$ \citep{Giocoli++08,Springel++08,Fiacconi++16,Benson20,Gannon++25}. These relatively narrow sampling distributions reflect the fact that the amplitude of the halo mass function and the logarithmic slope of the subhalo mass function are by now very robust predictions of collisionless dark matter above the free-streaming scale.
		
		Unlike $\delta_{\rm{LOS}}$ and $\alpha$, the amplitude of the subhalo mass function is subject to additional uncertainties related to tidal evolution. As we discuss in the following paragraphs, there remains some systematic uncertainty related to the CDM prediction for this parameter. We draw $\Sigma_{\rm{sub}} / \rm{kpc^{-2}}$ from a log-uniform sampling distribution in $\log_{10} \Sigma_{\rm{sub}}\sim \left[-2.2, 0.2\right]$, which spans a broad range around the theoretical predictions from the Symphony suite of $N$-body simulations. Relative to previous analyses \citep{Gilman++20,Keeley++24}, we extend the sampling distribution on $\Sigma_{\rm{sub}}$ to significantly higher infall mass function amplitudes to test the dependence of the results on the prior, and to demonstrate how imaging data helps break covariance between $\Sigma_{\rm{sub}}$ and $m_{\rm{hm}}$.
		
		In Section \ref{sec:results}, we will incorporate theoretically-motivated priors on $\Sigma_{\rm{sub}}$ from both {\tt{galacticus}} and the Symphony $N$-body simulations \citep{Nadler++23} when quoting constraints on $m_{\rm{hm}}$. We will also compare measurements of subhalo abundance in CDM with the theoretical predictions. We determine the theoretical predictions for $\Sigma_{\rm{sub}}$ by examining the bound mass function in dark-matter-only simulations in a 10 kpc annulus around the center of the hosts \citep{Gannon++25}. Specifically, we determine the prior on $\Sigma_{\rm{sub}}$, the amplitude of the mass function at infall, that results in the bound mass function predicted by either Symphony or ${\tt{galacticus}}$, given our tidal evolution model \citep{Du++25}. From our tidal evolution model, in CDM, for a group scale halo of 10$^{13}$ M$_\odot$ at redshift 0.5, the bound mass function is related to the infall mass function by a constant factor, which we refer to as the mean bound mass fraction, $\bar{f}_{\rm{bound}}=0.05$ \citep{Du++25}. When determining priors on $\Sigma_{\rm{sub}}$, we also take into account the additional $15\%$ suppression of the bound mass function amplitude caused by tidal stripping effects from a central elliptical galaxy \citep{Gannon++25}. 
		
		The {\tt{galacticus}} dark-matter-only prediction for the bound mass function amplitude corresponds to $\bar{f}_{\rm{bound}} \times\Sigma_{\rm{sub}} = 0.009 \ \rm{kpc^{-2}}$. Including an additional $15 \%$ suppression due to tidal stripping by the central galaxy gives $0.0075 \ \rm{kpc^{-2}}$. This gives $\Sigma_{\rm{sub}} = 0.15 \ \rm{kpc^{-2}}$ as the prediction corresponding to $\tt{galacticus}$. We note that because we use {\tt{galacticus}} to calibrate the tidal evolution model, by construction we also match the predicted infall mass function amplitude. This is not the case for the $N$-body simulations, for which we only have knowledge of the bound mass function amplitude. As discussed by \citet{Gannon++25}, the Symphony $N$-body simulations, analyzed using the latest generation of halo finders \citep{Mansfield++24}, predict a bound mass function amplitude suppressed by $\sim 50 \%$ relative to {\tt{galacticus}} in the inner regions of the host, corresponding to  $\bar{f}_{\rm{bound}}\times\Sigma_{\rm{sub}}=0.006 \ \rm{kpc^{-2}}$. Including the additional $15 \%$ suppression from the central galaxy, we find $\Sigma_{\rm{sub}}=0.1 \ \rm{kpc^{-2}}$ gives the amplitude of the bound mass function predicted by $N$-body simulations. Based on these considerations, we use theoretically-motivated priors based on {\tt{galacticus}} $\pi\left(\log_{10} \Sigma_{\rm{sub}} / \rm{kpc^{-2}}\right) = \mathcal{N}\left(-0.8, 0.2\right)$ and Symphony $\pi\left(\log_{10} \Sigma_{\rm{sub}} / \rm{kpc^{-2}}\right) = \mathcal{N}\left(-1.0, 0.2\right)$. The width of these priors is motivated by the 0.2 dex difference between the predicted mean $\Sigma_{\rm{sub}}$ values, which we take as a representative level of systematic uncertainty between different tools for examining non-linear structure formation.  
		
		Finally, for each lens we sample the host halo mass from a log-normal prior $\log_{10} m_{\rm{host}}/ M_{\odot}\sim\mathcal{N}\left(13.3, 0.3\right)$, based on the typical host halo masses for strong lens systems \citep{Lagattuta++10}. We marginalize over $m_{\rm{host}}$ for each lens before multiplying likelihoods, meaning the host halo mass is treated as a nuisance parameter in Equation \ref{eqn:likelihood}.  
		
		\subsection{Globular clusters}
		\label{ssec:globularclusters}
		The change to an image magnification from small-scale structure depends on the size, mass, and central density of a perturber. The high central density of globular clusters (GCs), together with the fact that their total masses are of order $10^5 \mathrm{M}_{\odot}$--$10^6 \mathrm{M}_{\odot}$, raises the possibility that these objects may contribute as a source of small-scale to flux ratios \citep{MaoSchneider98}. 
		
		We include GCs in our forward model via the following steps. First, we draw their masses $m_{\rm{gc}}$ from a log-normal mass function with a mean of $\log_{10} m_{\rm{gc}}/\mathrm{M}_{\odot} = 5.3$ and a standard deviation of 0.6 dex \citep{Jordan++07,Parmentier++07}. We assume a surface mass density in GCs of $10^{5.6} \mathrm{M}_{\odot} \rm{kpc^{-2}}$ \citep{Peng++08,He++18}. Second, we randomly distribute these objects around each lensed image in a circular aperture of radius $0.2 \ \rm{arcsec}$. We model the GCs as point masses. 
		
		While we include these objects in our analysis, we do not find that they have a significant effect on flux ratio statistics. Although their compact inner structure can lead to strong perturbations, these events occur only if the GC makes a ``direct hit'' on a lensed image, and the cross section for direct hits is much lower than the probability that a dark matter halo imparts a perturbation of equal strength. We provide further discussion of flux-ratio perturbations by globular clusters in Appendix~\ref{app:globularclusters}. 
		
		\section{Lens modeling}
		\label{sec:lensmodeling}
		This section describes aspects of the lens modeling, including assumptions and overall strategy. In Section~\ref{ssec:macromodel}, we discuss assumptions related to the lens macromodel, which comprises the combined projected stellar and dark matter mass profile of the main deflector from the galaxy and host dark matter halo. Section~\ref{ssec:luminoussats} discusses the modeling luminous satellite or companion galaxies. Section~\ref{ssec:lensmodeling} discusses how we reconstruct the imaging data, including how we choose the complexity in the source light profile and considerations of the PSF model. Section~\ref{sssec:quasarlight} discusses how we model the unresolved nuclear narrow-line and warm dust regions around the background quasar, and finite-source effects relevant for computing the image magnifications. In Section~\ref{ssec:lmspecific}, we review lens modeling considerations for the 24 systems for which we model the imaging data, providing additional discussion of individual cases, if necessary.
		
		\subsection{Main deflector mass profile}
		\label{ssec:macromodel} 
		Following standard practice, we model the combined projected mass in dark matter from the host halo and the stellar mass from the central galaxy in the region relevant for strong lensing using an elliptical power-law (EPL) profile \citep{TessoreMetcalf15}
		\begin{equation}
			\kappa_{\rm{EPL}}\left( R \right) = \frac{3-\gamma}{2}\left(\frac{\theta_{\rm{E}}}{R}\right)^{\gamma-1}
		\end{equation}
		where $R = \sqrt{qx^2 + y^2/q}$ is the elliptical radius, $q$ is the axis ratio, and $\gamma$ is the logarithmic profile slope. This mass profile can be rotated to any position angle $\phi_q$. We add external shear in the main lens plane with a strength $\gamma_{\rm{ext}}$ at an angle $\phi_{\rm{ext}}$. This class of models has been shown to provide an excellent description of the mass density profile of lens galaxies \citep[e.g.,][]{Millon++20,TDCOSMO25}.
		
		\begin{table*}
			\setlength{\tabcolsep}{11pt}
			\caption{\label{tab:tablenmax} From left: Lens system, deflector and source redshifts, instrument used to measure imaging data, quasar host galaxy model (notation ``ES+n" means elliptical S\'ersic (ES) plus shapelets of order $n_{\rm{max}}$), lens model (notation EPLSM means EPL+SHEAR+MULTIPOLES), and $N_{\rm{real}}$, the number of realizations generated when evaluating Equation \ref{eqn:likelihood}. We place systems without measured lens redshifts at $z_{\rm{d}}=0.5$, marked with $\star$.}
			\begin{tabular}{ccccccc}
				\hline
				Lens system & $z_{\rm{d}}$ & $z_{\rm{s}}$ & imaging data & source light & lens model & $N_{\rm{real}}/10^6$ \\
				\hline 
				PSJ0147+4630 & 0.68 & 2.38 & - & - & EPLSM & $4.5$
				\\\\
				J0248+1913 & 0.5$^\star$ & 2.44 & HST F814W &  ES only & EPLSM + SIS & 13.8 \\
				\\
				J0259-1635 & 0.91 & 2.16 &  HST F814W & ES + 10& EPLSM & 2.5\\
				\\
				J0405-3308 & 0.5$^\star$ & 1.70 &  HST F814W & ES + 3& EPLSM & $21.1$\\
				\\
				MG0414+0534 & 0.96 & 2.64 & MIRI F560W & ES only & EPLSM+SIS & $4.0$\\
				\\
				HE0435-1223 & 0.45 & 1.69 &  NIRCam F115W & ES + 28 & EPLSM+SIS  & $5.1$\\
				\\
				J0607-2152 & 0.56 & 1.30 & MIRI F560W &  ES + 3& EPLSM+SIS & $7.1$\\
				\\
				J0608+4229 & 0.5$^\star$  & 2.35  & MIRI F560W &  ES only & EPLSM & 15.0\\
				\\
				J0659+1629 & 0.77 & 3.10 &  MIRI F560W & ES + 1 & EPLSM+SIS & 6.9\\
				\\
				J0803+3908 & 0.5$^\star$ & 2.97 &  MIRI F560W & ES + 2 & EPLSM & 15.5\\
				\\
				RXJ0911+0551 & 0.77 & 2.76 &  - & - & EPLSM+SIS & 0.3\\
				\\
				J0924+0219 & 0.39 & 1.52 &  MIRI F560W & ES + 6& EPLSM & 6.0\\
				\\
				J1042+1641 & 0.59 & 2.50 &  HST F160W& ES + 5 + S & EPLSM+SIS & 10.1\\
				\\
				PG1115+080 & 0.31 & 1.71 & NIRCam F115W & ES only & EPLSM & $2.0$\\
				\\
				GRAL1131-4419 & 0.47 & 1.09 &  HST F814W& ES + 10& EPLSM & 2.2\\
				\\
				RXJ1131-1231 & 0.30 & 0.66  & HST F814W  & ES + 34& EPLSM+SIS & $3.6$ \\
				\\
				2M1134-2103 & 0.66 & 2.77 & MIRI F560W & ES only& EPLSM+SIS& 2.2 \\
				\\
				J1251+2935 & 0.40 & 0.80  & HST F814W& ES + 7 & EPLSM & 1.0\\
				\\
				H1413+117 & $1.15$ & 2.56  & MIRI F560W & ES only & EPLSM+SIS & 3.2
				\\
				\\
				B1422+231 & 0.34 & 3.62  & - & -& EPLSM & 1.2\\ 
				\\
				J1537-3010 & 0.59 & 1.71  & HST F814W& ES + 8& EPLSM & 3.1\\
				\\
				PSJ1606-2333 & 0.92 & 1.70  & HST F814W & ES + 8 & EPLSM+SIS & 7.0\\
				\\
				WFI2026-4536 & 0.5$^\star$ & 2.23 & - & - & EPLSM & 0.3\\
				\\
				WFI2033-4723 & 0.66 & 1.66  & NIRCam F115W & ES + 24  & EPLSM+2$\times$SIS & $6.4$\\
				\\
				J2038-4008 & 0.23 & 0.78  & HST F814W & ES + 16 & EPLSM & 0.7\\
				\\
				J2145+6345 & $0.50^{\star}$ & 1.56  & MIRI F560W & ES only & EPLSM+SIS & 11.1\\
				\\
				J2205-3727 & 0.63 & 1.85  & MIRI F560W & ES only & EPLSM & 2.6\\
				\\
				J2344-3056 & 0.5$^\star$ & 1.3 & MIRI F560W & ES only & EPLSM & 16.4\\
				\\
				\hline
			\end{tabular}
		\end{table*}
		
		We include deviations from ellipticity in the projected mass profile of the main deflector by adding elliptical multipoles \citep{PaugnatGilman25}. In the context of strong lensing analyses, multipoles refer to perturbations in 2 dimensions around projected isodensity contours. Given an elliptical iso-density contour $R=constant$, elliptical multipoles introduce deviations $\delta R$ to the elliptical radius that follow 
		\begin{equation}
			\delta R = a_{\rm{m}} \cos \left(m \left(\varphi - \varphi_{\rm{m}}\right)\right)
		\end{equation}
		where $\varphi = \arctan\left(qx, y\right)$ is the eccentric anomaly. Due to the dependence of $\delta R$ on the axis ratio $q$, the expressions for the convergence and deflection angles become considerably more complicated than the expressions for the more commonly used circular multipoles, and we refer to \citet{PaugnatGilman25} for further details. We include elliptical multipole terms of order $m=1$, $m=3$, and $m=4$, each characterized by a strength $a_m$ and an orientation $\varphi_m$ relative to the ellipse semi-major axis. The multipole terms are concentric with the EPL profile, with orientations $\varphi_{\rm{m}}$ allowed to rotate freely with respect to $\phi_q$. This representation includes an overall amplitude $a_m$ and an angle $\phi_m$, and is equivalent to using two different amplitudes for an expansion with both sine and cosine terms \citep[e.g.][]{Xu++15,Oh++24}. 
		
		Many analyses have investigated how overly simplistic modeling of the main deflector mass profile could lead to spurious detections of substructure \citep[e.g.][]{Trotter++2000,Congdon++05,Hsueh++18,Cohen++24,Riordan++24}. Our inclusion of multipole perturbations is crucial for avoiding overly simplistic lens modeling assumptions, which could bias the model-predicted flux ratios and our interpretation of the data. Our analysis improves recent treatments of this challenge in strong lens modeling by simultaneously varying the multipole strengths alongside the parameters describing entire halo populations. This accounts for potential correlations between azimuthal structure and the positions, masses, and density profiles of (sub)halos near lensed images, allowing us to statistically distinguish between various sources of small scale perturbation \citep{Gilman++24}.
		
		As discussed in Section \ref{sssec:astrometriclike}, we solve for a subset of macromodel parameters $\xlens$ such that the macromodel satisfies the lens equation for the observed image positions. Given the 8 constraints from the four lensed image positions\footnote{Recall that we do not use the flux ratios or the imaging data for this step in the analysis.}, we vary only $\theta_{\rm{E}}$, $G_{1,x}$, $G_{1,y}$, $\phi_q$, $\gamma_{\rm{ext}}$, and $\phi_{\rm{ext}}$, $\beta_x$, $\beta_y$, i.e. the Einstein radius, main deflector mass centroid, ellipticity position angle, external shear strength and orientation, and the source position, respectively. The remaining macromodel parameters, which include the EPL axis ratio, $q$, the power-law slope $\gamma$, the multipole terms, and satellite galaxy positions and masses (see next subsection) are held fixed to the values drawn from a macromodel prior $p\left(\xlens\right)$. 
		
		For each lens, we draw $\gamma$ from $\mathcal{N}\left(2.1, 0.1\right)$ \citep{Auger++10}. We make two exceptions to this prior when the data exhibit a strong preference for a significantly different $\left(> 2 \sigma\right)$ logarithmic slope, as discussed further in Section~\ref{ssec:lmspecific} where we comment on individual lens systems. This choice primarily serves to make the sampling for these systems more efficient. We draw $q$ from a truncated normal distribution, assuming that the ellipticity of the main deflector mass profile is not significantly more elongated than that of the light. In particular, we draw $q$ from a truncated normal distribution with a mean equal to $q_{\rm{light}}$, a lower bound $\min \left(0.4, q_{\rm{light}}-0.15\right)$, and a standard deviation of 0.2 \citep{Keeley++25}. For the elliptical multipole perturbations, we allow their orientations to vary freely within $\pm \pi / \left(2m\right)$, unless $a_4 > 0.02$, in which case we force $\varphi_{4}=0$, because very disky isodensity contours tend to align with the semi-major axis \citep{Hao++06}. We draw the strength $a_m$ from a Gaussian $\mathcal{N}\left(0, \sigma_m\right)$, with $\sigma_1 = 0.005$, $\sigma_3 = 0.005$, $\sigma_4 = 0.01$. The priors on $a_m$ are based on measurements of isophote shapes of massive elliptical galaxies \citep{Hao++06,Oh++24,Amvrosiadis++25}. This approach directly ties the prior distribution for the multipole amplitudes to the measured deviations from ellipticity in the light on the population level. One caveat of our implementation is that the priors on the Fourier coefficients were derived for circular multipole perturbations, rather than elliptical perturbations. However, we expect the strength of elliptical multipole perturbations be comparable, and perhaps smaller~\citep{PaugnatGilman25}.  
		
		\subsection{Luminous satellites}
		\label{ssec:luminoussats}
		We render dark matter halos up to a maximum mass of $5\times 10^{10} \mathrm{M}_{\odot}$. We include more-massive perturbers as satellite galaxies when we detect them from their stellar light. We model these luminous satellites as singular isothermal spheres (SIS), and place them at their observed locations in the plane of the lens with uncertainties in their positions of 50 m.a.s. We make two exceptions in the case of HE0435 and WFI2033, for which the redshifts of companion galaxies have been measured to be different from the main deflector. We discuss this topic in more detail in Sections~\ref{sssec:he0435} and~\ref{sssec:wfi2033}. 
		
		Where possible, we use priors on the Einstein radius of the perturber $\theta_{\rm{E},G2}$ derived from a velocity dispersion measurement. If no velocity dispersion is available, we use, as a starting point, a prior derived from from the total integrated flux of the satellite galaxy light: $\theta_{\rm{E},G2} = \theta_{E} \sqrt{L_{\rm{G2}}/{L_{\rm{G1}}}}$, $\theta_{E}$ is the main deflector's Einstein radius and $L_{\rm{G1}}$ is the luminosity of the main deflector light profile and $L_{\rm{G2}}$ is luminosity of the satellite light profile. This relation follows from the Faber-Jackson relation $L \propto \sigma^4$  \citep{FaberJackson76}, and assumes $\theta_{E} \propto \sigma^2$, which is a good approximation for near-isothermal mass profiles. In most cases, the imaging data and flux ratios will constrain further the Einstein radius of the satellite. To make the sampling of the parameter space required to evaluate Equation~\ref{eqn:likelihood} more efficient, for a subset of systems, we adjust priors on the satellite Einstein radius to align with the values that best-reproduce the flux ratios and imaging data in the lens modeling. In Section~\ref{ssec:lmspecific}, where we discuss the lens modeling of individual systems, we state the assumed priors on satellite galaxy Einstein radii for each individual system. 
		
		\subsection{Lens and source light modeling}
		\label{ssec:lensmodeling}
		This section describes the modeling of the lens and source light, which comprise the parameters $\xlight$ that appear in Equation \ref{eqn:likelihood}. Three ingredients enter the calculation of the imaging data likelihood: First, the lens light from the main deflector and nearby satellite galaxies (Section~\ref{sssec:lenslight}). Second, the PSF model, which is particularly relevant for reconstructing the lensed arcs near images of the lensed quasar (Section~\ref{sssec:psf}). Third, the source light profile used to model the quasar host galaxy (Section~\ref{sssec:sourcelight}), which becomes deformed into lensed arcs. 
		
		\subsubsection{Lens light model}
		\label{sssec:lenslight}
		We model the lens surface brightness as an elliptical S{\'e}rsic profile \citep{Sersic63}. In some cases, a single elliptical S{\'e}rsic profile does not fully capture the surface brightness near the center of the main deflector. If we observe a clear residual pattern in the lens light near the center of the main deflector, we mask out the inner $0.2$ arcseconds. This is acceptable because the inner regions of the deflector surface brightness do not affect the inference based on the lensed arc light, which is typically an arcsecond away. The light from each satellite galaxy, if they are present, is modeled as a circular S{\'e}rsic profile concentric with the position of the satellite in the lens model. 
		
		\subsubsection{Point spread function}
		\label{sssec:psf}
		We model imaging data in HST F160W, HST F814W, MIRI F560W, and NIRCam F115W filters, using the appropriate point spread function for each band. For the systems with HST data, we obtain initial estimates of the PSF from nearby stars in the field, and estimate a PSF uncertainty map by stacking the individual PSF estimates from different stars. Using the initial PSF estimated from stars and the corresponding uncertainty map, we model each lens system (without dark substructure), and iteratively reconstruct the PSF simultaneously with the lensed image \citep{Shajib++19,Birrer++19}. For these tasks, we use PSFr \citep{Birrer++19,Birrer++22}. For the three systems with NIRCam data (PG1115, HE0435, WFI2033), we use the NIRCam F115W PSF model presented by \citet{Williams++25}. This PSF model was computed using the {\tt{starred}} software package \citep{Millon++24}, and provides an excellent representation, including an uncertainty map, of the NIRCam PSF.  
		
		The MIRI imaging data does not have stars from which we can estimate the PSF. Instead, we use the PSF model developed by \citet{Keeley++24}, which is based on the \texttt{webbpsf} software.  The lack of a PSF uncertainty map results in strong residuals at the image positions due to small imperfections in the PSF model and the bright quasar images. When computing the imaging data likelihood, we therefore mask the quasar images to remove artifacts near the image positions. These masks appear in the imaging data residual maps associated with each lens model, which we present in Appendix~\ref{app:baselinelensmodels}. The resulting PSF models obtained for the HST, MIRI, and NIRcam data are stored for future use in the dark matter inference. 
		
		\subsubsection{Quasar host galaxy model}
		\label{sssec:sourcelight}
		To model the quasar host galaxy, we use a combination of an ellipical S{\'e}rsic profile and circular shapelets \citep{Refregier03,Birrer++15}. Shapelets form an orthonormal basis set for analyzing one and two-dimensional images. They are characterized by an angular scale $\beta^{\prime}$, and provide an increasingly accurate representation of small-scale features in an image as one increases the number of basis functions via the parameter $n_{\rm{max}}$. The choice of $n_{\rm{max}}$ depends on the resolution of the imaging data, the signal to noise ratio in the lensed arcs, and the complexity of the quasar host galaxy. 
		
		As both $n_{\rm{max}}$ and $\beta^{\prime}$ depend on the specific properties of each lens system we must determine them on a case-by-case basis. We reproduce the observed data as well as possible without overfitting by choosing the value of $n_{\rm{max}}$ that minimizes the Bayesian information criterion (BIC)
		\begin{equation}
			\label{eqn:bic}
			\rm{BIC} = k \log \left(n\right) - \max \left[\log \mathcal{L}\right],
		\end{equation} 
		where $n$ represents the number of data points, $k$ is the number of model parameters, and $ \max \left[\log \mathcal{L}\right]$ represents the maximum likelihood of the imaging data given the model. We calculate the BIC using the same forward modeling process described in Section \ref{sec:inference}, but omit dark matter substructure, and do not include the flux ratios in the log-likelihood. We use a uniform prior on $\beta^{\prime}$ with a lower bound $\beta^{\prime}_{\rm{min}} = \left(d_{\rm{pix}}/2.5\right) \sqrt{n_{\rm{max}}+1}$ and an upper bound $5\beta^{\prime}_{\rm{min}}$, where $d_{\rm{pix}}$ is the angular size of a pixel. The lower bound on $\beta^{\prime}$ is intended to prevent the shapelets from fitting artifacts in the imaging data on smaller angular scales than the angular size of a pixel \citep{Birrer++15}. In practice, the inferred values of $\beta^{\prime}$ are not significantly affected by the prior. 
		\begin{figure*}
			\centering
			\includegraphics[trim=5cm 1cm 5cm
			0cm,width=0.95\textwidth]{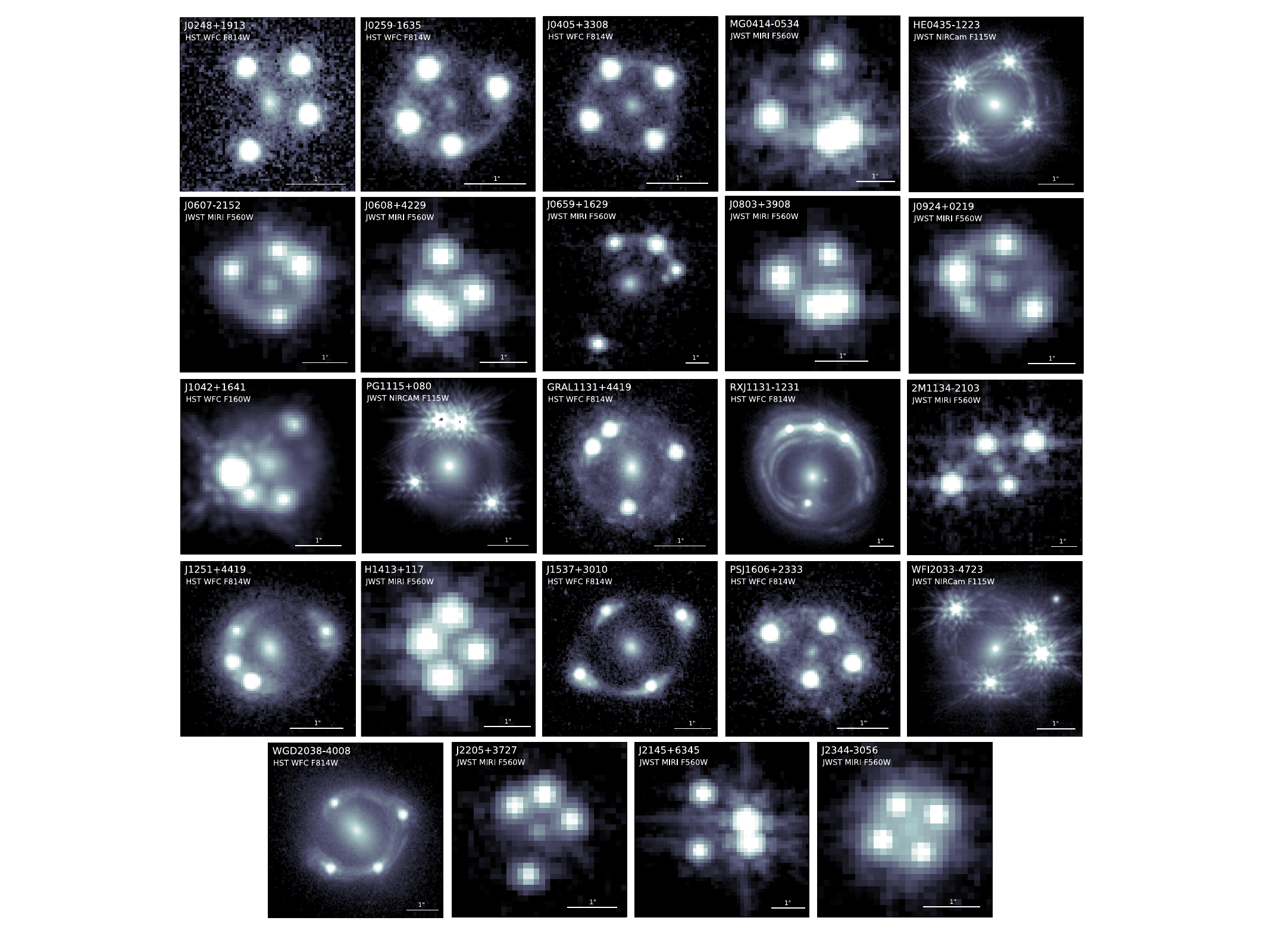}
			\caption{\label{fig:mosaic} The 24 systems for which we jointly model flux ratios and imaging data. Each panel shows the lens system as it appears in the imaging band used in the analysis, as written under the target name. A white line in the bottom right of each panel spans 1 arcsecond, and the color scale spans a factor of 2 in $\log_{10}$flux. Lens models for these systems that form the starting point of the dark matter inference are presented in Appendix \ref{app:baselinelensmodels}.}
		\end{figure*}

		There are two scenarios for which we choose a different $n_{\rm{max}}$ than the one that minimizes the BIC. First, the lensed arcs may exhibit so much small-scale complexity that the BIC continues to drop for every simulated $n_{\rm{max}}$, up to $n_{\rm{max}}=40$. In these cases, however, the flux ratios predicted by the lens model converge at some $n_{\rm{max}}$, and we choose the convergence value of $n_{\rm{max}}$ for the dark matter analysis. In other cases, the BIC continues dropping even when considering lensed arcs with relatively low angular resolution and signal to noise ratio, for which we do not expect a significant amount of structure in the lensed arcs to warrant a large $n_{\rm{max}}$. This occurs only when considering the MIRI imaging data, which suggests the shapelets are compensating for residuals near the quasar images that arise from our imperfect model for the MIRI PSF, and the lack of a PSF error map to down-weight these features. In these cases, we use an elliptical S{\'e}rsic profile to model the source light. The value of $n_{\rm{max}}$ used to reconstruct the imaging data for each system is listed in the 5th column of Table~\ref{tab:tablenmax}. 
		
		\subsection{Quasar warm dust and narrow-line regions}
		\label{sssec:quasarlight}
		Our analysis requires measurements of image fluxes that emanate from a spatially-extended region around the background quasar. The extended source removes contamination from stellar microlensing, and washes out time variability in the quasar flux on timescales less than the light crossing time of the region. 
		
		The strength of the perturbation by a halo depends on the angular size of the source relative to the deflection angle produced by a halo \citep{Dobler++06,Nierenberg++24}, so the size of the emission region where we measure the image fluxes must be accounted for in the dark matter inference. To compute image magnifications for a given lens model, we ray trace around each image coordinate to the source plane on a finely-sampled grid, and compute the magnification of each image by integrating the total flux in the image plane (see the bottom row of Figure \ref{fig:1131zooms}). For these calculations, we model the surface brightness of the emission region around the background quasar as a circular Gaussian. For the warm dust measured by JWST, we sample the full-width at half-maximum of the source size from a uniform distribution between 1-10 pc \citep{Sluse++13}. For the more spatially-extended nuclear narrow-line emission, we sample the source size from a uniform distribution between 20-80 pc \citep{Muller-Sanchez++11}. 
		
		\subsection{Discussion of individual systems}
		\label{ssec:lmspecific} 
		In this section, we will discuss considerations that affect the lens modeling decisions for the 28 systems analyzed in this work, including the 24 depicted in Figure~\ref{fig:mosaic}, for which we jointly model flux ratios with imaging data. 
		
		We have developed what will be referred to as \textit{baseline} lens models for the 24 systems shown in Figure~\ref{fig:mosaic}. These baseline models consist of the macromodel fit to the image positions and imaging data, with the degree of source light complexity specified in Table~\ref{tab:tablenmax}. The baseline models include the same lens mass profile and source light model used in the substructure lensing analysis, but omit small-scale perturbations from multipoles, dark matter halos, or globular clusters. The most-probable lens models produced in the dark matter analysis reproduce the imaging data to the same level of fidelity as these baseline lens models. The full lens reconstruction, the reconstructed arc imaging data (subtracting main deflector and quasar light), the magnification model, and normalized residuals, are shown for each baseline model in Figures~\ref{fig:bmodel1}-\ref{fig:bmodel5} in Appendix \ref{app:baselinelensmodels}. 
		
		The lens mass, lens light, and source light models for each system, as well as the reconstructed PSFs, are stored in the open-source repository {\tt{samana}}. Table \ref{tab:tablemacroparams} in Appendix \ref{app:baselinelensmodels} lists macromodel parameters inferred for each system using the image positions, flux ratios, and lensed arcs, after marginalizing over substructure. Notebooks that reproduce the baseline lens models may serve as useful starting points for future modeling of these systems for other scientific purposes. We note that, for some systems, we have identified several satellite galaxies near the main deflector that affect the lens model, but do not appear in other published lens models. Unless otherwise noted in the discussion of each system, we model the flux ratios from the warm dust region observed through JWST GO-2046 \citep{Nierenberg++20,Keeley++24,Keeley++25}. 
		
		\subsubsection{PSJ0147+4630}
		This system has multi-band HST imaging \citep{Shajib++19}, but more prominent lensed arcs than those visible in the HST data appear in the bluest MIRI filter, F560W. However, the cruciform PSF structure in the MIRI data extends across most of the image plane and dominates the surface brightness of the lensed arcs, which makes the imaging data for this system more susceptible to biases associated with an imperfect PSF model. We therefore do not incorporate constraints from lensed arcs for this system, and use only the image positions and flux ratios to constrain the lens model. 
		
		This system has a flux-ratio anomaly associated with image D, which is de-magnified by a factor of $\sim 2$ relative to the predictions of the macromodel. Image D appears $1.05$ arcseconds away from the main deflector, while the closest among images A-C in the merging triplet is $2.21$ arcseconds away. The apparent flux-ratio anomaly in this system likely derives from the macromodel parameterization, which has a constant logarithmic profile slope at all radii, in contrast to composite parameterizations that model stars and dark matter separately, which tend to exhibit power-law slopes that change across angular scales comparable to the Einstein radius \citep{Gilman++17,Shajib++21}. Given that a logarithmic profile slope that changes across an angular scale $\sim \theta_{\rm{E}}$ could lead to a differential de-magnification of image D relative to images A-C, we do not use the D/A flux ratio for this system, and rely on the more robust smooth lens model predictions for the relative magnifications among the merging triplet,images A-C. 
		
		\subsubsection{J0248+1913}
		%Using 3-band photometry of the main deflector light \citep{Shajib++19}, we estimate a main deflector redshift $\zlens = 0.96 \pm 0.06$. 
		We model the imaging data obtained with HST WFC3 through filter F814W. Preliminary analysis revealed an extreme flux-ratio anomaly ($> 50 \%$) in this system, significantly larger than milli-lensing perturbations expected in CDM. While investigating possible origins for this anomaly, we identified a faint luminous feature in the imaging data $\sim 1$ arcsec away from image A. Including an SIS at the position of the feature with an Einstein radius $\theta_{\rm{E},G2} \sim 0.1 \ \rm{arcsec}$ improves the fit to the imaging data, and brings the flux ratios into better agreement with those predicted by lens models without substructure. We therefore identify this object as a probable satellite galaxy relevant for the flux ratio analysis, and include it in the lens model. We sample the mass $\theta_{\rm{E,G2}} = \mathcal{N}\left(0.1, 0.1\right)$. 
		
		\subsubsection{J0259-1635}
		We model the imaging data obtained with HST WFC3-UVIS through filter F814W. It has no detectable satellite galaxies within several Einstein radii of the lensed images. 
		
		\subsubsection{J0405-3308}
		%We obtain a photometric redshift estimate $\zlens = 0.85 \pm 0.07$ using 3-band photometry of the main deflector \citep{Shajib++19}. 
		We model the imaging data obtained with HST WFC3-UVIS through filter F814W. A near-complete Einstein ring is visible. 
		
		\subsubsection{MG0414+0534}
		Discovered nearly 30 years ago \citep{Hewitt++92}, this system has been observed and modeled extensively, but lensed arcs do not appear in the available imaging data. Prominent lensed arcs do appear, however, in MIRI F560W. 
		
		Using VLBI images of the lens system, \citet{Ros++00} identify a probable satellite galaxy ``GX'' between images C and D. The lensed arcs provide additional evidence for the presence of this satellite -- including an SIS at the coordinates of GX reproduces the structure of the lensed arc, in particular, the splitting of the arc north of image C, as is clearly visible in the second column of Figure \ref{fig:bmodel1}. We include the satellite in our lens models with a prior on the the mass $\theta_{\rm{E,G2}} = \mathcal{N}\left(0.1, 0.1\right)$ and astrometric uncertainties of 50 m.a.s. relative to the positions quoted by \citet{Ros++00}. Section~\ref{ssec:casestudies} includes additional discussion and illustrations regarding the role of lensed arcs in constraining the mass and position of the satellite. 
		
		\subsubsection{HE0435-1223}
		\label{sssec:he0435}
		This system has been modeled extensively, including for time-delay cosmography \citep{Wong++17}, using both HST data and adaptive optics \citep{Chen++19}. We model the imaging data for this systems using new observations from JWST-NIRCam in F115W \citep{StiavelliMorishita17}. Relative to HST and adaptive optics imaging, the lensed arcs appear spectacularly bright in the NIRCam data, with numerous secondary arcs appearing around the primary Einstein ring (see Figure~\ref{fig:bmodel1} in Appendix \ref{app:baselinelensmodels}). 
		
		This system has a background galaxy at $z_{\rm{G,2}}=0.78$, located $\sim 4.2$ arcseconds from the main deflector at observed position $\left(-2.45, -3.60\right)$ relative to the main deflector centroid. As this object is behind the main deflector, its actual location along the line of sight will differ from the observed position due to foreground lensing effects by the main deflector. We use the baseline lens model to estimate the physical location of the galaxy, given the macromodel and the observed position, and infer a physical location $\left(-1.82, -3.08\right)$. The background galaxy is placed at this estimated physical location during the substructure inference with astrometric uncertainties of 0.1 arcsecond, and we sample the mass $\theta_{\rm{E},G2} \sim \mathcal{N}\left(0.35,0.05\right)$ \citep{Wong++17}. 
		
		\subsubsection{J0607-2152}
		We estimate a deflector redshift $z_{\rm{d}} \sim 0.56$ from features in the quasar spectrum observed by the Low Resolution Imaging Spectrometer (LRIS) \citep{Keeley++25}. We model the imaging data obtained with JWST MIRI through filter F560W. A near-complete Einstein ring is visible. A satellite galaxy is visible near the merging image pair, which we include in the lens model with a prior on the mass $\theta_{\rm{E},G2} \sim \mathcal{N}\left(0.05,0.05\right)$. Initially, we assumed a prior based on the flux of the satellite and the Faber-Jackson relation centered on $\theta_{\rm{E},G2} \sim 0.1$, but the heavier satellite masses are ruled out by the combined imaging data and flux ratio likelihoods. Lens models that match the flux ratios and imaging data often have very high magnification ($\gtrsim 50$) in image D, which sits almost on top of the critical curve (see Figure \ref{fig:bmodel2}). 
		
		\subsubsection{J0608+4229}
		Lacking HST data, we model the imaging data obtained with JWST MIRI through filter F560W.
		
		\subsubsection{J0659+1629}
		We model the imaging data obtained with JWST MIRI through filter F560W. The MIRI image shows more prominent lensed arcs than the available HST data \citep{Schmidt++23}. A satellite galaxy splits the lensed arc near image C (see Figure \ref{fig:bmodel2}) In the dark matter inference, we draw the mass of the satellite from $\theta_{\rm{E},G2} \sim \mathcal{N}\left(0.25,0.2\right)$. Given the proximity of the satellite to the merging triplet, it has a significant effect on the flux ratios, and the constraints on the luminous satellite's mass and position obtained from modeling the imaging data aid in isolating the effects of dark matter substructure from uncertainties associated with the satellite mass and position. 
		
		\subsubsection{J0803+3908}
		%We estimate a deflector redshift $z_{\rm{d}} = 1.11$ from LRIS.
		Lacking HST imaging, we model the imaging data obtained with JWST MIRI through filter F560W. A near-complete Einstein ring is visible. 
		
		\subsubsection{RXJ0911+0551}
		We do not model imaging data for this system because it lacks prominent lensed arcs in the available HST imaging. We model the image positions and flux ratios using measurements presented by \citet{Nierenberg++20}. The system has a luminous satellite near the main deflector, which we include in the lens model with a prior on the Einstein radius $\theta_{\rm{E},G2} \sim \mathcal{N}\left(0.25,0.15\right)$ \citep{Nierenberg++20}. 
		
		\subsubsection{J0924+0219}
		We model the imaging data obtained with JWST MIRI through filter F560W. For a smooth macromodel with perturbations on angular scales comparable to the image separation, the merging pair of images A and D should have a flux ratio $\sim 1$. For years, stellar microlensing has been invoked to explain the demagnification of image D \citep{Badole++20}. The persistence of this flux-ratio anomaly in the mid-IR, which should not experience microlensing, together with the non-detection of a flux-ratio anomaly in ALMA data, which probes a large physical region around the quasar \citep{Badole++20}, suggests that the purported microlensing could actually be a case of extreme millilensing. However, the extreme flux-ratio anomaly between the merging images is difficult to explain with millilensing perturbations by NFW halos, and appears to require denser perturbers. We use the ABC inference methodology discussed in Section \ref{ssec:frlike} to compute the likelihood for this system. 
		
		\subsubsection{J1042+1641}
		\label{sssec:j1042}
		The merging triplet in this system straddles the critical curve, and therefore the lensed arc in this vicinity becomes highly magnified. The high image magnifications also introduce significant diffraction spikes in the PSF model that blend with the arcs. These features complicate the modeling of the imaging data obtained with WFC3-UVIS through filter F814W, which yields the highest signal to noise ratio among the 3 HST bands. These issues are lessened to some degree when modeling the imaging data obtained with WFCR-IR through filter F160W, which has a lower signal to noise ratio and spatial resolution than the WFC3-UVIS F814W image. 
		
		As noted by \citet{Glikman++23}, there is a second source component, offset in the source plane from the quasar, that produces a feature in the lensed arc between images C and D (see the top row of Figure \ref{fig:bmodel3}). To model this feature, we add an additional S{\'e}rsic profile in the source plane, but define its position in the image plane, such that one image always appears at the observed location in the lensed arc between images C and D. The imaging data likelihood then quantifies how well a proposed lens model reproduces the position and brightness of any counter image(s). 
		
		We include the satellite galaxy identified by \citet{Glikman++23}, which is located $\sim 2$ arcseconds away from the main deflector, near the merging triplet. We include this object in the lens model with a prior on the Einstein radius $\mathcal{N}\left(0.05, 0.05\right)$. 
		
		\subsubsection{PG1115+080}
		This system has been subject to extensive analyses over the course of the past 30 years, including for time-delay cosmography \citep{TK02,Chen++19}. We model the imaging data using new observations from JWST NIRCam in F115W. Unlike the other systems with NIRCam imaging data (HE0435 and WFI2033), PG1115 does not exhibit significant complexity in the lensed arc, which we reproduce with a single elliptical S{\'e}rsic model for the quasar host galaxy (see second row of Figure \ref{fig:bmodel3}). 
		
		\subsubsection{GRAL1131-4419}
		This system has an unpublished (at the time of writing) deflector redshift $\zlens = 0.47$\footnote{D.Sluse, 2025, Private communication.}. Our baseline lens model fits the large-scale features of the lensed arcs (see Figure \ref{fig:bmodel3}, and also the top row of Figure \ref{fig:1131zooms}). However, the imaging data in F814W display numerous compact features, most likely star forming regions in the quasar host galaxy, that are not captured by a single shapelet basis set with $n_{\rm{max}}=10$. 
		
		\subsubsection{RXJ1131-1231}
		We model the imaging data obtained with HST through filter F814W. The imaging data in this band has been modeled extensively for time-delay cosmography \citep{Suyu++13,Birrer++16} and substructure inferences \citep{Birrer++17}. The extremely high degree of complexity in the lensed arc is difficult to fully reproduce, even with our quasar host galaxy model having $n_{\rm{max}}=34$. By computing the model-predicted flux ratios as a function of $n_{\rm{max}}$, however, we find the flux ratios predicted by the lens model converge for $n_{\rm{max}} \gtrsim 30$ while the BIC keeps dropping. We interpret this as a sign that the additional degrees of freedom improve the fit to the data through the source light while leaving the lens model unchanged. We therefore set $n_{\rm{max}}=34$ when reconstructing the imaging data in the dark matter inference. 
		
		The mass of the satellite estimated from the Faber-Jackson relation (see Section \ref{ssec:luminoussats}) corresponds to $\theta_{E,G2}\sim 0.1$, but both the imaging data and flux ratios strongly disfavor satellites this small. We draw the mass of the luminous satellite from $\mathcal{N}\left(0.3, 0.2\right)$. 
		
		\subsubsection{2M1134-2103}
		This system resides near a galaxy group, which is likely responsible for the large $\left(\gamma_{\rm{ext}} \gtrsim 0.3\right)$ external shear in the lens model and the peculiar diamond-like image configuration. The system has an unpublished deflector redshift $z_{\rm{d}}=0.66$\footnote{D. Sluse, 2025, Private communication}.  We model the lensed arc observed by JWST MIRI through filter F560W, which is brighter and has higher signal to noise ratio than the image taken with HST-F814W. We include the closest luminous satellite in the lens model with a prior on its mass $\theta_{\rm{E},G2} \sim \mathcal{N}\left(0.1,0.1\right)$. The imaging data also exhibit a clear preference for steeper logarithmic profiles slopes, so we draw $\gamma$ from a prior $\mathcal{N}\left(2.3,0.1\right)$. 
		
		\begin{figure*}
			\centering
			\includegraphics[trim=0cm 2.cm 0cm
			0cm,width=0.95\textwidth]{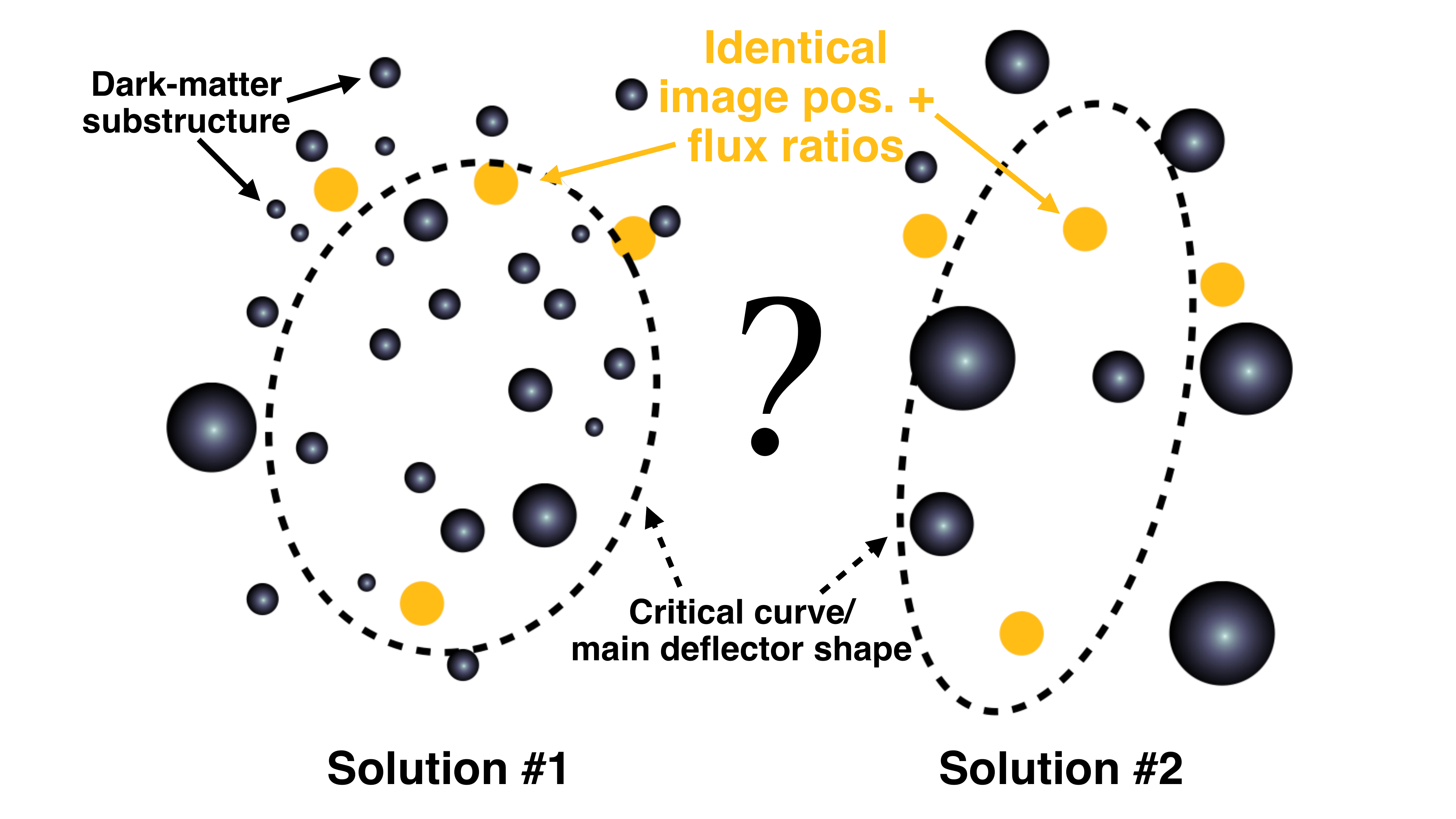}
			\caption{\label{fig:simpleillustration} A schematic that conveys the challenges associated with modeling image positions and flux ratios without complementary information from lensed arcs. The illustrations depict two possible configurations of a quadruple-image lens system with identical (to within measurement uncertainties) image positions and flux ratios. By incorporating constraints on the macromodel from lensed arcs, we can better determine the mass profile of the main deflector and differentiate between Solutions $\#1$ and $\#2$, effectively isolating perturbation by dark matter substructure from uncertainties associated with the macromodel.}
		\end{figure*}
		
		\subsubsection{J1251+2935}
		We model the imaging data obtained with HST WFC3-UVIS through filter F814W. We find no evidence for nearby satellite or companion galaxies. 
		
		\subsubsection{H1413+117}
		\citet{Jean++98} estimate a deflector redshift $\zlens \sim 1.15$. The somewhat high inferred redshift for this system is corroborated by observations presented by \citet{Kneib++98}, who show the lens appears in a group of galaxies at $z \approx 0.9$. While HST imaging is available for this system, the field of view has too few stars from which to estimate the PSF, and PSF models constructed for other systems with HST data performed poorly. We therefore use the imaging data obtained with JWST MIRI through filter F560W. We include a nearby galaxy in the lens model with $\theta_{\rm{E},G2} \sim \mathcal{N}\left(0.5, 0.1\right)$, which \citet{MacLeod++09} noted could affect the flux ratios for this system. 
		
		\subsubsection{B1422+231}
		This system does not present extended lensed arcs, and we therefore model only the point-source coordinates and flux ratios. We use the narrow-line flux ratios and astrometry presented by \citep{Nierenberg++14}. 
		
		\subsubsection{J1537-3010}
		We model the imaging data obtained with HST WFC3-UVIS through filter F814W.  We find no evidence for nearby satellites or companion galaxies. 
		
		\subsubsection{PSJ1606-2333}
		We model the imaging data obtained with HST WFC3-UVIS through filter F814W. This system has an unpublished (at the time of writing) deflector redshift $\zlens = 0.96$\footnote{D.Sluse, 2025, Private communication.}. We include the satellite galaxy near image C with a prior on the Einstein radius $\theta_{\rm{E},G2} \sim \mathcal{N}\left(0.15,0.1\right)$.   
		
		\subsubsection{WFI2026-4536}
		This system does not present extended lensed arcs, and we therefore model only the point-source coordinates and the flux ratios. We find no evidence for nearby satellite or companion galaxies. There is no confirmed spectroscopic redshift or nearby galaxy group with confirmed redshift, although \citet{Cornachione++20} note the time delays are consistent with a deflector at $z \approx 1$. 
		
		\subsubsection{WFI2033-4723}
		\label{sssec:wfi2033}
		We model the imaging data obtained with JWST NIRCam through filter F115W. \citet{Williams++25} also modeled the NIRCam data for this system, and developed a model for the PSF using the software package {\tt{starred}} \citep{Millon++24}. We use the PSF model presented by \citet{Williams++25} for the other two systems with NIRCam imaging data (HE0435 and PG1115). The lens model parameters we infer for this system are in excellent agreement with the results presented by \citet{Williams++25}.   
		
		We include the two nearby satellite galaxies with priors on their masses $\theta_{\rm{E},G2} \sim \mathcal{N}\left(0.05,0.05\right)$ and $\theta_{\rm{E},G3} \sim \mathcal{N}\left(0.6,0.1\right)$. Following the same approach used for HE0435, we  correct the observed position of the background galaxy, G3, for foreground lensing effects by the main deflector, and insert the galaxy as an SIS profile at the corrected angular position. As also noted by \citet{Williams++25}, the inferred logarithmic slope of the main deflector is $\gamma \sim 1.9$, which differs significantly from the default prior on the logarithmic slope $\gamma \sim \mathcal{N}\left(2.1, 0.1\right)$. Given the clear preference in the likelihood for shallower slopes we draw $\gamma$ from $ \mathcal{N}\left(1.9, 0.1\right)$ in the dark matter analysis. 
		
		\subsubsection{J2038-4008}
		We model the HST F814W imaging data. These data have also been modeled extensively for time-delay cosmography \citep[e.g.][]{Shajib++22,Wong++24}. 
		
		\subsubsection{J2145+6345}
		%We use 3-band photometry of the main deflector light to estimate a photometric redshift $\zlens = 0.5 \pm 0.14$. 
		We model the lensed arcs using observations taken with JWST MIRI through filter F560W, for which the lensed arcs appear more prominently than in HST imaging. From the HST imaging in F814W we identify a satellite galaxy between images A and D. Assuming both satellites are at the same redshift and follow the Faber-Jackson relation gives an estimated Einstein radius $\theta_{\rm{E},G2} \sim 0.3$, but both the imaging data and flux-ratio likelihoods rule out satellites this massive. We assume a prior in the forward model $\mathcal{N}\left(0.2, 0.1\right)$.  
		
		\subsubsection{J2205-3727}
		We model the imaging data taken with JWST MIRI through filter F560W, which show a more prominent lensed arc than the available HST data. We find no evidence for nearby satellites or companion galaxies. This system was observed with longslit spectroscopy using the Faint Object Camera and Spectrograph (FOCAS; \citet{Kashikawa02}) on the Subaru telescope on October 16, 2022. We extract the lens galaxy spectrum using a custom routine based on the method developed by \citet{Mozumdar++23} and measure a deflector redshift of $z_{\rm{d}}=0.63 \pm 0.01$.
		
		\subsubsection{J2344-3056}
		Given the low signal to noise ratio of the lensed arcs in the HST F814W image, we model the imaging data obtained with MIRI through filter F560W.
		
		\begin{figure*}
			\centering
			\includegraphics[trim=0cm 1.cm 0cm
			0cm,width=0.95\textwidth]{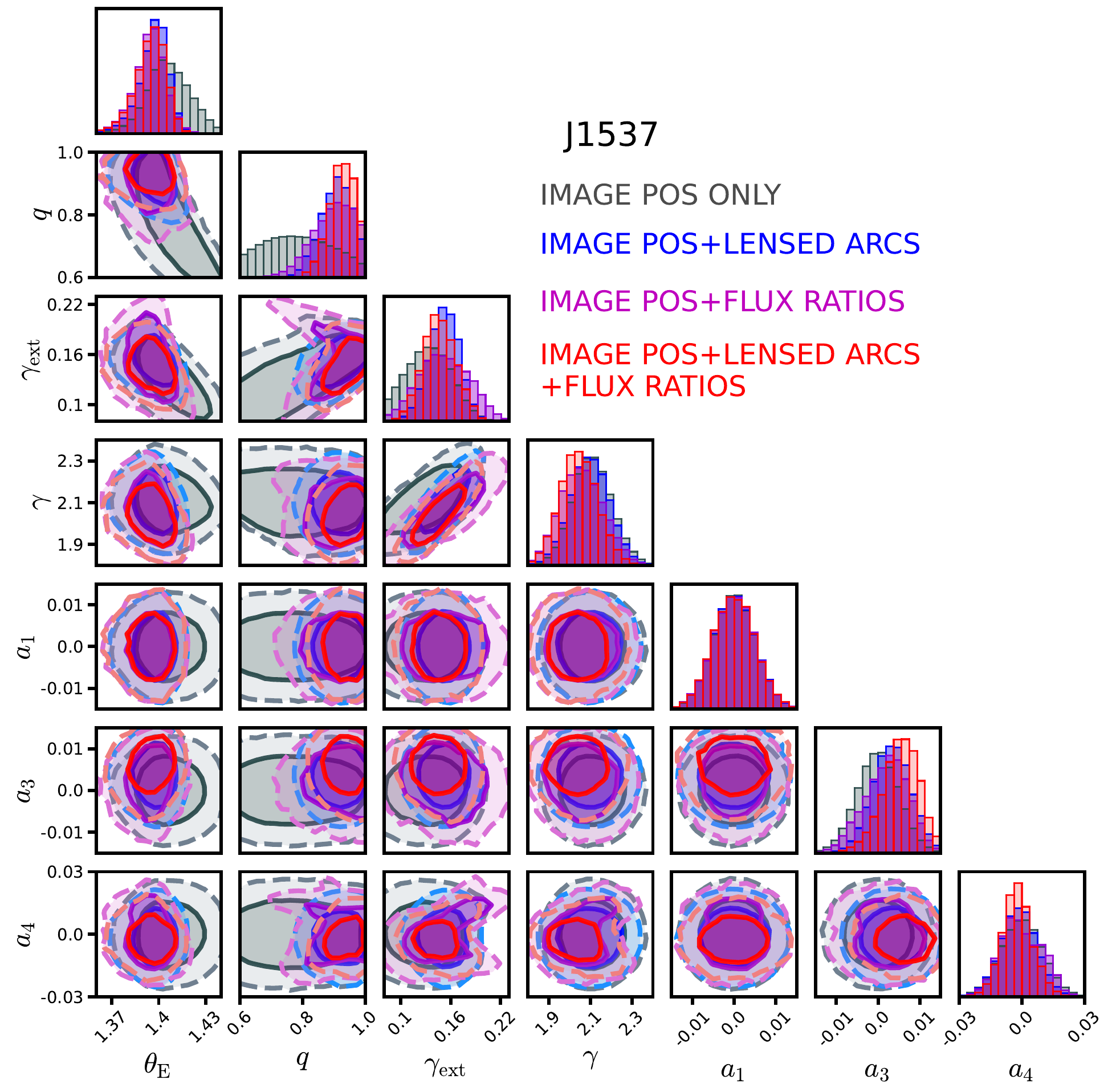}
			\caption{\label{fig:1537macro} Improvements in the inferred macromodel parameters for the system J1537 obtained from modeling only relative image positions (light gray), image positions and lensed arcs (blue), image positions and flux ratios (magenta), and the full combination of image positions, flux ratios, and lensed arcs (red). Contours correspond to $68\%$ and $95\%$ confidence intervals. Reading from the left along the lower axes, the lens model parameters are the Einstein radius $\theta_{E}$, the axis ratio $q$, the external shear strength $\gamma_{\rm{ext}}$, the logartihmic slope of the mass profile $\gamma$, and the strength of the $a_1$, $a_3$, and $a_4$ multipole perturbations. Note how the additional information constrains some of the parameters significantly better than just image positions, e.g. axis ratio $q$, breaking some of the degeneracies in the macromodel.}
		\end{figure*}
		
		\begin{figure*}
			\centering
			\includegraphics[trim=0cm 0.7cm 0cm
			2cm,width=0.48\textwidth]{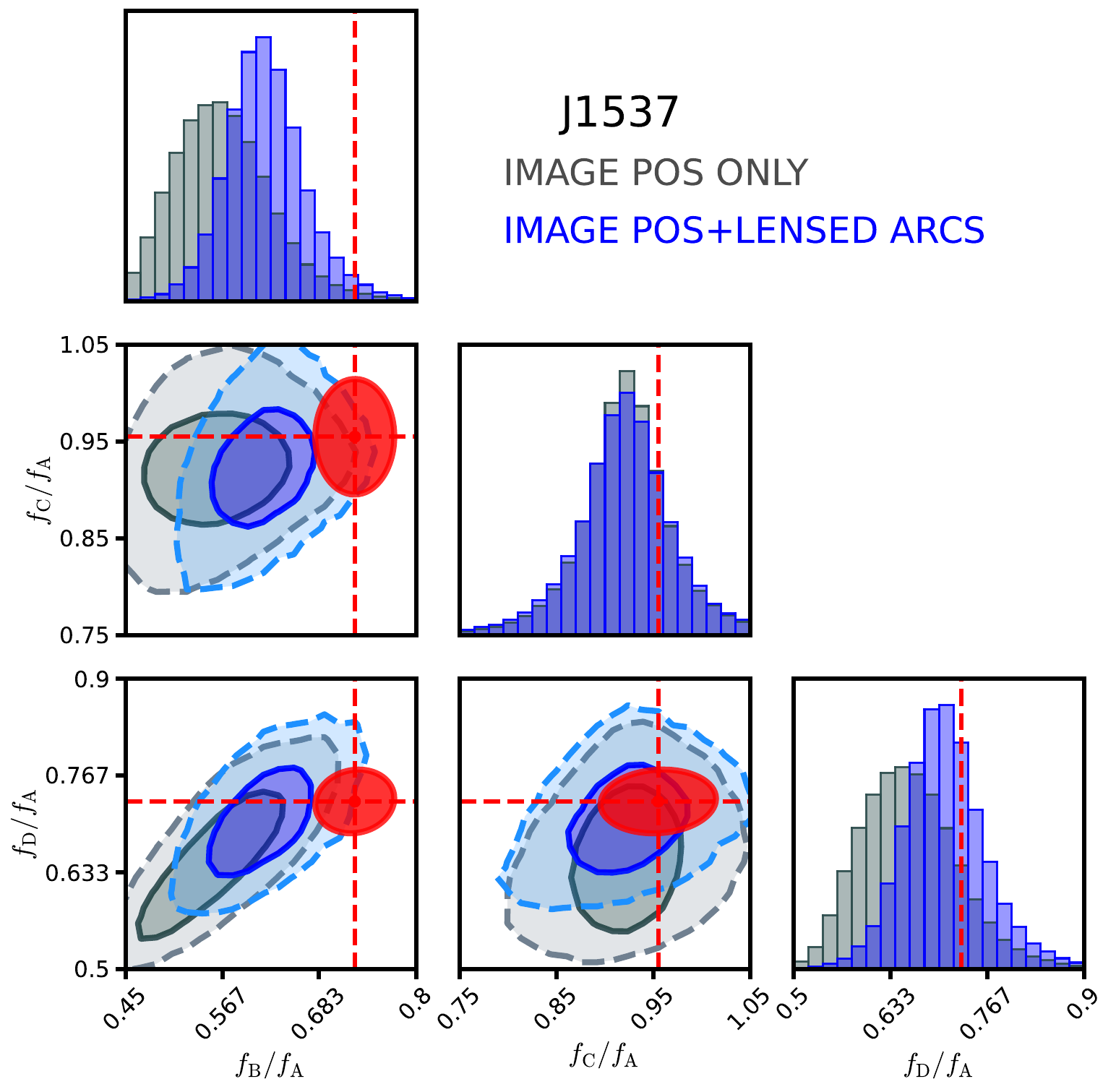}
			\includegraphics[trim=0cm 0.7cm 0cm
			2cm,width=0.48\textwidth]{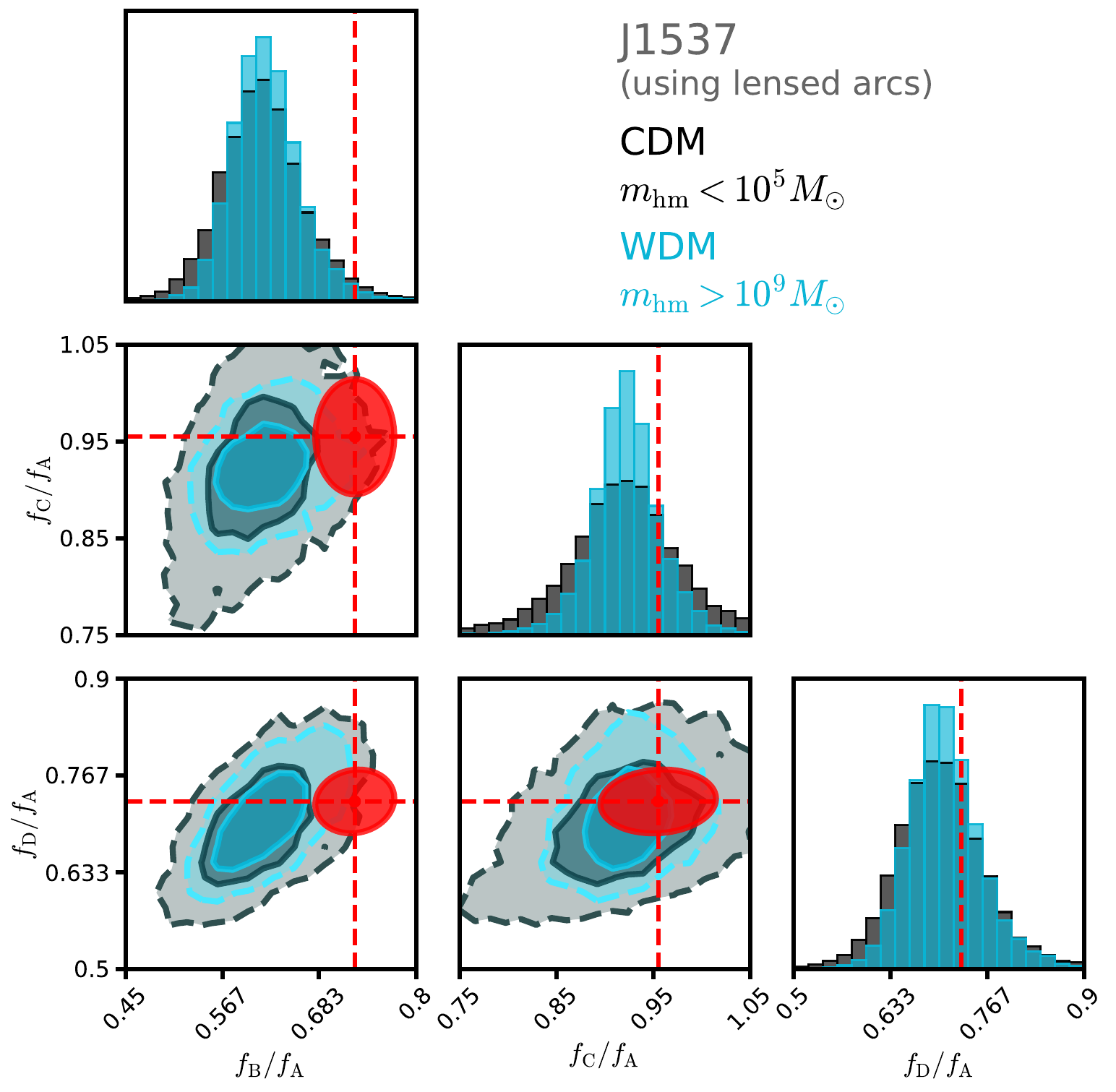}
			\caption{\label{fig:1537fr} {\bf{Left:}} Flux ratios predicted by the lens model for J1537, including both CDM and WDM models, from modeling only the relative image positions (light gray), and after incorporating constraints from the lensed arcs (blue), with contours corresponding to $68\%$ and $95\%$ confidence regions. Red ellipses show the measured flux ratios and $95\%$ confidence regions. {\bf{Right:}} Flux ratios predicted by the lens model using information from the lensed arcs when including only CDM-like realizations with $m_{\rm{hm}} < 10^{5} \mathrm{M}_{\odot}$ (black), and WDM-like realizations with $m_{\rm{hm}} > 10^{9} \mathrm{M}_{\odot}$ (cyan), as well globular clusters and multipole perturbations. Distributions peak at the median flux ratio predicted by the macromodel, and differences between CDM and WDM manifest as increased scatter around the smooth-model prediction.}
			\includegraphics[trim=0cm 0.2cm 0cm
			0.2cm,width=0.32\textwidth]{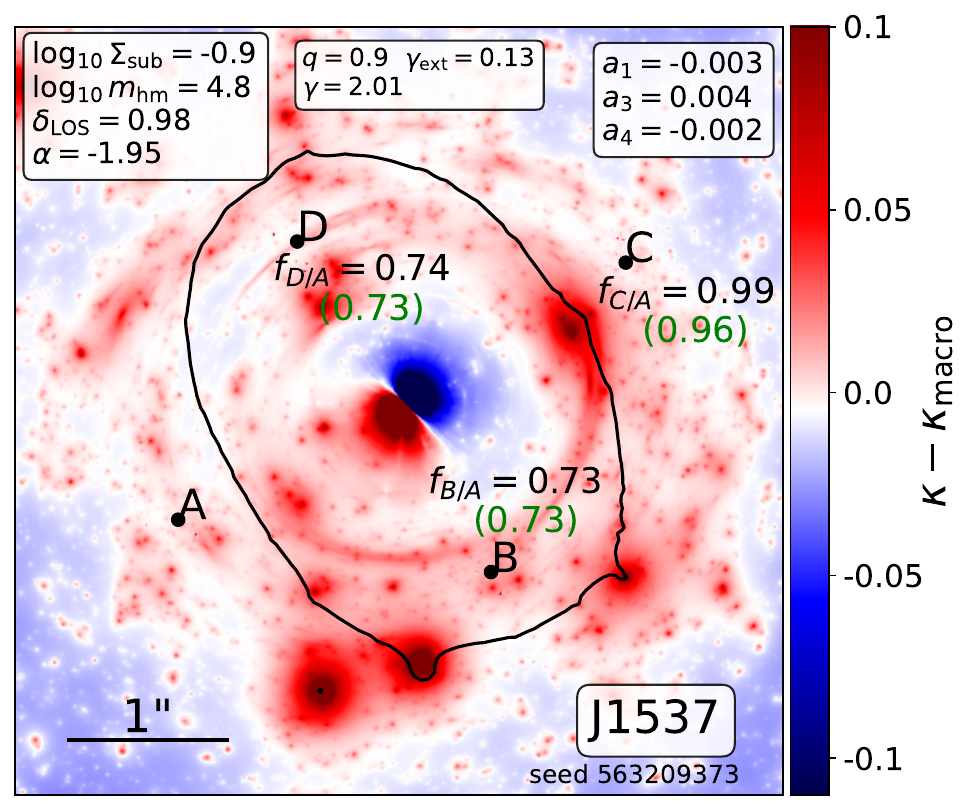}
			\includegraphics[trim=0cm 0.2cm 0cm
			0.2cm,width=0.34\textwidth]{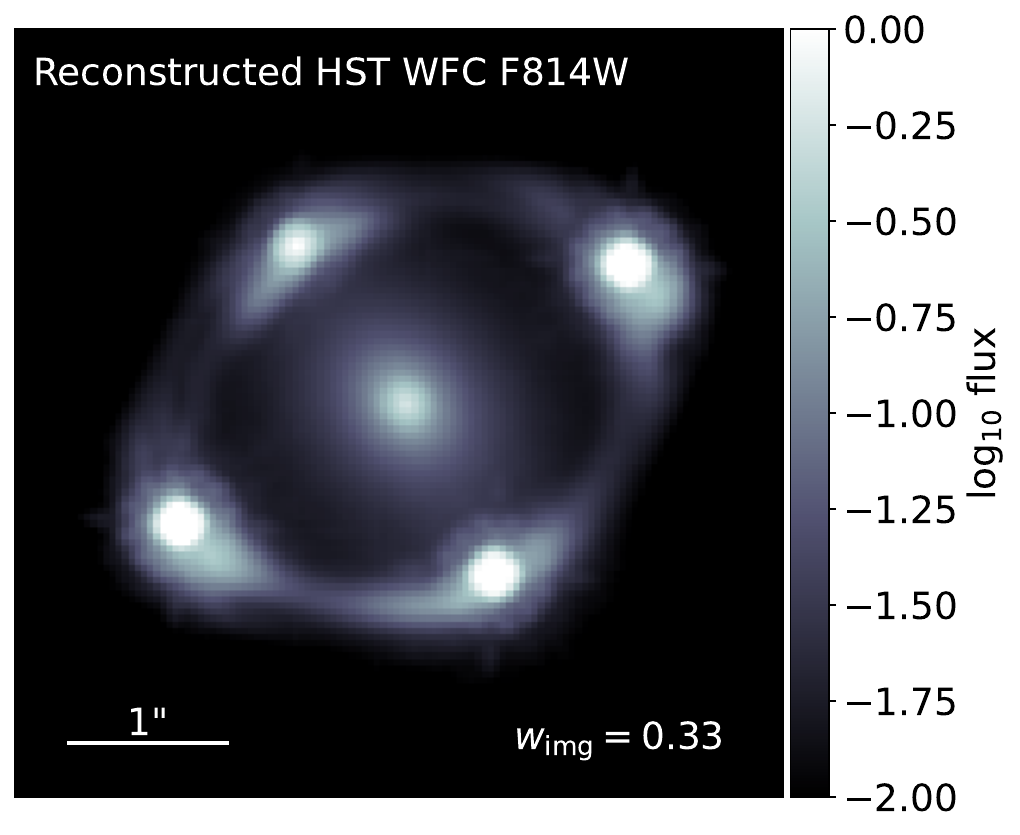}
			\includegraphics[trim=0cm 0.2cm 0cm
			0.2cm,width=0.32\textwidth]{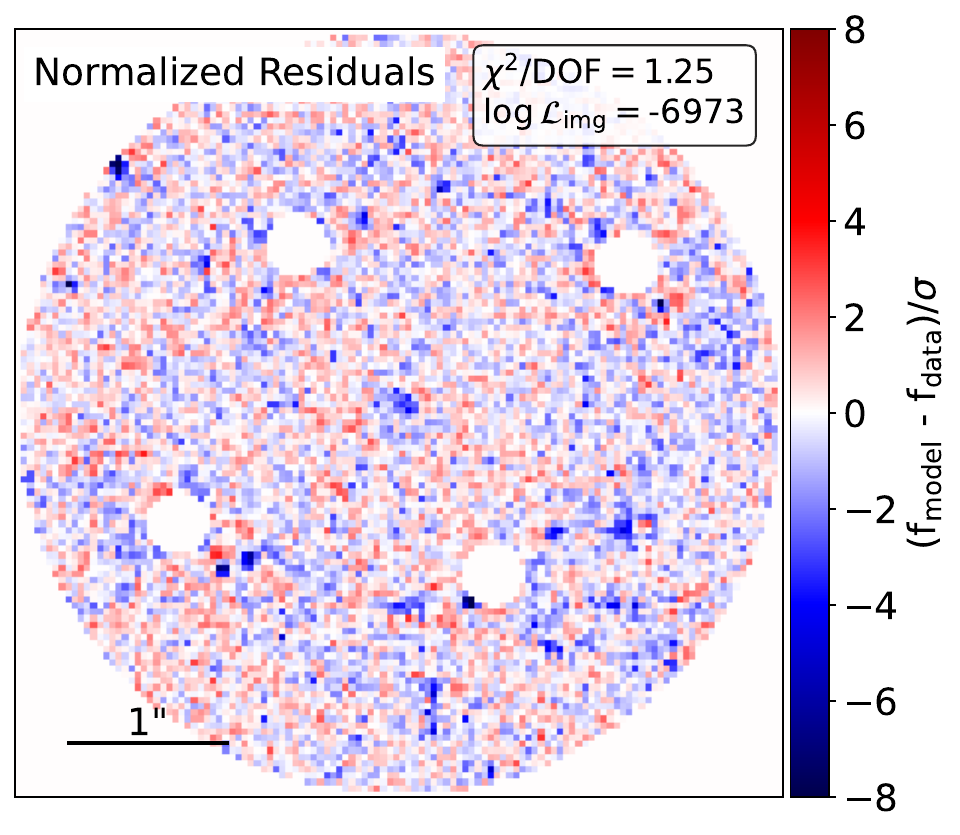}
			\includegraphics[trim=0cm 0.2cm 0cm
			0.2cm,width=0.32\textwidth]{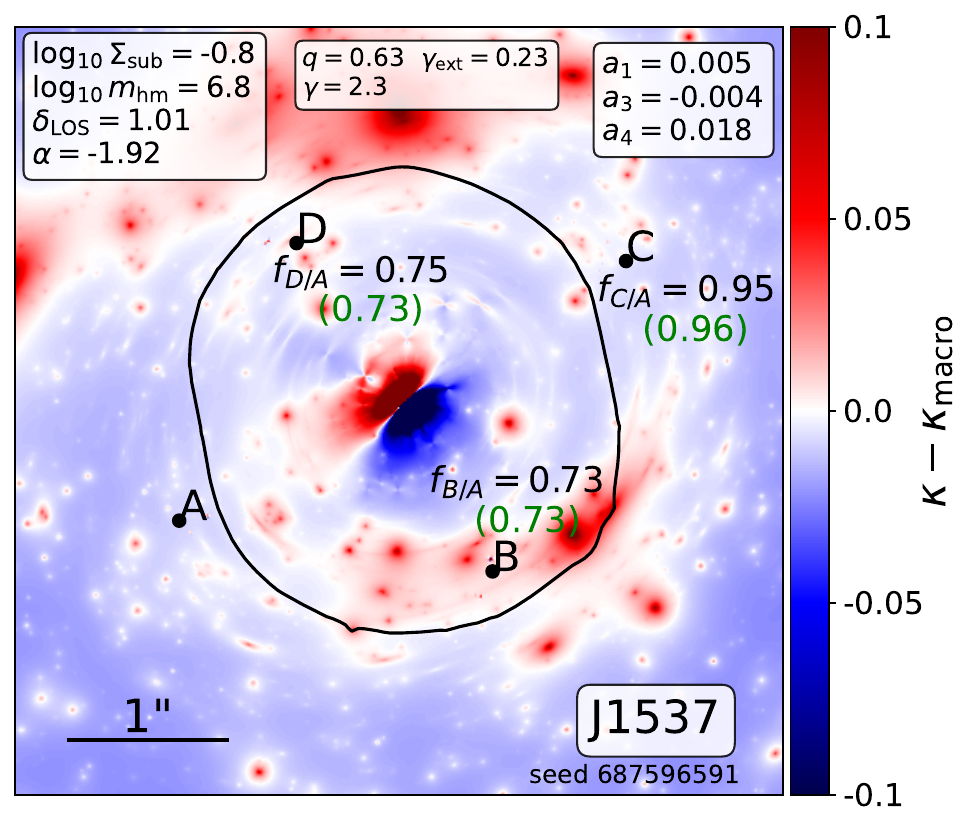}
			\includegraphics[trim=0cm 0.2cm 0cm
			0.2cm,width=0.34\textwidth]{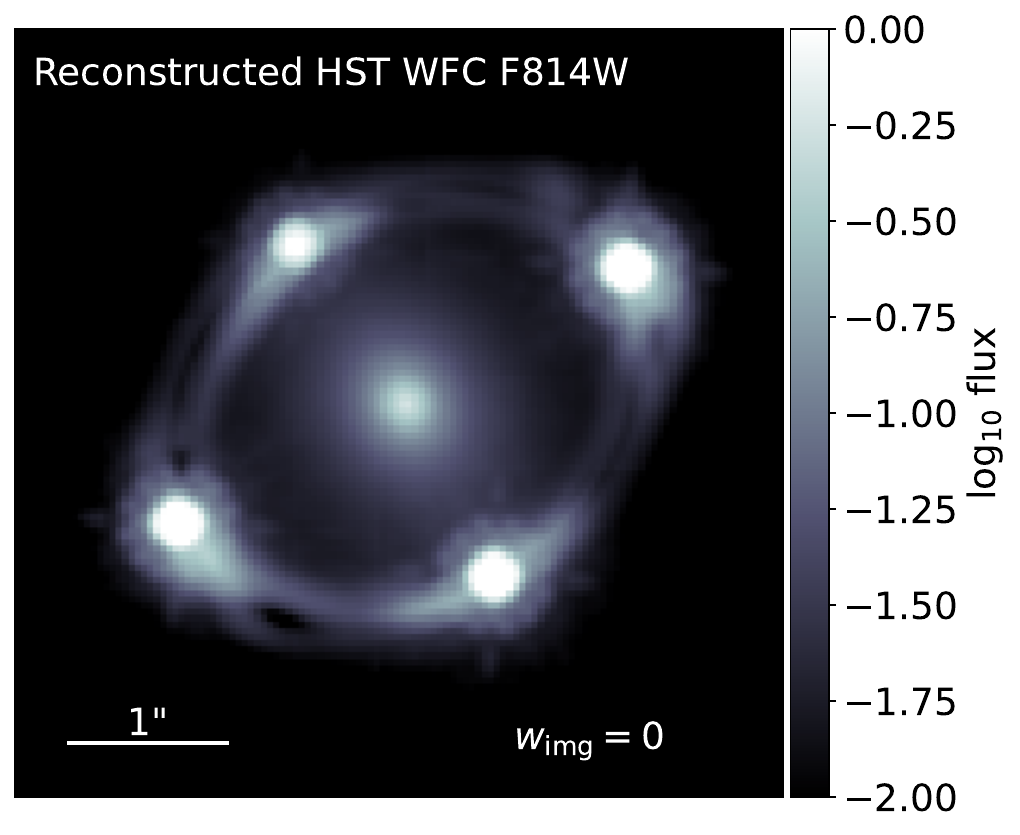}
			\includegraphics[trim=0cm 0.2cm 0cm
			0.2cm,width=0.32\textwidth]{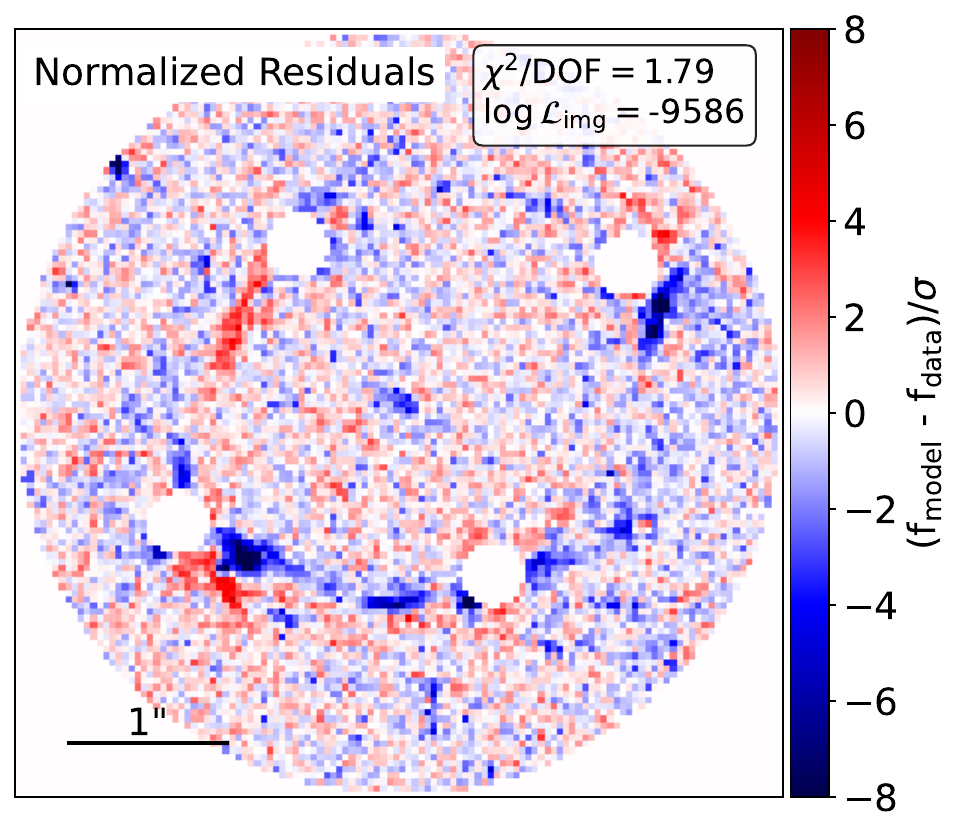}\caption{\label{fig:1537fromseedaccepted} Rows correspond to lens models of J1537 that reproduce observed image positions and flux ratios. The model in the top row also reproduces the lensed arcs, while the model in the bottom row is rejected because it cannot reproduce the lensed arcs. {\bf{Left:}} Effective convergence in dark matter substructure (Equation \ref{eqn:effectivekappa}), relative to mean projected mass density. The black curve is the critical curve. Lensed images are labeled A--D, alongside measured (green) and model-predicted (black) flux ratios. The three insets list the dark matter hyper-parameters (left), axis ratio $q$, external shear strength $\gamma_{\rm{ext}}$, and logarithmic profile slope $\gamma$ (center), and the strength of the multipole terms (right). {\bf{Center:}} Reconstructed image plane surface brightness, including the lens light, lensed arcs, and point sources. The image data importance weight, $w_{\rm{img}}$ (Equation~\ref{eqn:neff}), is specified for each model. {\bf{Right:}} Normalized residuals of the imaging data likelihood, with reduced $\chi^2$, and log-likelihood listed in the upper right.}
		\end{figure*}
		
		\begin{figure*}
			\centering
			\includegraphics[trim=0cm 1.cm 0cm
			0cm,width=0.48\textwidth]{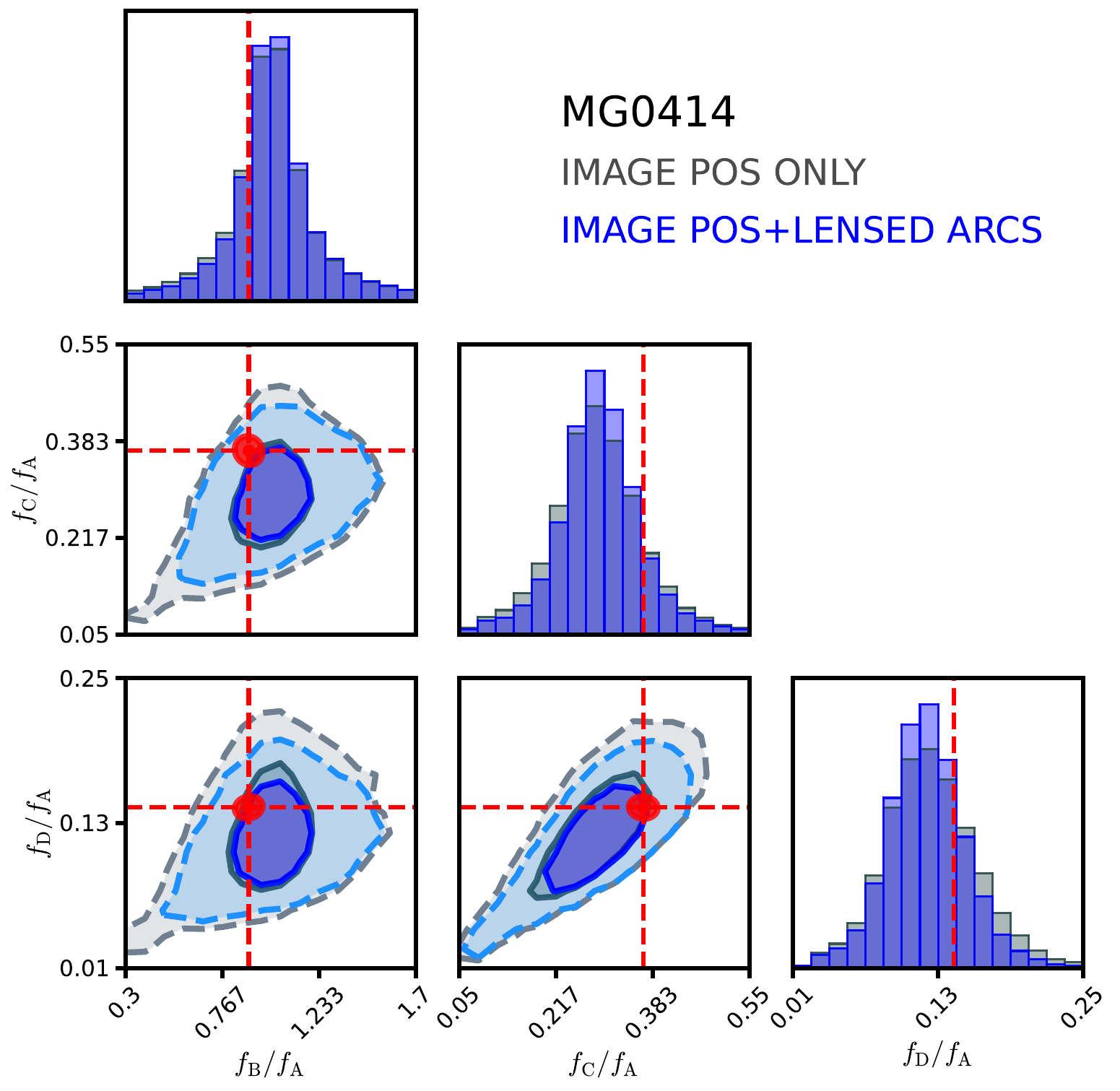}
			\includegraphics[trim=0cm 1.cm 0cm
			0cm,width=0.48\textwidth]{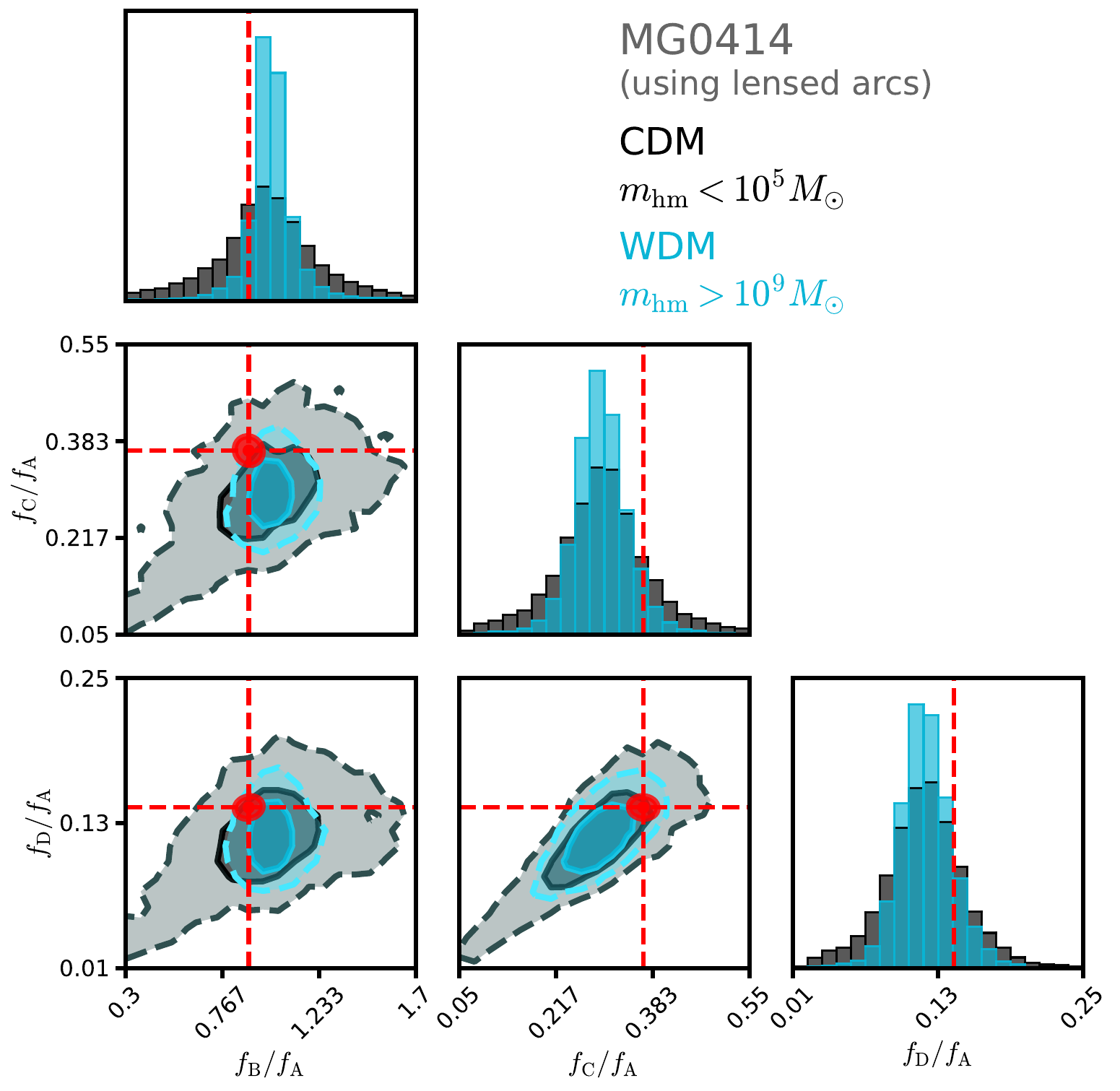}
			\caption{\label{fig:mg0414fr} The same as Figure \ref{fig:1537fr}, but for the lens system MG0414.}
		\end{figure*}
		
		\begin{figure*}
			\centering
			\includegraphics[trim=0cm 0.5cm 0cm
			0cm,width=0.32\textwidth]{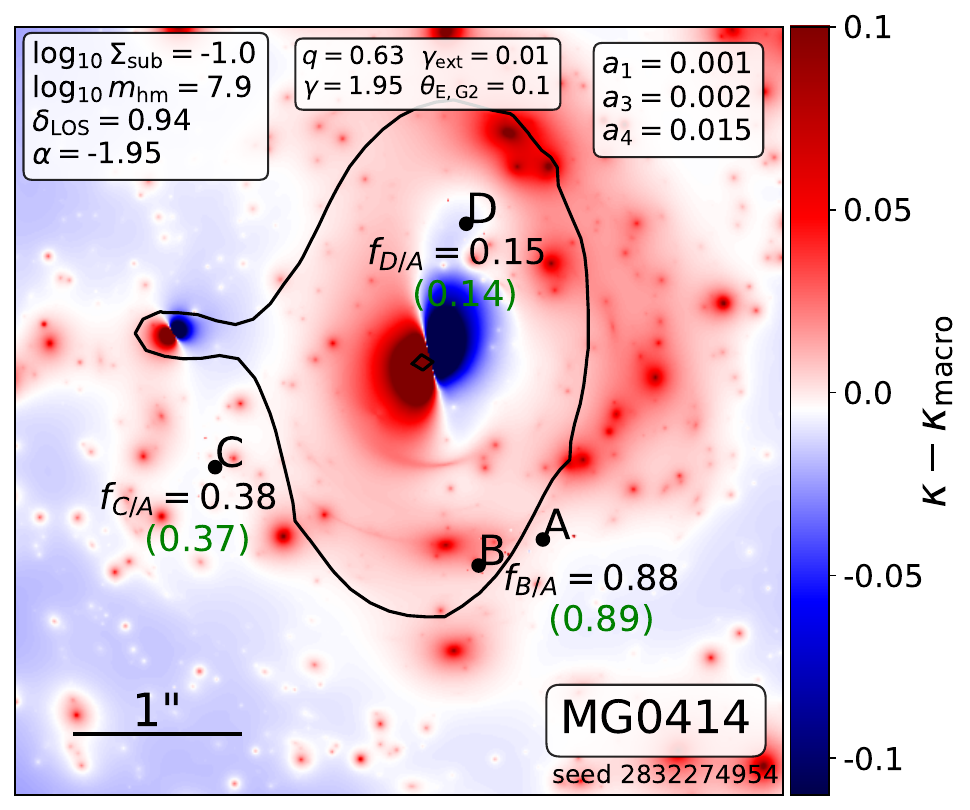}
			\includegraphics[trim=0cm 0.5cm 0cm
			0cm,width=0.32\textwidth]{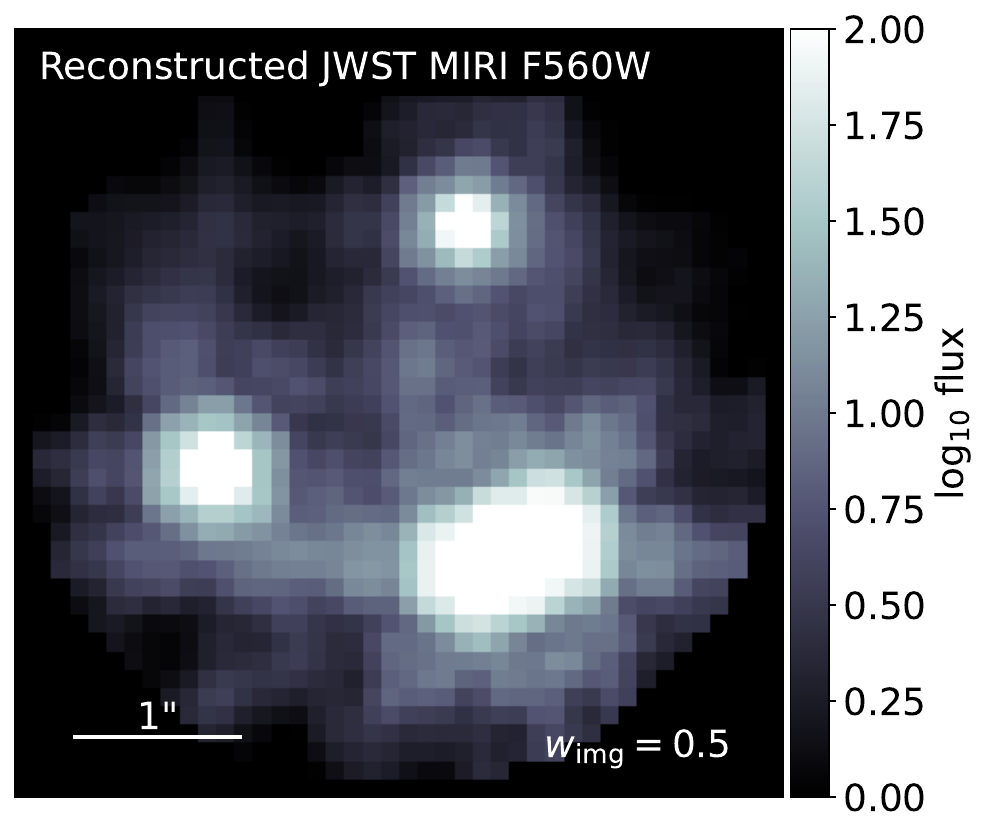}
			\includegraphics[trim=0cm 0.5cm 0cm
			0cm,width=0.32\textwidth]{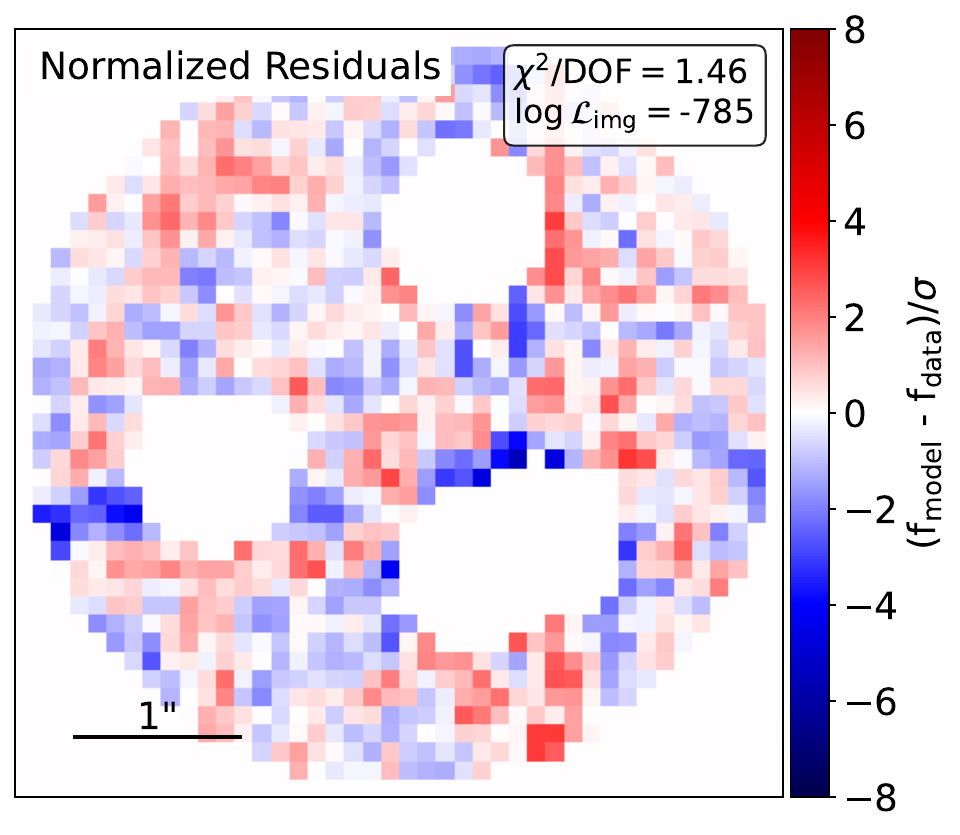}
			\includegraphics[trim=0cm 0.5cm 0cm
			0cm,width=0.32\textwidth]{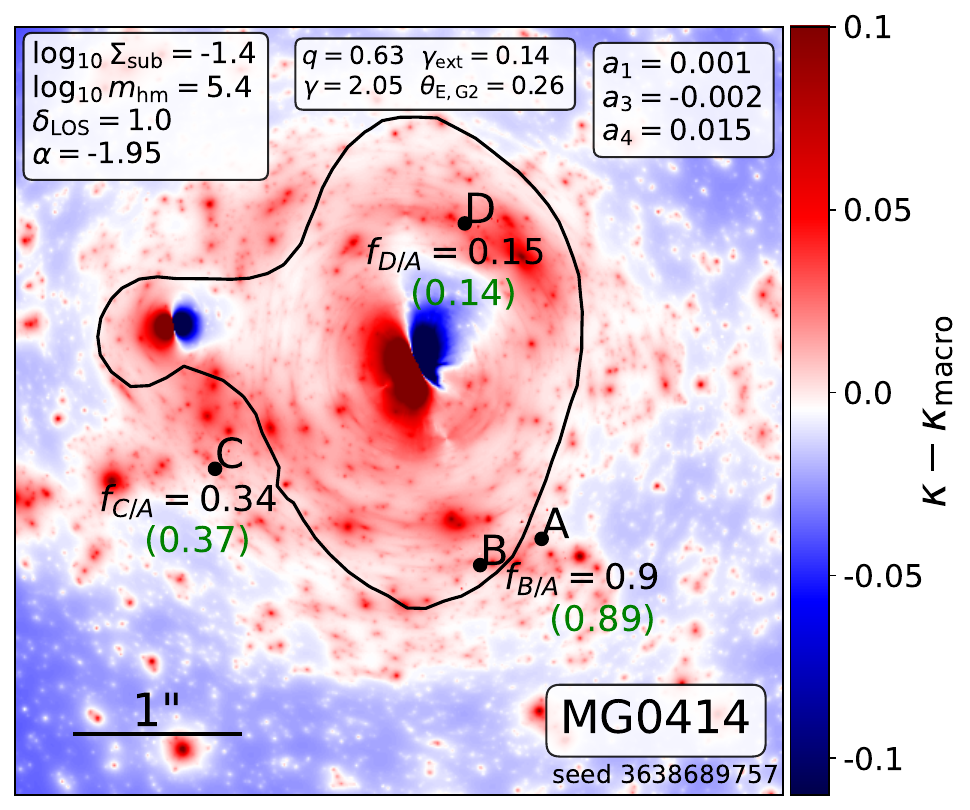}
			\includegraphics[trim=0cm 0.5cm 0cm
			0cm,width=0.32\textwidth]{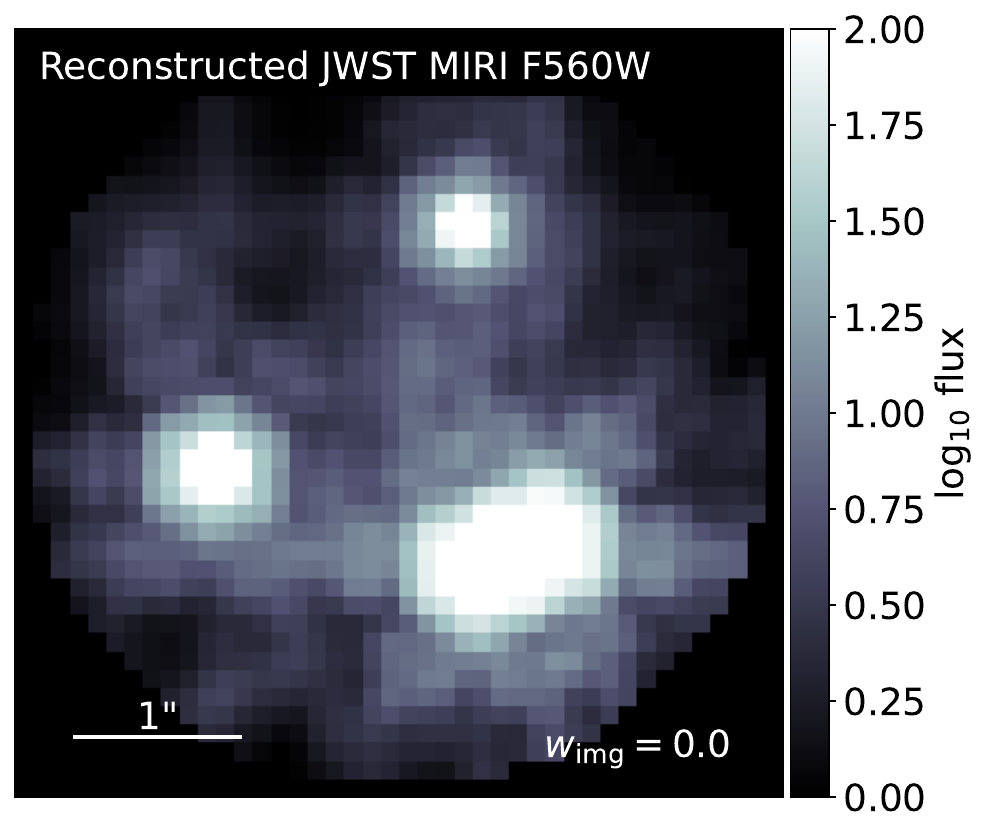}
			\includegraphics[trim=0cm 0.5cm 0cm
			0cm,width=0.32\textwidth]{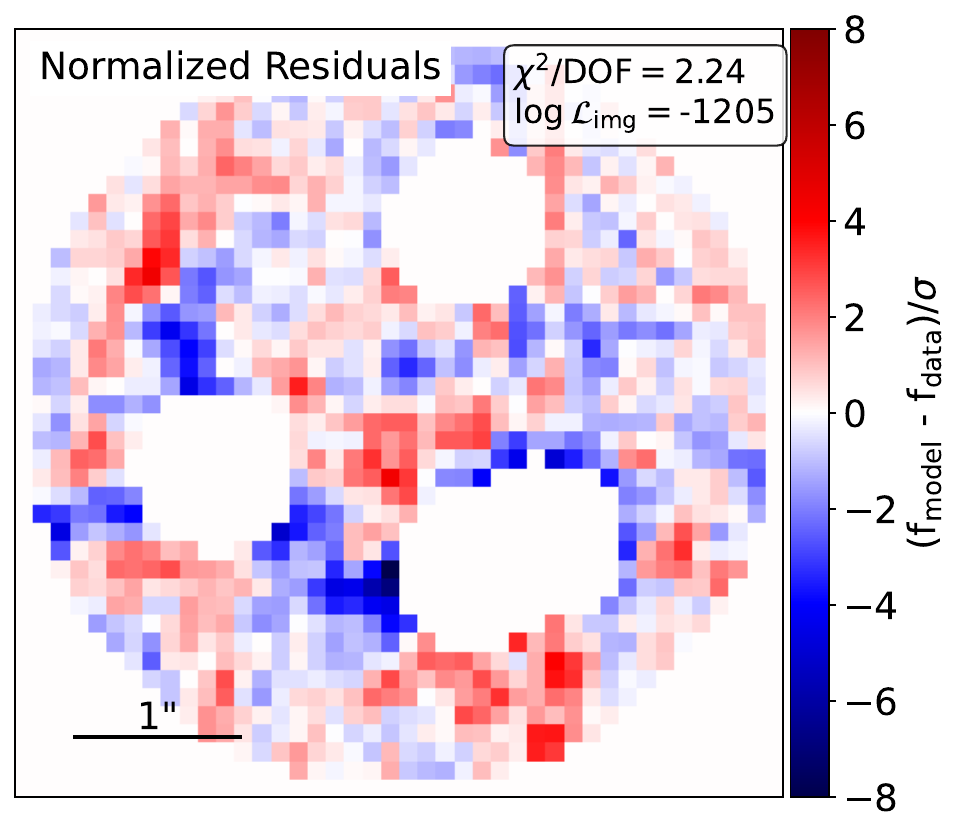}
			\caption{\label{fig:0414fromseedaccepted} The same as Figure \ref{fig:1537fromseedaccepted}, but for the lens system MG0414.}
		\end{figure*}
		
		\begin{figure*}
			\centering
			\includegraphics[trim=0cm 0.cm 0cm
			0.cm,width=0.32\textwidth]{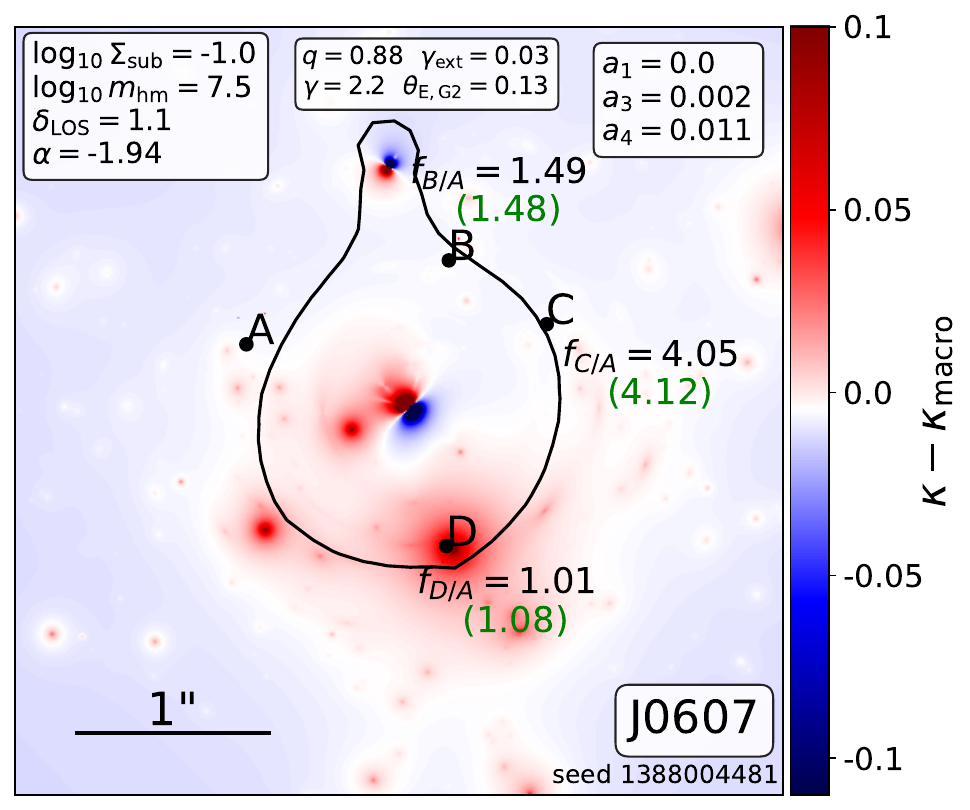}
			\includegraphics[trim=0cm 0.cm 0cm
			0.cm,width=0.32\textwidth]{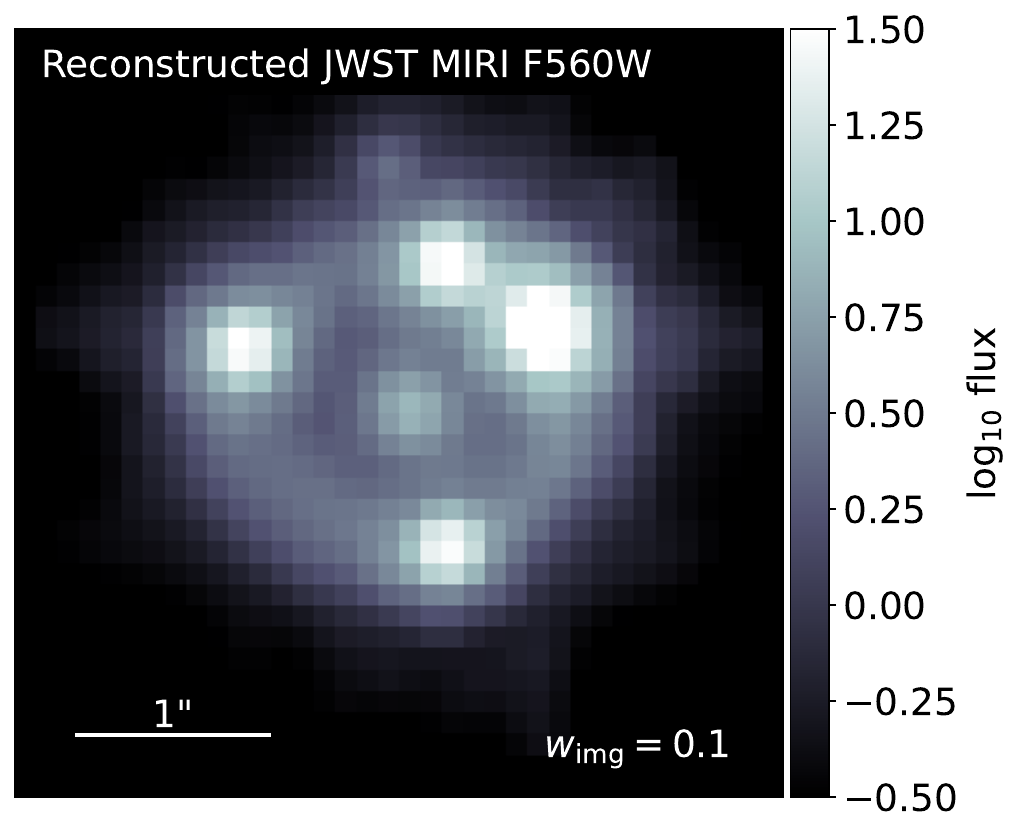}
			\includegraphics[trim=0cm 0.cm 0cm
			0.cm,width=0.32\textwidth]{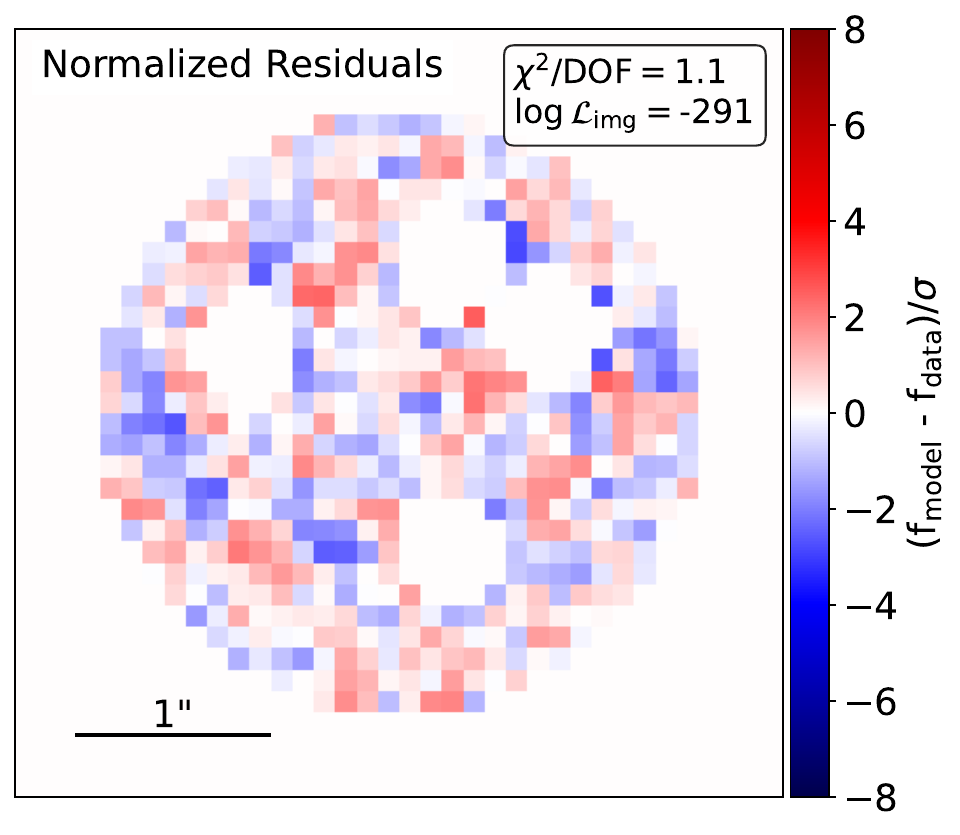}
			\includegraphics[trim=0cm 0.cm 0cm
			0.cm,width=0.32\textwidth]{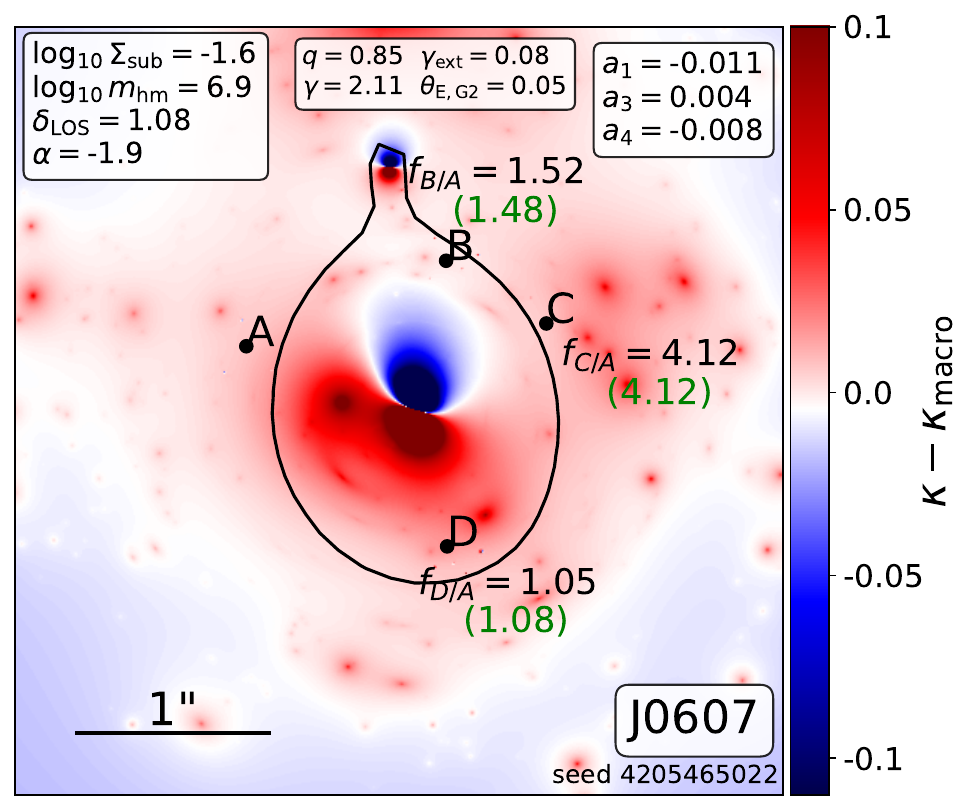}
			\includegraphics[trim=0cm 0.cm 0cm
			0.cm,width=0.32\textwidth]{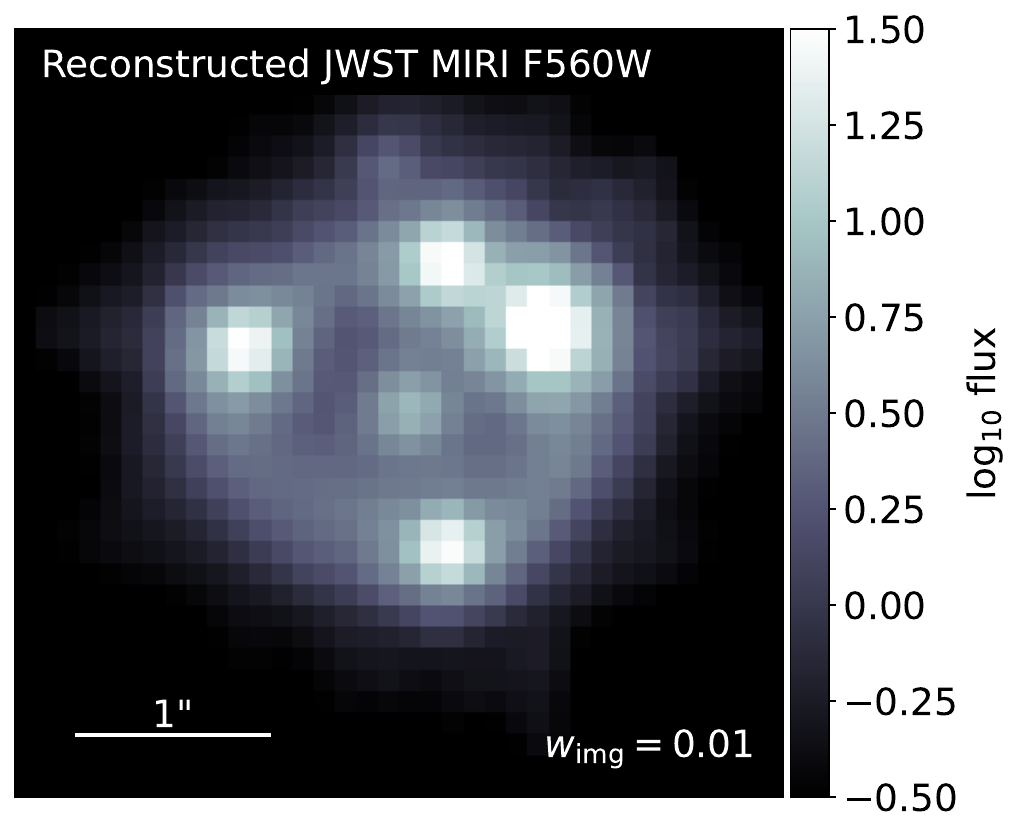}
			\includegraphics[trim=0cm 0.cm 0cm
			0.cm,width=0.32\textwidth]{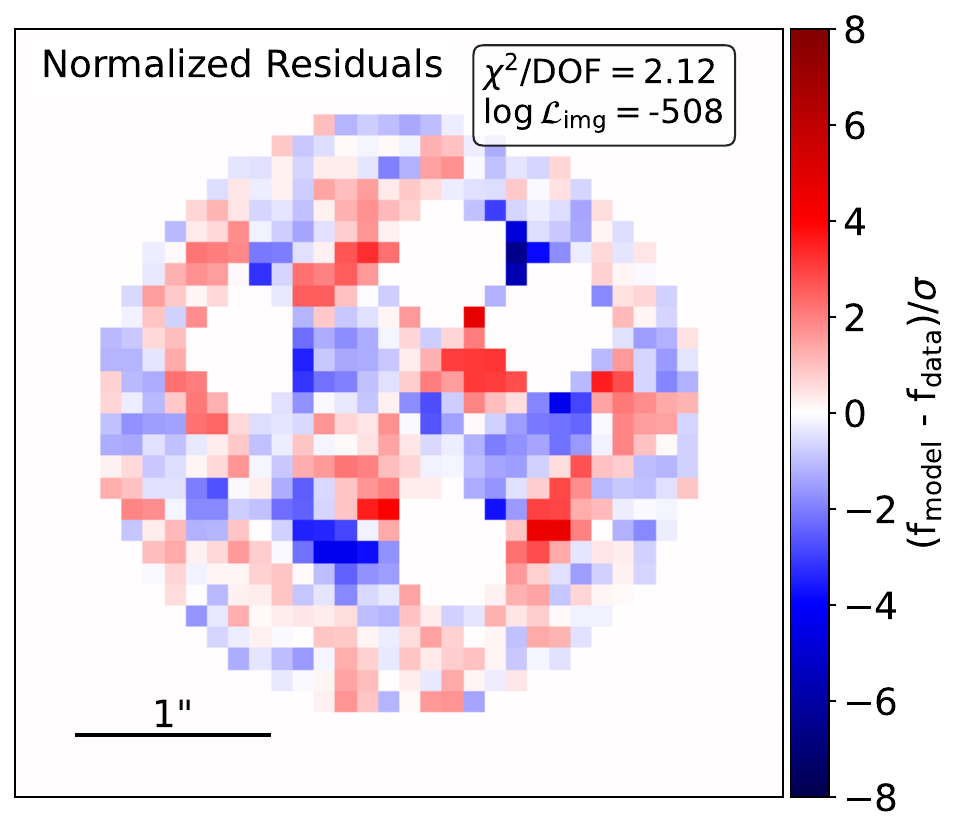}
			\caption{\label{fig:j0607fromseedaccepted} The same as Figure \ref{fig:1537fromseedaccepted}, but for the lens system J0607.}
		\end{figure*}
		
		\begin{figure}
			\centering
			\includegraphics[trim=0cm 1.cm 0cm
			0cm,width=0.48\textwidth]{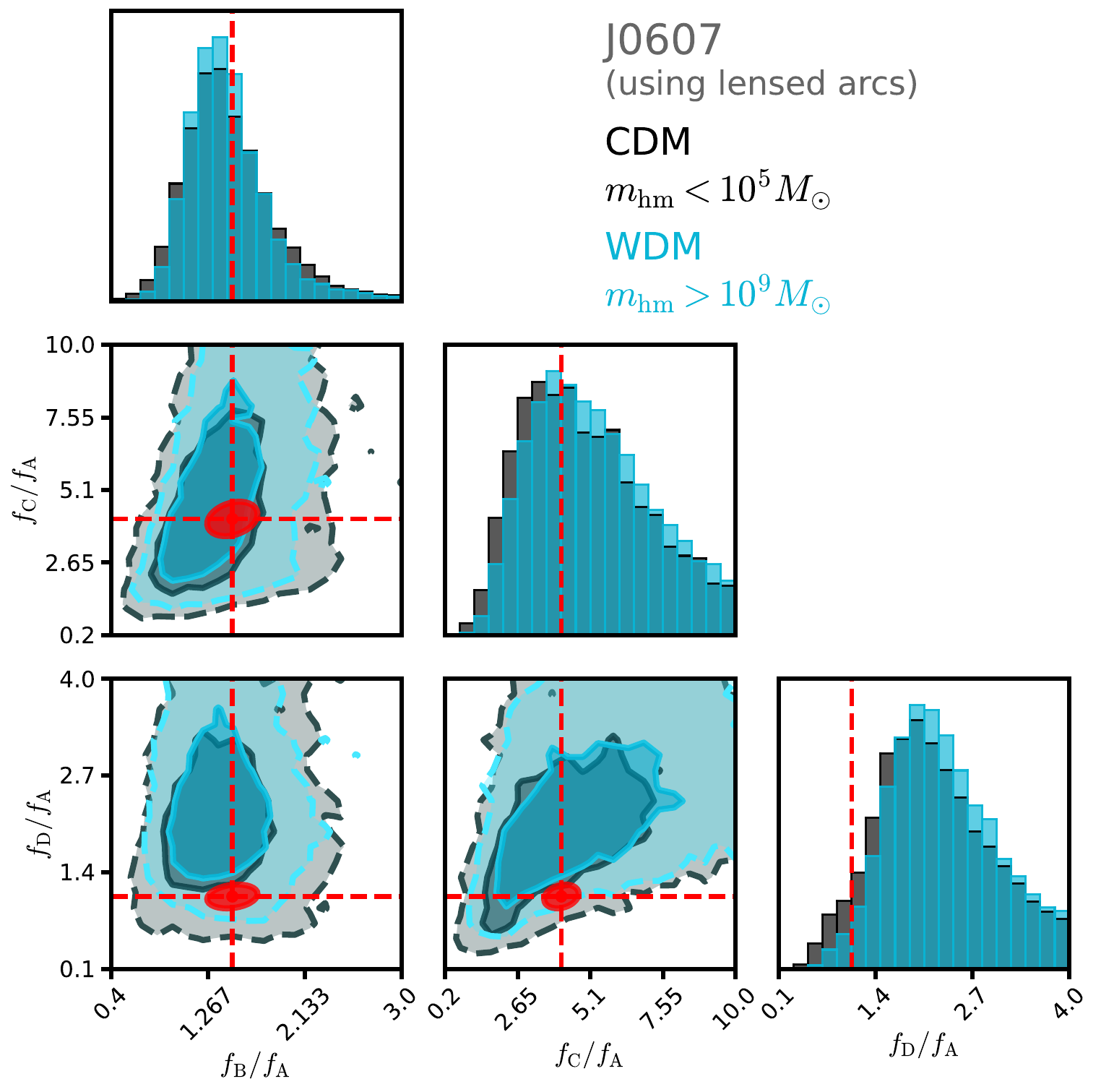}
			\caption{\label{fig:j0607fr} The model-predicted flux ratios for J0607 in CDM (gray) and WDM (cyan). Figure \ref{fig:frwimg} shows the model-predicted flux ratios before and after incorporating imaging data.}
		\end{figure}
		
		\section{Results}
		\label{sec:results}
		This section presents the results of applying the inference methodology discussed in Section~\ref{sec:inference} to the dark matter model presented in Section~\ref{sec:dmmodel}, following the lens modeling approach discussed in Section~\ref{sec:lensmodeling}. We begin in Section~\ref{ssec:casestudies} with demonstrations of how the simultaneous modeling of imaging data and flux ratios enables stronger constraints on dark matter properties than each dataset individually. Sections~\ref{ssec:resultswdm} and \ref{ssec:resultscdm} present the main results of this paper, including constraints on the free-streaming length of WDM and a measurement of the amplitude of the subhalo mass function, assuming CDM. 
		
		\subsection{The superior constraining power of combining image positions, lensed arcs, and flux ratios}
		\label{ssec:casestudies}
		Figure~\ref{fig:simpleillustration} illustrates the main challenge associated with using only image positions and flux ratios to constrain dark matter substructure. Two populations of dark matter halos, in combination with two different solutions for the lens macromodel, result in identical (to within measurement uncertainties) image positions and flux ratios. Without additional information, we therefore assign equal likelihood to the two substructure realizations. By incorporating constraints on the macromodel from lensed arcs, we can rule out configurations of the main deflector mass profile and dark matter substructure that may reproduce the measured flux ratios, but which cannot reproduce the arcs. This leads to more precise model-predicted flux ratios, and ultimately enables improved constraining power over dark matter models. 
		
		To illustrate this concept in practice, we consider detailed results for J1537. 
		%in the 4th row and 4th column of Figure \ref{fig:mosaic}. 
		In Figure~\ref{fig:1537macro} we show the macromodel parameters inferred from modeling only the relative image positions (light gray), the relative image positions and lensed arcs using the importance weights $w_{\rm{img}}$ (blue), the relative image positions and flux ratios (magenta), and the simultaneous modeling of all three datasets (red). For J1537, the lensed arcs require $q > 0.75$ and $0.12 \lesssim \gamma_{\rm{ext}} \lesssim 0.19$. Noting the difference between the blue and gray distributions, the imaging data also improve the constraints on the multipole strengths $a_3$ and $a_4$ for this system. In turn, these constraints on the macromodel improve the precision of the model-predicted flux ratios, as illustrated by Figure \ref{fig:1537fr}. In both the left and right panels, red ellipses represent the measured flux ratios for this system and their $2 \sigma$ uncertainties. The left panel shows how the improved constraints on the macromodel affect the model-predicted flux ratios, including populations generated with values of $m_{\rm{hm}}$ in the full prior range (i.e., with both CDM-like and WDM-like populations). In the right panel, we show the flux ratios after incorporating constraints from the lensed arcs, and down-selecting on halo populations with $m_{\rm{hm}}<10^5 \mathrm{M}_{\odot}$ (black), and $m_{\rm{hm}}>10^9 \mathrm{M}_{\odot}$ (cyan). Relative to the WDM-like halo populations ($m_{\rm{hm}}>10^9 \mathrm{M}_{\odot}$), models with CDM-like substructure ($m_{\rm{hm}}<10^5 \mathrm{M}_{\odot}$) produce more frequent and stronger perturbations to the flux ratios than WDM. 
		
		Lens models with both CDM and WDM-like halo populations can explain the flux ratios in J1537, but the relative likelihood differs between these competing explanations of the data. The difference in relative likelihood drives our constraints on the dark matter models. We emphasize that these considerations apply to any other source of small-scale perturbation, including globular clusters, or angular structure in the main deflector mass model. Analyses that attempt to explain observations within the context one particular model --- for example, one model that considers only CDM-like halo populations, or a model that considers only multipole perturbations --- may succeed in doing so. However, meaningful statements regarding the nature of dark matter must take into account simultaneously all known sources of small-scale perturbation.
		\begin{figure*}
			\centering
			\includegraphics[trim=0cm 1.cm 0cm
			0cm,width=0.95\textwidth]{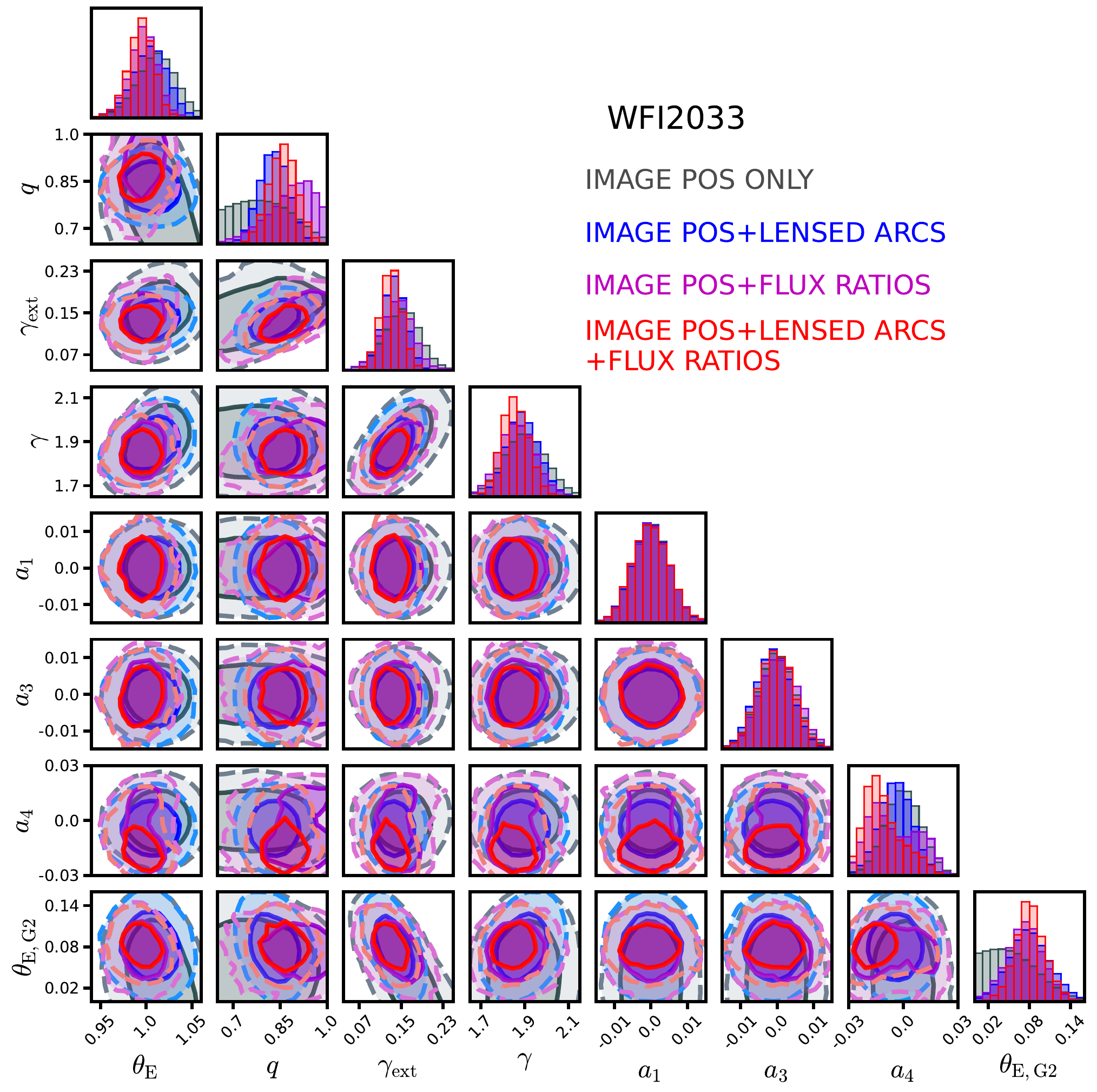}
			\caption{\label{fig:2033macro}  Inferred macromodel parameters for the system WFI2033. The color scheme is the same as Figure~\ref{fig:1537macro}. The simultaneous modeling of flux ratio and imaging data reveals a preference for boxy isodensity contours (negative $a_4$).}
		\end{figure*} 
		\begin{figure*}
			\centering
			\includegraphics[trim=0cm 0.cm 0cm
			0cm,width=0.48\textwidth]{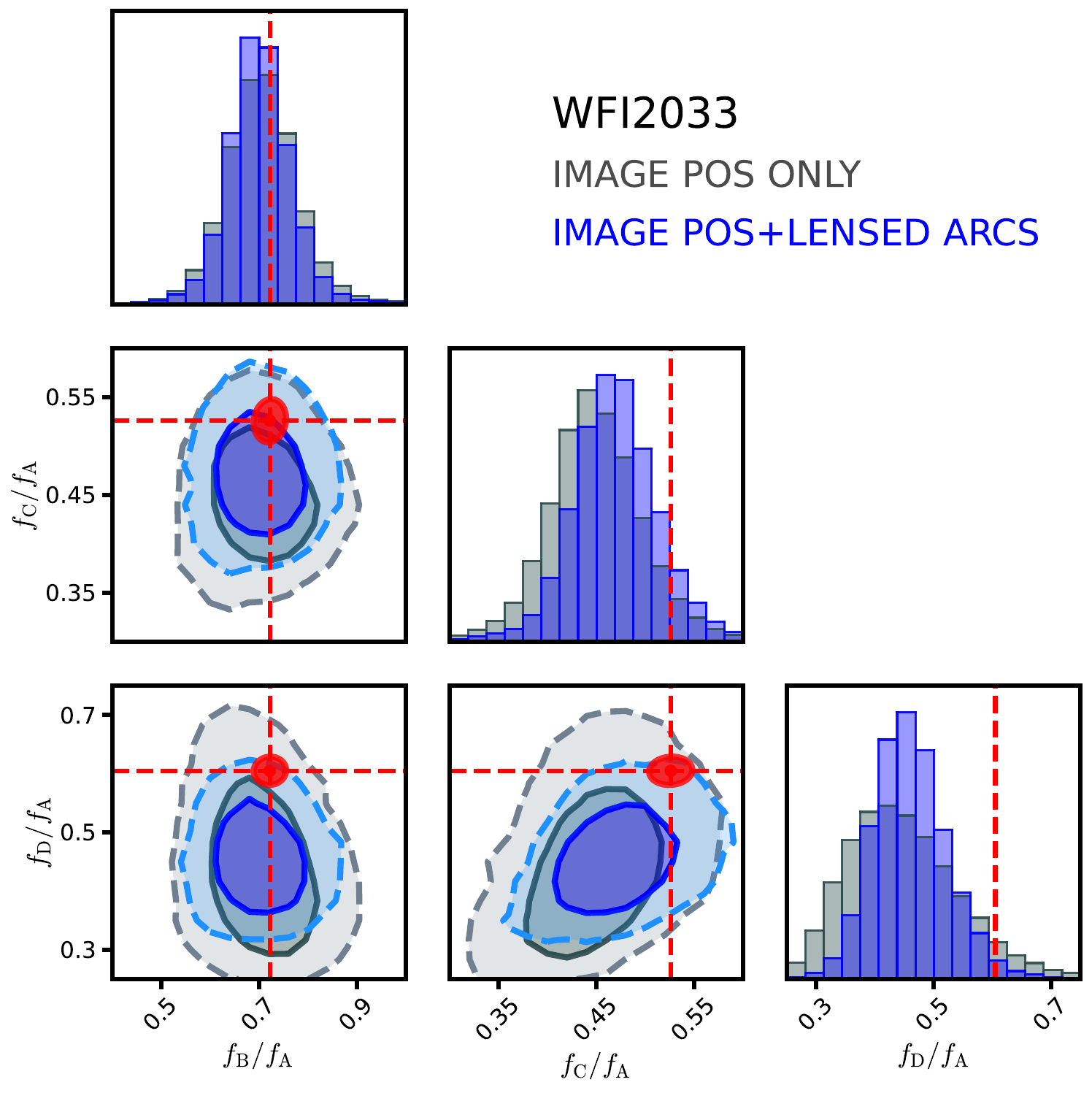}
			\includegraphics[trim=0cm 0.cm 0cm
			0cm,width=0.48\textwidth]{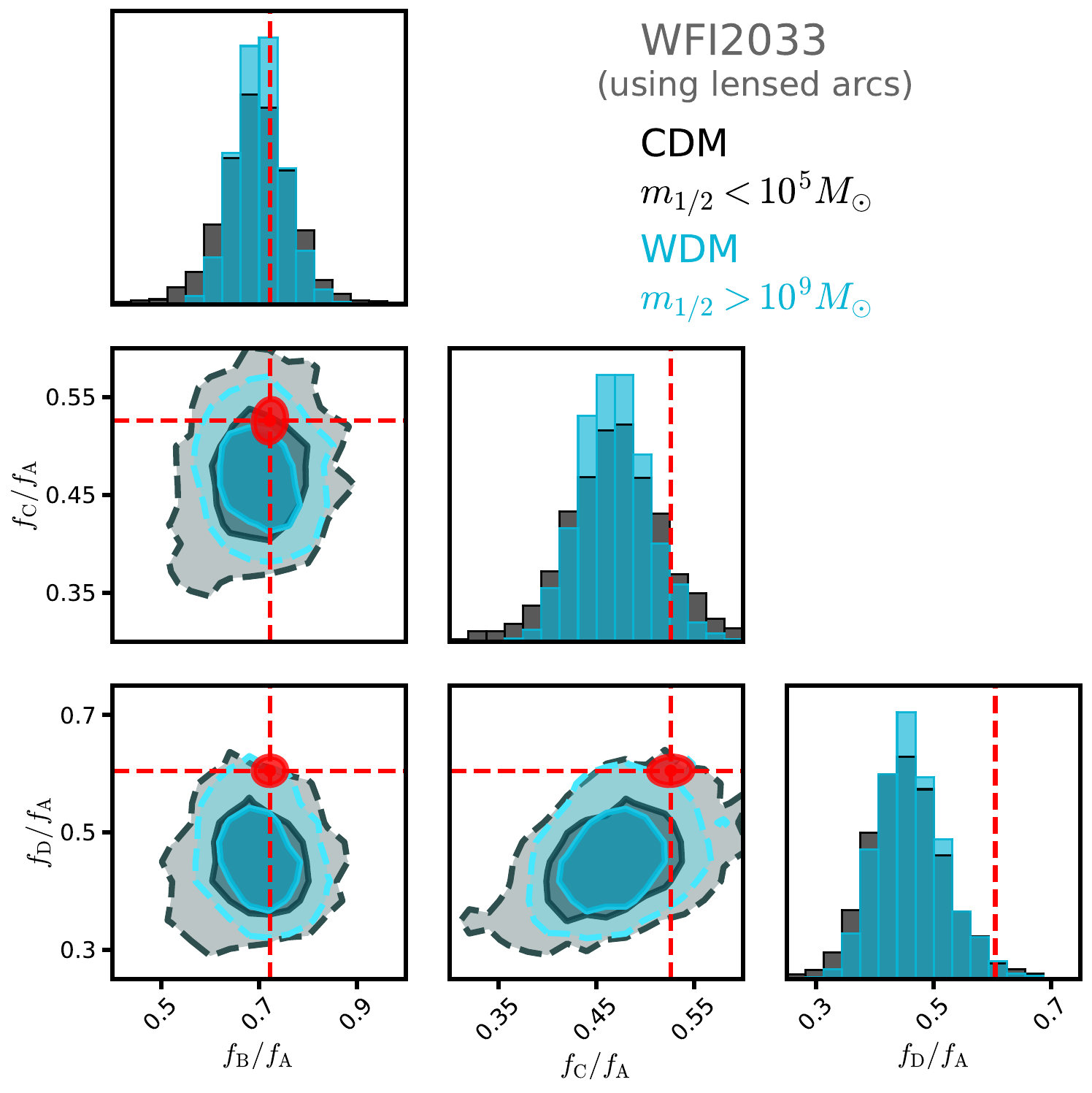}
			\caption{\label{fig:2033fr} The same as Figures \ref{fig:1537fr} and \ref{fig:mg0414fr}, but for the lens system WFI2033.}
		\end{figure*}
		\begin{figure*}
			\centering
			\includegraphics[trim=0cm 0.cm 0cm
			0cm,width=0.32\textwidth]{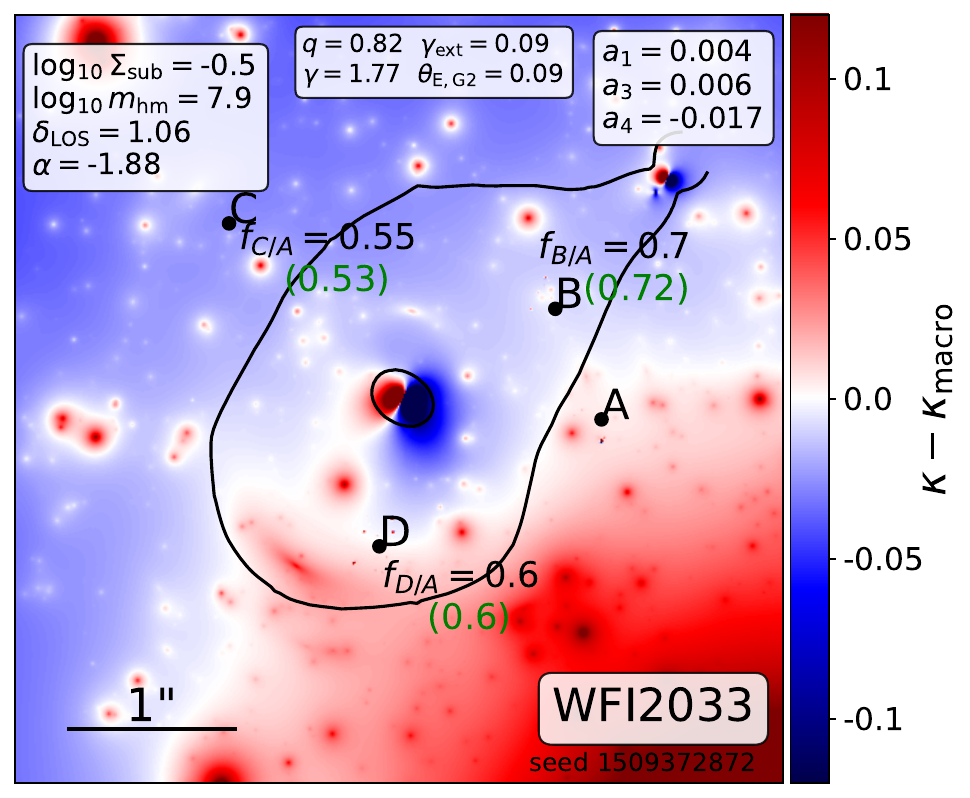}
			\includegraphics[trim=0cm 0.cm 0cm
			0cm,width=0.34\textwidth]{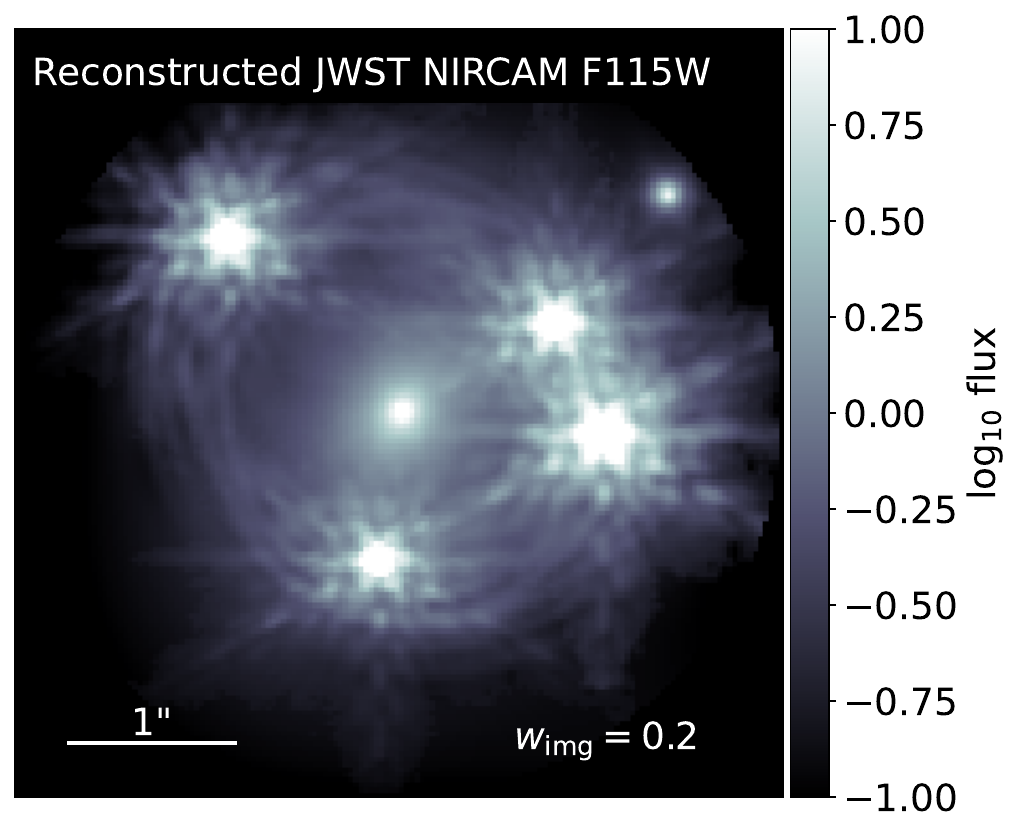}
			\includegraphics[trim=0cm 0.cm 0cm
			0cm,width=0.32\textwidth]{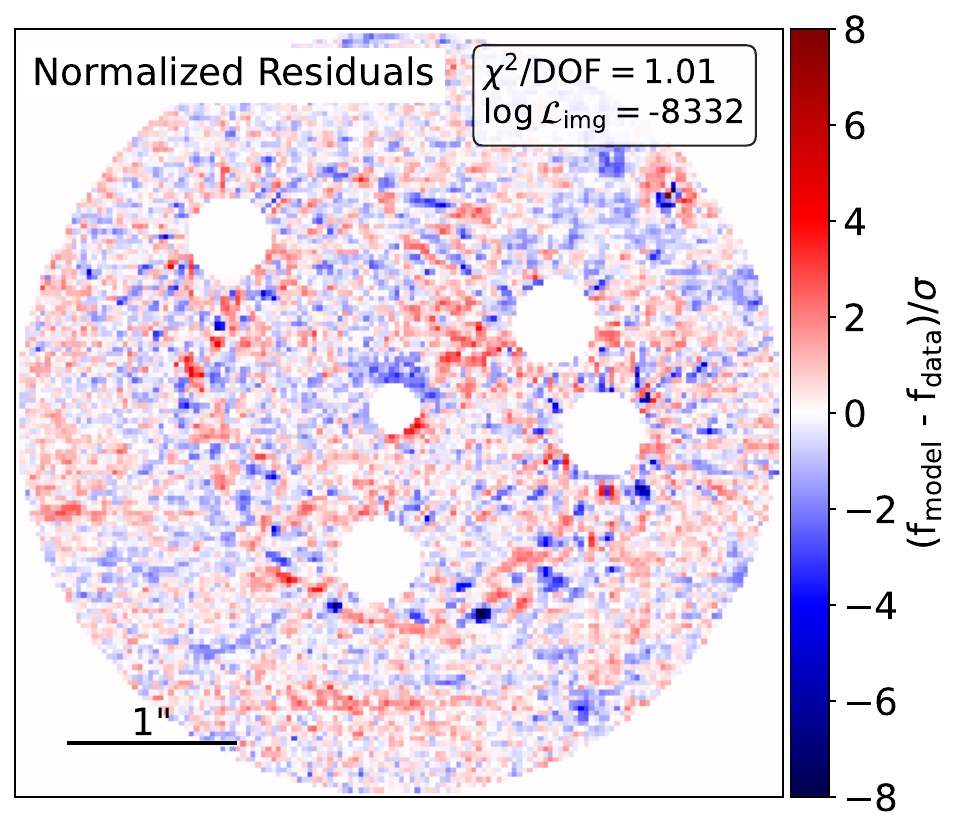}
			\includegraphics[trim=0cm 0.cm 0cm
			0cm,width=0.32\textwidth]{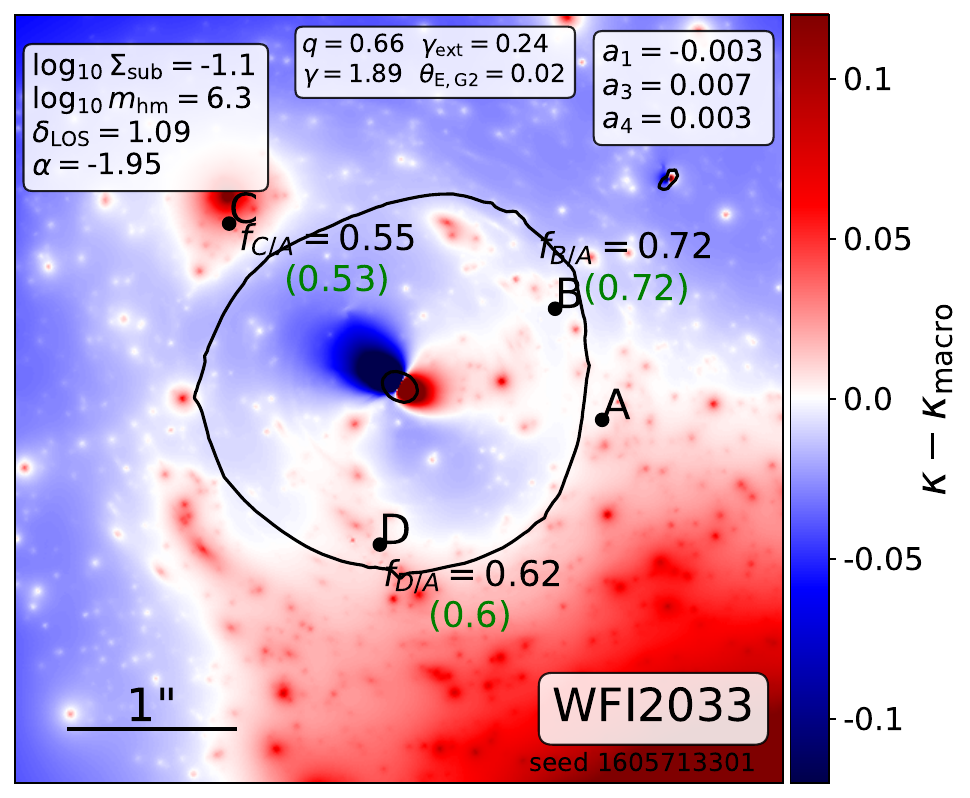}
			\includegraphics[trim=0cm 0.cm 0cm
			0cm,width=0.34\textwidth]{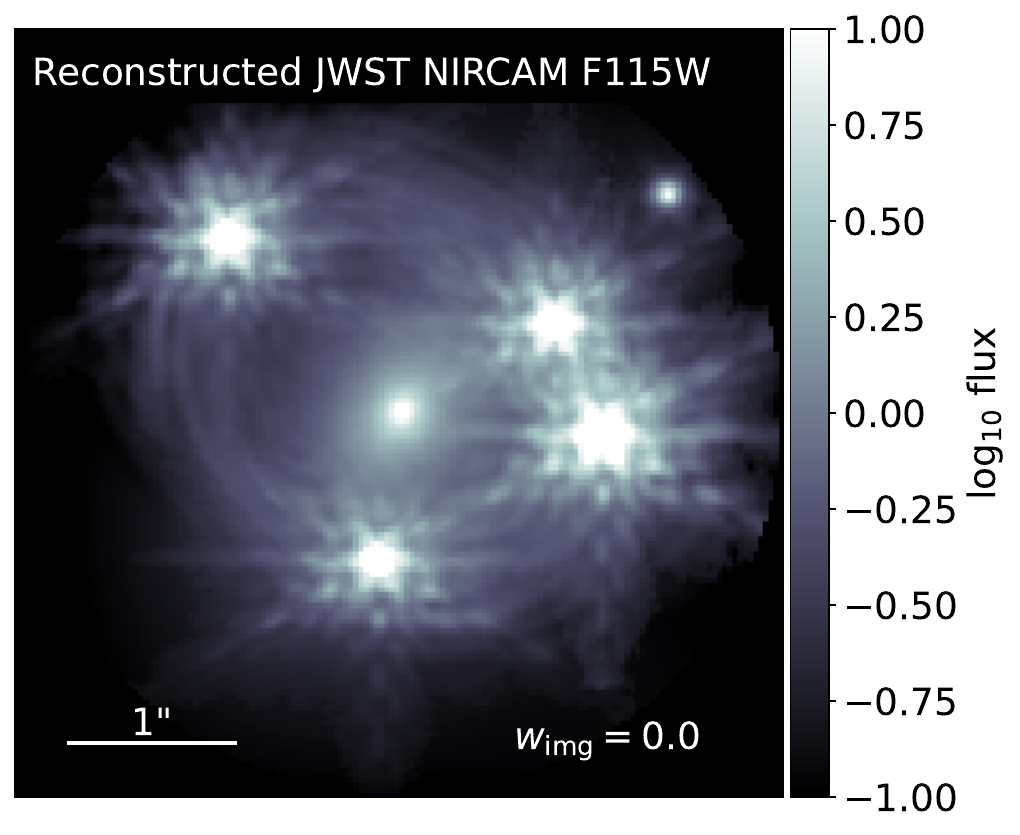}
			\includegraphics[trim=0cm 0.cm 0cm
			0cm,width=0.32\textwidth]{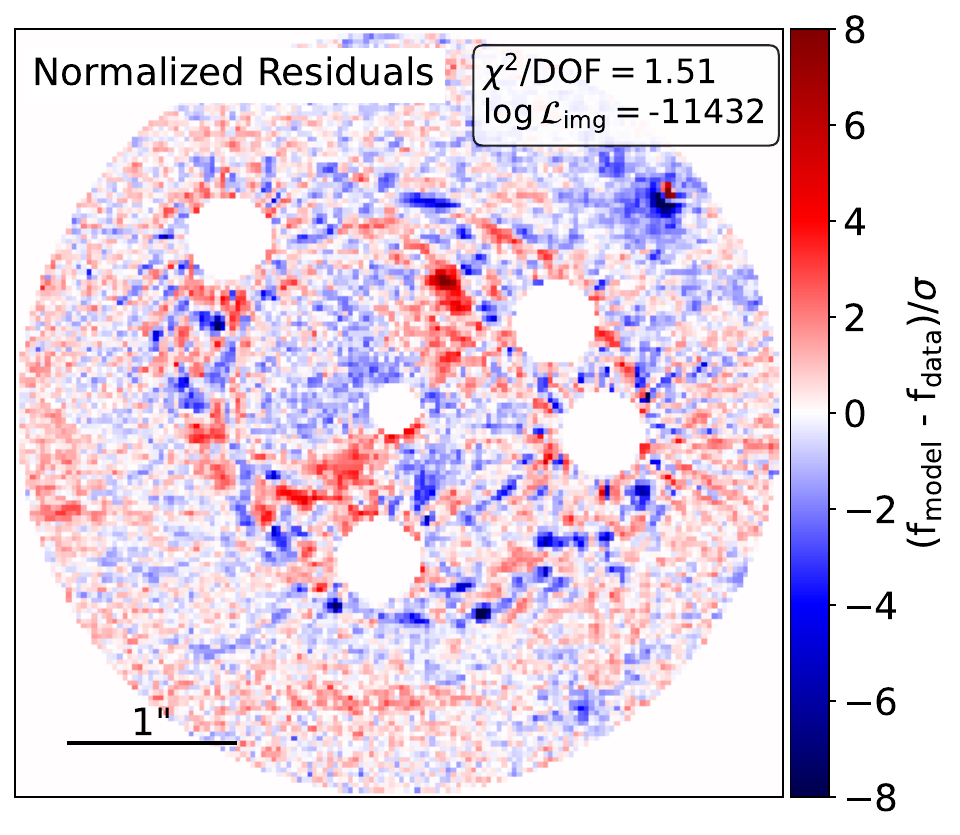}
			\caption{\label{fig:2033fromseedaccepted} The same as Figures \ref{fig:1537fromseedaccepted} and \ref{fig:0414fromseedaccepted}, but for the lens system WFI2033. The dipole structure of the convergence map appears due to the inclusion of the massive $\left(\theta_{\rm{E},G3} \sim 0.7 \ \rm{arcsec}\right)$ galaxy at $z = 0.745$ between the main deflector ($z_{\rm{d}} = 0.66$) and the source.}
		\end{figure*}  
		
		Figures \ref{fig:1537macro} and \ref{fig:1537fr} demonstrate how modeling of lensed arcs improves sensitivity to perturbations by dark matter substructure on a statistical level. We can also illustrate this concept on a realization-by-realization basis by reconstructing individual lens models from random seeds. In Figure \ref{fig:1537fromseedaccepted}, the top and bottom rows show two lens models produced for J1537 in our forward modeling pipeline. The left panels show the convergence in dark matter substructure (Equation \ref{eqn:effectivekappa}). Images are labeled A-D, with the measured (green) and model-predicted (black) flux ratios listed alongside. The solid black curve is the critical curve. White boxes show the dark matter hyper-parameters, a subset of macromodel parameters, and the strength of the multipole terms (from left to right). The center panels show the reconstructed imaging data, including the quasar images, lensed arcs, and the main lens light. In the center panels we also quote the image data importance weight, $w_{\rm{img}}$, defined in Equation \ref{eqn:neff}. The panels on the right show normalized residuals of the imaging data together with the log-likelihood and the reduced $\chi^2$ of the fit to the imaging data. 
		
		The lens models shown in the top and bottom row of Figure \ref{fig:1537fromseedaccepted} match the observed image positions and flux ratios to within $1 \sigma$. Using only these data, we would therefore assign nearly equal probability to these solutions. However, the lens model shown in the bottom row cannot reproduce the structure of the lensed arcs. We therefore reject this solution ($w_{\rm{img}}=0$), and the dark matter hyper-parameters associated with it. These figures clearly illustrate how lensed arcs break degeneracies between small-scale structure in the lens model and the structure of the main deflector mass profile. The lens model in the bottom row that does not fit the arcs happens to have a WDM-like mass function $m_{\rm{hm}} = 10^{6.8} \mathrm{M}_{\odot}$, but we stress that the lens models shown in Figure \ref{fig:1537fromseedaccepted} are just two examples out of 3.1 million created for this system. The constraining power comes from considering millions of realizations. Whether the additional constraints from lensed arcs favor a WDM or a CDM explanation for the data varies from one lens to another. 
		
		The degree to which lensed arcs aid in disentangling small-scale perturbation by halos from uncertainties associated with the macromodel depends on the number of degrees of freedom in the macromodel. With increasing complexity, the imaging data plays a more important role in isolating perturbations by halos from other sources of uncertainty. Figures \ref{fig:0414fromseedaccepted} and \ref{fig:j0607fromseedaccepted} illustrate this concept for two lens systems, MG0414 and J0607, that have nearby luminous satellite galaxies. The former is a lens system that has been the subject of lens modeling analyses for $\sim 30$ years since its discovery in 1992 \citep{Hewitt++92}. This system has a luminous feature near image C that \citet{Ros++00} identify as a probable satellite galaxy associated with the main deflector. Although the MIRI F560W image has lower signal to noise ratio than the HST or NIRCam data, the lensed arc contains enough information to constrain the mass and position of the satellite. As seen in the bottom row of Figure \ref{fig:0414fromseedaccepted}, an SIS satellite with $\theta_{E}\sim 0.25$ cannot reproduce the surface brightness of the lensed arc. In particular, the satellite appears to split the arc, as seen from the dipole pattern in the imaging data residuals to the north of image C. The top row of Figure~\ref{fig:0414fromseedaccepted} shows a more probable lens model in which the satellite has $\theta_{\rm{E}} \sim 0.15$, providing a more probable recreation of the lensed arc near image C. The model-predicted flux ratios for MG0414 are shown in Figure \ref{fig:mg0414fr}. In the case of J0607, which was discussed in Section~\ref{ssec:imglike} in relation to computing the importance sampling weights $w_{\rm{img}}$ (see Figures~\ref{fig:wimgpdfs} and~\ref{fig:frwimg}), the MIRI imaging data also constrain the mass of the satellite to the north of image B. Two lens models that fit the observed flux ratios, but which we can distinguish when incorporating constraints on the lensed arcs, are shown in Figure~\ref{fig:j0607fromseedaccepted}. We show the flux ratio distributions in CDM and WDM for J0607 in Figure~\ref{fig:j0607fr}, the distributions showing the effect of incorporating imaging data appeared in Figure~\ref{fig:frwimg}. 
		
		In addition to constraining the mass of nearby satellites, imaging data also aid in disentangling small-scale perturbation by halos from angular complexity in the main deflector mass profile. Figure~\ref{fig:2033macro} shows the inferred macromodel parameters for WFI2033, and provides a clear demonstration. Considering only the image positions and lensed arcs provides no significant evidence for a $m=4$ multipole perturbation. Some evidence for a significant $m=4$ component appears when considering the image positions and flux ratios (magenta). The strongest evidence for a non-zero $m=4$ term comes when combining all three datasets (red). While the imaging data does not directly constrain the $m=4$ component, as shown from the blue distribution, it does rule out combinations of other lens model parameters ($q$, $\gamma_{\rm{ext}}$, $\gamma$, etc) that reproduce the observed flux ratios in combination with $a_4 \sim 0$. The most probable solutions, those that simultaneously match all available data, include dark matter substructure and boxy isodensity contours ($a_4 \lesssim -0.015$). Figures \ref{fig:2033fr} and \ref{fig:2033fromseedaccepted} show the flux ratios predicted by the lens model and two examples of lens model reconstructions for WFI2033, respectively. The lens system shown in the bottom row fits the flux ratios with a particular configuration of dark matter substructure, $a_4 \approx 0$, and an Einstein radius of the nearby satellite galaxy $\theta_{\rm{E,G2}}=0.02$, but this combination of macromodel parameters cannot reproduce the lensed arcs. 
		
		In the previous paragraphs we have summarized some of the key advantages of using imaging data in a substructure lensing analysis. In particular, the imaging data leads to more precise model-predicted flux ratios, isolating small-scale perturbations by halos from uncertainties associated with the lens macromodel. The imaging data also constrains the mass and position of satellite galaxies, as illustrated by the cases of MG0414, J0607, and WFI2033. In the case of WFI2033, simultaneous reconstruction of the lensed arcs and flux ratios reveals the presence of angular complexity in the main deflector mass profile. The concepts discussed throughout this section apply to each of the 24 lens systems shown in Figure~\ref{fig:mosaic} for which we jointly model lensed arcs with image positions and flux ratios. The additional constraints from lensed arcs enable more robust inferences of dark matter properties, as discussed in the next section, which presents our inference on WDM from the full sample of lenses. 
		
		\subsection{Constraints on WDM}
		\label{ssec:resultswdm}
		Following Equation~\ref{eqn:likelihood}, we marginalize over the macromodel parameters and individual substructure realizations to compute the likelihood of the $n$th dataset given the dark matter hyper-parameters, $\mathcal{L}\left(\bf{\datan}|\qsub\right)$. Taking the product of the likelihoods and multiplying by the prior $\pi\left(\bf{q}\right)$ yields the posterior. The posterior distributions presented in this section result from generating millions of realizations per lens, and over 174 million for the full sample. Appendix~\ref{app:convergence} presents convergence tests that confirm that we have generated enough samples to overcome shot noise in individual likelihood functions. 
		
		Throughout this section, we will quote the results of our analysis in terms of both the one-sided $95 \%$ exclusion limit on $m_{\rm{hm}}$ obtained from the posterior, and the value of $m_{\rm{hm}}$ that corresponds to a Bayes factor (or relative likelihood) of 10:1 with respect to the peak of the posterior distribution. The motivation for quoting a Bayes factor is that this metric does not depend on the $m_{\rm{hm}}$ prior. On the other hand, for an inference that is consistent with CDM, we expect $m_{\rm{hm}}$ to be unconstrained from below, because the data are insensitive to $m_{\rm{hm}} \lesssim 10^5 \mathrm{M}_{\odot}$. Confidence intervals and exclusion limits will therefore depend on the lower bound of the $m_{\rm{hm}}$ prior. Any future comparison with our bounds on $m_{\rm{hm}}$ in terms of a one-sided exclusion limit should be made using the same prior. 
		
\begin{figure*}
			\centering
			\includegraphics[trim=0cm 0.5cm 0cm
			0cm,width=0.48\textwidth]{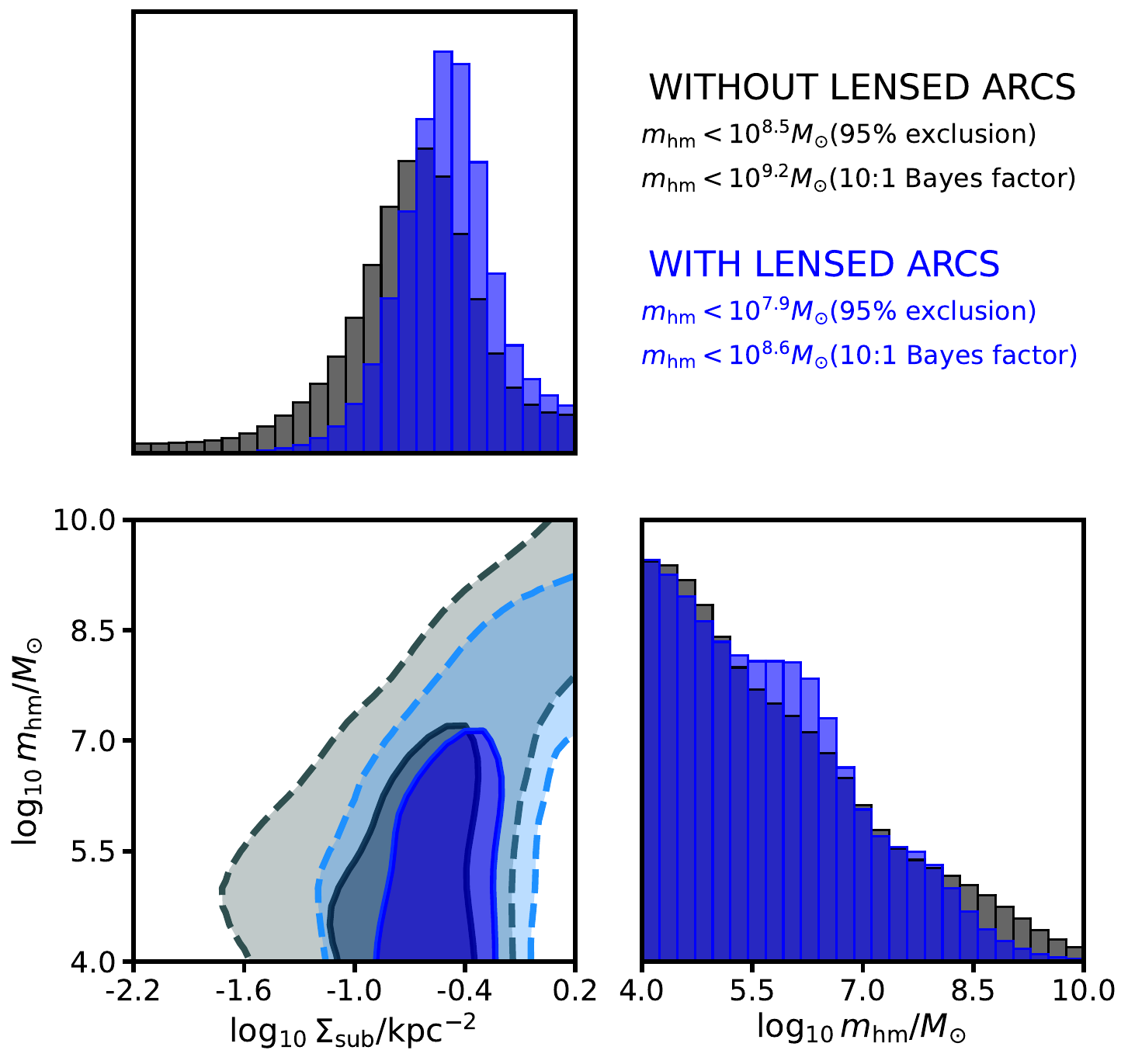}
			\includegraphics[trim=0cm 0.5cm 0cm
			0cm,width=0.48\textwidth]{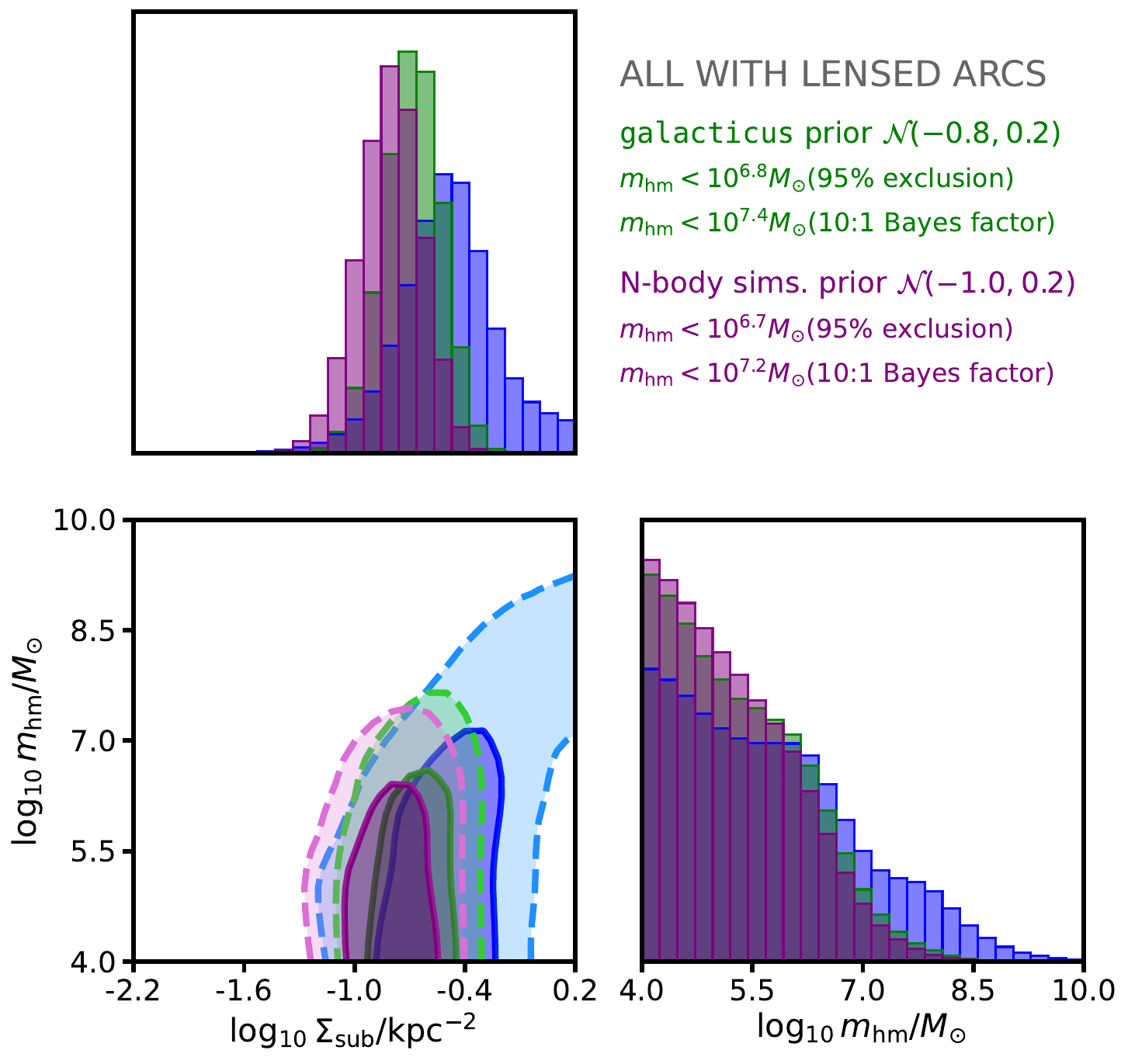}
			\caption{\label{fig:uniformpriors} {\bf{Left:}}Posterior distribution of the hyper-parameters $\Sigma_{\rm{sub}}$ and $m_{\rm{hm}}$, the normalization of the differential subhalo mass function (Equation \ref{eqn:subhalomfunc}) and the half-mode mass (Equation \ref{eqn:mhm}), respectively. We marginalize over $\alpha$ and $\delta_{\rm{LOS}}$, the logarithmic slope of the subhalo mass function and the amplitude of the field halo mass function, respectively, as these parameters are unconstrained within the range of their priors. The black distribution uses only image positions and flux ratios, and the blue distribution shows the inference after incorporating imaging data. {\bf{Right:}} Inference on $\Sigma_{\rm{sub}}$ and $m_{\rm{hm}}$ using flux ratios and lensed arcs, showing the effect of including a prior on $\Sigma_{\rm{sub}}$. The blue distribution is the same as the blue distribution in the left panel, and uses a log-uniform prior. The green distribution uses a prior $\log_{10}\Sigma_{\rm{sub}} / \rm{kpc^{-2}}\sim \mathcal{N}\left(-0.8, 0.2\right)$ based on predictions from the semi-analytic model {\tt{galacticus}}. The purple distribution shows the result when using a prior based on $N$-body simulations $\log_{10}\Sigma_{\rm{sub}} / \rm{kpc^{-2}}\sim\mathcal{N}\left(-1.0, 0.2\right)$, as discussed in Section \ref{ssec:priors}}.
		\end{figure*}

		The left panel of Figure~\ref{fig:uniformpriors} shows the posterior $p\left(\log_{10}\Sigma_{\rm{sub}}, \log_{10} m_{\rm{hm}} | \data \right)$ with a log-uniform prior on each hyper-parameter. We marginalize over $\alpha$ and $\delta_{\rm{LOS}}$, which are unconstrained given the width of the prior on these quantities. In black we show the inference obtained by using only the relative image positions and flux ratios, while in blue we incorporate constraints on the macromodel from the lensed arcs. 
		
		From Figure~\ref{fig:uniformpriors} we see that $\Sigma_{\rm{sub}}$ and $m_{\rm{hm}}$ are correlated. The physical interpretation of this correlation is that adding additional more-massive subhalos to warmer (larger $m_{\rm{hm}}$) dark matter models can mimic the lensing signal of a colder dark matter model with fewer subhalos. Incorporating constraints from imaging data aids in breaking this covariance, strengthening the bound on $m_{\rm{hm}}$ by 0.6 dex, as summarized in the figure caption.
		
		The right panel of Figure~\ref{fig:uniformpriors} shows the posterior distribution with informative priors on $\Sigma_{\rm{sub}}$ based on $N$-body simulations (purple) and semi-analytic model {\tt{galacticus}} (green), as indicated in the figure caption. These priors break covariance between $\Sigma_{\rm{sub}}$ and the half-mode mass $m_{\rm{hm}}$. As discussed in Section \ref{ssec:priors}, we derive these priors by determining the value of $\Sigma_{\rm{sub}}$ (the infall mass function amplitude) that produces the bound mass function amplitude predicted by ${\tt{galacticus}}$ and the Symphony $N$-body simulations \citep{Nadler++23}. For a prior $\pi\left(\log_{10}\Sigma_{\rm{sub}} / \rm{kpc^{-2}}\right)= \mathcal{N}\left(-0.8, 0.2\right)$, as predicted by {\tt{galacticus}}, the bound on $m_{\rm{hm}}$ strengthens to $m_{\rm{hm}} < 10^{7.4} \mathrm{M}_{\odot}$. This constraint is 0.5 dex stronger than the limit obtained without using lensed arcs, presented in a companion paper, \citet{Keeley++25}. Assuming $\pi\left(\log_{10}\Sigma_{\rm{sub}} / \rm{kpc^{-2}}\right)= \mathcal{N}\left(-1.0, 0.2\right)$, as predicted by $N$-body simulations, we infer $m_{\rm{hm}} < 10^{7.2} \mathrm{M}_{\odot}$, an improvement of 0.4 dex relative to omitting imaging data. 
		
		These upper limits on $m_{\rm{hm}}$ correspond to (see Equation \ref{eqn:mhm}) lower bounds on the mass of a thermal relic WDM particle of $6.5 \ \rm{keV}$ and $7.4 \ \rm{keV}$ for the {\tt{galacticus}} and $N$-body priors, respectively. The corresponding one-sided $95 \%$ exclusion limits are $m_{\rm{hm}}< 10^{6.8} M_{\odot}$ and $m_{\rm{hm}} < 10^{6.7} M_{\odot}$, respectively, which correspond to thermal relic particle constraints $m_{\rm{thermal}}> 9.6 \ \rm{keV}$ and $m_{\rm{thermal}} > 10.2 \ \rm{keV}$, respectively. As discussed in Section \ref{ssec:wdm}, to compute these thermal relic particle masses we use the conversion presented by \citet{Vogel++23}, who calibrate the $m_{\rm{hm}} \rightarrow m_{\rm{therm}}$ mapping for colder WDM models than those presented by \citet{Viel++05}. The thermal relic particle mass corresponding to a given $m_{\rm{hm}}$ is lighter by $\sim 20\%$ than the value derived from \citet{Viel++05}, who considered warmer models than \citet{Vogel++23}. Using the conversion factors quoted by \citet{Viel++05}, our 10:1 Bayes factor limits are 7.4 keV and 8.4 keV for $m_{\rm{hm}}=10^{7.4} M_{\odot}$ and $m_{\rm{hm}}=10^{7.2} M_{\odot}$, respectively.  
		
		\subsection{Measurement of the subhalo mass function in CDM}
		\label{ssec:resultscdm}
\begin{figure}
			\centering
			\includegraphics[trim=0.0cm 0.2cm 0.0cm
			0cm,width=0.48\textwidth]{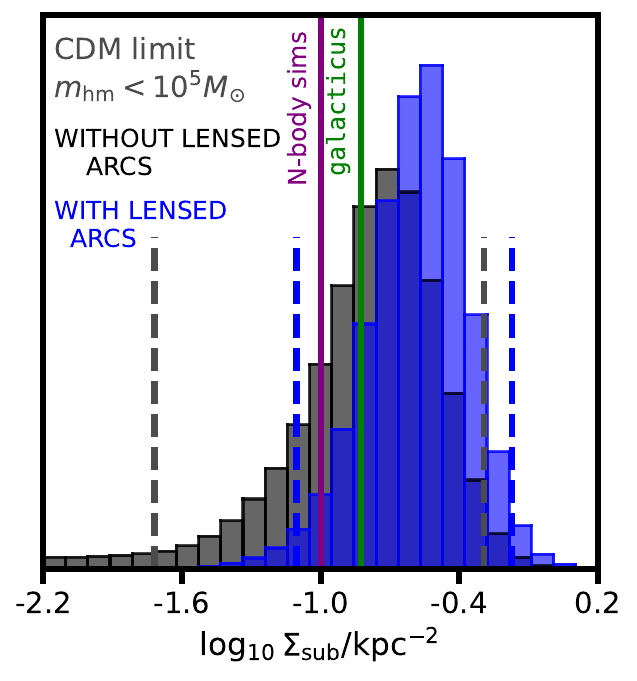}
			\caption{\label{fig:cdmprior} The posterior distribution of $\Sigma_{\rm{sub}}$ obtained from taking the CDM limit through a prior on the half-mode mass $\pi\left(\log_{10} m_{\rm{hm}} / \mathrm{M}_{\odot}\right) = \mathcal{N}\left(4.0, 0.5\right)$. The color scheme is the same as the left panel of Figure \ref{fig:uniformpriors}. Vertical dashed lines represent $95\%$ confidence intervals. Predictions from the semi-analytic model $\tt{galacticus}$ and $N$-body simulations (see discussion in Section \ref{ssec:priors}) are marked with vertical lines. The upper x-axis shows the corresponding projected mass density in subhalos with infall masses in the range $10^6 < m/M_{\odot}<10^{10.7}$.}
		\end{figure}

		Restricting to regions of parameter space with $m_{\rm{hm}} < 10^5 \mathrm{M}_{\odot}$ removes most of the effects of WDM free streaming in the model-predicted datasets and flux ratio statistics. This prior therefore takes the CDM limit of the joint posterior distribution, which we can use to infer the amplitude of the differential subhalo mass function in CDM (Equation~\ref{eqn:subhalomfunc}). In Figure~\ref{fig:cdmprior}, we show the posterior distribution $p\left(\Sigma_{\rm{sub}} | \data\right)$. At $95\%$ confidence we infer $\log_{10} \Sigma_{\rm{sub}} / \rm{kpc^{-2}} = -0.57_{-0.55}^{+0.42}$, a more precise measurement of subhalo abundance relative to omitting imaging data $\log_{10} \Sigma_{\rm{sub}} / \rm{kpc^{-2}} = -0.76_{-0.95}^{+0.46}$. As discussed in Section \ref{ssec:priors}, to convert these values to the amplitude of the bound mass function in CDM, for which tidal stripping is approximately independent of subhalo infall mass, one can use Equation \ref{eqn:subhalomfunc} with a normalization at $10^8 M_{\odot}$ given by $\bar{f}_{\rm{bound}} \times \Sigma_{\rm{sub}}$, with $\bar{f}_{\rm{bound}}=0.05$ \citep{Du++25}. 
		
		Subhalo abundance is often quoted in terms of a projected mass in substructure, or as a projected mass fraction in substructure \citep[e.g.][]{Dalal++02,Vegetti++14,Hsueh++20,Banik++21,Nibauer++25}. To enable comparison with previous work, we recast the inferences on $\Sigma_{\rm{sub}}$ in terms of a projected mass density in subhalos, $\Sigma_{\rm{DM,sub}}$, and a projected mass fraction in subhalos, $f_{\rm{sub}}$. To calculate $\Sigma_{\rm{DM,sub}}$, we integrate the subhalo mass function (Equation \ref{eqn:subhalomfunc}) over the range $m_{L} < m < m_{H}$
		\begin{equation}
			\label{eqn:sigmadm}
			\Sigma_{\rm{DM,sub}} = \frac{\Sigma_{\rm{sub}} \bar{f}_{\rm{bound}} m_0}{2+\alpha} \left[\left(m_{\rm{H}}/m_0\right)^{2+\alpha} - \left(m_{\rm{L}}/m_0\right)^{2+\alpha} \right].
		\end{equation}
		Here, we assume a typical host halo mass $10^{13} \mathrm{M}_{\odot}$ for a deflector at $z_{\rm{d}}=0.5$, such that $\mathcal{F}\left(m_{\rm{host}},z_{\rm{d}}\right)=1$ in Equation \ref{eqn:shmfscaling}. Given the width of the prior on $\alpha$, and the fact that the posterior on $\alpha$ is not constrained within the range of this prior, we set $\alpha = -1.9$. Our inference used $m_{H} = 10^{10.7} \mathrm{M}_{\odot}$ and $m_L = 10^6 \mathrm{M}_{\odot}$, respectively, with a pivot scale $m_0=10^8 \mathrm{M}_{\odot}$. With these numbers we find $\Sigma_{\rm{DM,sub}} = 6.2 \times 10^7 \left(\Sigma_{\rm{sub}} / \rm{kpc^{-2}}\right) $. The upper x-axis of Figure \ref{fig:cdmprior} shows the value of $\Sigma_{\rm{DM,sub}}$ that corresponds to $\Sigma_{\rm{sub}}$. Our inference on $\Sigma_{\rm{sub}}$ corresponds to $\Sigma_{\rm{DM,sub}} = 1.7_{-1.2}^{+2.6} \times 10^7 \ \mathrm{M}_{\odot} \rm{kpc^{-2}}$ at $95 \%$ confidence. This is statistically consistent with the predictions from {\tt{galacticus}}, $\Sigma_{\rm{DM,sub}} = 0.9 \times10^7 \ \mathrm{M}_{\odot} \rm{kpc^{-2}}$. It is in mild $\sim 1.8 \sigma$ tension with the predictions from the recent Symphony $N$-body simulations presented by \citet{Nadler++23}, $\Sigma_{\rm{DM,sub}} = 0.6 \times10^7 \mathrm{M}_{\odot} \rm{kpc^{-2}}$. These predictions from $N$-body simulations and {\tt{galacticus}} follow from the theoretical priors on $\Sigma_{\rm{sub}}$ discussed in Section \ref{ssec:priors}. 
		
		We now cast our inference on $\Sigma_{\rm{sub}}$ as an inference on $f_{\rm{sub}}$, the fraction of mass bound to dark matter subhalos in projection near the Einstein radius, $f_{\rm{sub}} \equiv \Sigma_{\rm{DM,sub}} A /  M$. Here, $A$ represents an area near lensed images, and $M$ is the total mass in dark matter in this area. For early-type galaxies with near-isothermal mass profiles, the convergence, or projected mass divided by the critical surface mass density for lensing, $\Sigma_{\rm{crit}}$, will be approximately $0.5$. Assuming a projected mass fraction in dark matter, $f_{\rm{DM}}$, at the Einstein radius, we can therefore write $M = (1/2) f_{\rm{DM}} \Sigma_{\rm{crit}} A$. We then obtain $f_{\rm{sub}}$ in terms of $\Sigma_{\rm{DM,sub}}$, $f_{\rm{DM}}$, and $\Sigma_{\rm{crit}}$
		\begin{equation}
			\label{eqn:fsub}
			f_{\rm{sub}} = \frac{2 \Sigma_{\rm{DM,sub}}}{f_{\rm{DM}} \Sigma_{\rm{crit}}}
		\end{equation}
		
		For $f_{\rm{DM}}=0.5$ \citep{Auger++10}, and $\Sigma_{\rm{crit}} = 2.0 \times 10^9 \mathrm{M}_{\odot} \ \rm{kpc^{-2}}$, the critical density for lensing assuming typical lens and source redshifts $z_{\rm{d}} = 0.5$ and $z_{\rm{s}}=2.0$, Equation \ref{eqn:fsub} gives $f_{\rm{sub}} = 0.012 \left(\Sigma_{\rm{sub}} / 0.1 \ \rm{kpc^{-2}}\right)$. Our inference on $\Sigma_{\rm{sub}}$ using imaging data therefore corresponds to a projected mass fraction bound to dark matter subhalos $f_{\rm{sub}} = 3.2_{-2.3}^{+5.1} \%$ at $95 \%$ confidence.  
		
		\section{Discussion and conclusions}
		\label{sec:discussion}
		We present an analysis of 28 quadruply imaged quasars with the objective of measuring the free-streaming length of dark matter and the normalization of the subhalo mass function. Our analysis combines observations from the Hubble and James Webb Space Telescopes, including the full complement of quadruply imaged quasars with flux ratios measured through JWST GO-2046, and archival HST and JWST observations of extended lensed arcs. Using a state-of-the-art forward modeling framework, we simultaneously reconstruct the lensed image positions, flux ratios, and extended lensed arcs in the presence of full populations of globular clusters, dark matter subhalos, and field halos along the line of sight. In total, we generated over 174 million realizations of dark matter substructure, which we use to characterize halo properties on the population level. We summarize our main results as follows: 
		\begin{itemize}
			\item Assuming a log-uniform prior on the normalization of the subhalo mass function $\Sigma_{\rm{sub}}$ that spans a factor of 100 in subhalo abundance, incorporating constraints from lensed arcs improves bounds on dark matter warmth by 0.6 dex, leading to $m_{\rm{hm}} < 10^{8.6} \mathrm{M}_{\odot}$ (Bayes factor 10:1). These weaker constraints relative to some previous work are the result of extending the $\Sigma_{\rm{sub}}$ prior to higher values, which reveals a significant covariance between $\Sigma_{\rm{sub}}$ and $m_{\rm{hm}}$. Breaking this covariance with a prior on $\Sigma_{\rm{sub}}$ based on the semi-analytic model {\tt{galacticus}} and $N$-body simulations results in $m_{\rm{hm}} < 10^{7.4} \mathrm{M}_{\odot}$ and $m_{\rm{hm}} < 10^{7.2} \mathrm{M}_{\odot}$, respectively. These bounds are stronger than those obtained without imaging data by 0.4 dex. The half-mode mass upper limits of $10^{7.4} M_{\odot}$ and $10^{7.2} M_{\odot}$ correspond to lower bounds on the mass of a thermal relic dark matter particle $6.5$ keV and $7.4$ keV, respectively, using Equation \ref{eqn:mhm}.
			\item Assuming CDM, we infer a projected surface mass density in subhalos $1.7_{-1.2}^{+2.6} \times 10^7 \ \mathrm{M}_{\odot} \rm{kpc^{-2}}$, and a projected mass fraction near the typical Einstein radius of a strong lens system $f_{\rm{sub}} = 3.2_{-2.3}^{+5.1} \%$. These measurements of subhalo abundance are statistically consistent with the predictions of the semi-analytic model {\tt{galacticus}}, but in mild tension (higher) than the predictions from recent $N$-body simulations. 
		\end{itemize}
		
		\subsection{Comparison with previous work}
		This analysis incorporates numerous improvements to the modeling framework. First, we use a model for subhalo tidal evolution based on the framework presented by \citet{Du++25}, which effectively changes the physical interpretation of the $\Sigma_{\rm{sub}}$ parameter relative to previous work \citep{Gilman++20,Keeley++24}. This improved tidal evolution model predicts additional suppression in the bound mass function of WDM, relative to CDM, as discussed in Section~\ref{ssec:mfuncs}, and computes $\rho_\mathrm{s}$ and $r_\mathrm{s}$ for subhalos at the infall redshift. This differs from the calculation of $\rho_\mathrm{s}$ and $r_\mathrm{s}$ at the lens redshift, as was done by \citet{Keeley++25}, or at $z=0$, as done by \citet{Gilman++20} and \citet{Hsueh++20}. Evaluating subhalo properties at infall makes subhalos denser by a factor of $10-200$ than evaluating $\rho_s$ and $r_s$ with respect to the critical density at $z=0$, due to the typical infall redshift $z \sim 4$ for subhalos accreted onto a group-scale host \citep{Du++25}, and the increase of the background density with redshift. The second improvement in the modeling framework pertains to the treatment of angular structure in the main deflector mass profile. Previous WDM studies either did not include angular structure \citep{Gilman++20,Hsueh++20}, or implemented it with fewer degrees of freedom \citep{Keeley++24} than in this work. Our priors on the $m=1, 3, 4$ multipole terms includes freely-varying orientations and amplitudes, which impart additional small-scale perturbation from halos alongside those caused by dark matter substructure. Furthermore, following \citet{PaugnatGilman25}, we have used the more appropriate elliptical multipole perturbations, rather than the standard but unphysical circular perturbations. Third, we have extended the upper mass range of subhalos rendered in our simulations to $10^{10.7} \mathrm{M}_{\odot}$ from the value of $10^{10} \mathrm{M}_{\odot}$ used in previous works. While we explicitly include more-massive satellites in the lens model if we detect their stellar light, we increase the upper limit to $10^{10.7} \mathrm{M}_{\odot}$ because many subhalos of this mass would not be visible in the MIRI imaging data that make up most of the lens sample. Fourth, we have used the model for the suppression of the halo mass function presented by \citep{Lovell++20}, which is less suppressed for a given $m_{\rm{hm}}$ than the model by \citep{Schneider++13} used by \citet{Gilman++20}. Finally, we include globular clusters alongside dark matter substructure as a source of small-scale perturbation to the image magnifications. With the exception the use of elliptical multipoles, which tend to impart weaker perturbations to flux ratios for highly-flattend systems, and the improved tidal stripping model, these changes add additional sources of flux-ratio perturbation for a fixed $m_{\rm{hm}}$. Additional sources of perturbation besides halos will tend to weaken constraints on WDM. Regarding the tidal stripping model, dependence of the subhalo bound mass fraction on the infall concentration causes additional additional suppression of small-scale structure for a fixed $m_{\rm{hm}}$, and therefore would tend to strengthen constraints. 
		
		In addition to the differences in the modeling, we use a different approach when computing the likelihood for each lens system. To generate model-predicted datasets, previous analyses initialized a non-linear solver using a particle swarm optimization (PSO), and then solved for the parameters of the elliptical power law profile and shear such that they satisfy the lens equation. In this work, we follow the same overall approach of solving for a set of macromodel parameters that satisfy the lens equation, but we enforce an explicit prior on the mass axis ratio $q$ based on the observed stellar light. This improvement uses observations of the main deflector to reduce the number of unconstrained degrees of freedom in the macromodel when applying the non-linear solver, and avoids an issue in which the initial size of the PSO search region can lead to an underestimation of uncertainties in the model-predicted flux ratios. Regarding the calculation of the likelihood function, previous analyses used Approximate Bayesian Computing, rendering statistical measurement uncertainties onto model-predicted fluxes or flux ratios, and then accepting or rejecting models based on a summary statistic. One of the motivating factors for using ABC is that uncertainties on the flux ratios are non-Gaussian, and ABC provides a means to deal with non-Gaussian errors. In this work, we have instead computed the flux ratio likelihood assuming a Gaussian covariance matrix. As discussed by \citet{Keeley++25}, the flux ratio measurements by JWST are sufficiently precise that the uncertainties are well-approximated by a multi-variate Gaussian. 
		
		Our results are broadly consistent with previous inferences from gravitational imaging \citep{Vegetti++14,Birrer++17,Hezaveh++16,Ritondale++19}, although the uncertainties from imaging analyses are significantly larger than those presented in this work. The larger uncertainties mainly result from extrapolating individual halo detections to statements on the full halo population. This complication aside, results from gravitational imaging offer a complementary tool to identify aspects of halo substructure that could impact population-level constraints. For example, the recent detection of an extremely compact $\sim 10^6 M_{\odot}$ object in an extended arc \citep{Powell++25} could suggest the existence of a population of similar objects. Assuming the perturber is not a globular cluster (which we already include in our model), if the number density of such objects is high enough, we may detect their presence in the flux ratio statistics of lensed quasars.  
		
		\subsection{Interpretation}
		Our constraints on the free-streaming length are stronger than those obtained from other individual small-scale probes, including previous lensing studies \citep{Birrer++17,Gilman++20,Hsueh++20,Keeley++24}, the Lyman-$\alpha$ forest \citep{Villasenor++23,Irsic++24}, dwarf galaxies \citep{Newton++21,Nadler++21,Dekker++22,Tan++25} and stellar streams \citep{Banik++21,Nibauer++25}. Our constraints surpass the strongest recent bound from strong lensing ($m_{\rm{hm}} < 10^{7.6} M_{\odot}$ at 10:1 odds) presented by \citet{Keeley++24}. Direct comparison with most other probes, such as the Lyman-$\alpha$ forest constraints, is complicated by the fact that these analyses quote bounds in terms of a confidence interval, which depends on the prior. The most recent Lyman-$\alpha$ forest constraints quote $m_{\rm{hm}} < 10^{7.6} M_{\odot}$ at $95 \%$ confidence \citep{Irsic++24}. Recent dwarf galaxy analyses also quote $m_{\rm{hm}} < 10^{7.6} M_{\odot}$ at $95 \%$ confidence \citep{Nadler++25cozmic}. The strongest existing bound \citep{Nadler++21} comes from a combination between strong lensing bounds from \citet{Gilman++20} and dwarf galaxies \citep{Nadler++20b}. These limits would likely improve considerably if the strong lensing constraints from ref. \citep{Gilman++20} were updated with the results from this work. Our limits are to be considered robust, as the numerous improvements to our modeling framework overall weaken the bounds on WDM for a given dataset with respect to previous analyses. 
		
		Our measurement of the projected mass density in CDM subhalos is consistent with the predictions from the semi-analytic model {\tt{galacticus}}, but in mild tension (inferring a higher value) with the predictions from the Symphony suite of $N$-body simulations \citep{Nadler++23}. The differences between subhalo abundance predicted by semi-analytic models and $N$-body simulations could suggest that recent $N$-body simulations still suffer from artificial subhalo disruption, or that halo finders do not locate all subhalos in the inner regions of the host, despite recent improvements on this front \citep[e.g.][]{Mansfield++24} that reveal more substructure than previously reported. 
		
		Alternatively, our inference of higher-than-predicted subhalo abundance could point to differences in the internal structure of subhalos and line-of-sight halos, with respect to the standard assumptions. For example, if halos had higher central densities than expected in CDM, we would infer a higher normalization of the halo mass function, because the inferred halo abundance is anti-correlated with the inferred concentration \citep{Gilman++22}. Investigating whether more-concentrated halos can reconcile our measurements with theoretical predictions for $\Sigma_{\rm{sub}}$ is left for future work. As another example, we have sampled host halo masses from a distribution $\log_{10} m_{\rm{host}}/ \mathrm{M}_{\odot} \sim \mathcal{N}\left(13.3, 0.3\right)$, which is typical for strong lensing deflectors \citep{Lagattuta++10}. If this value were systematically lower than the true population average in our sample this assumption would lead us to infer a higher $\Sigma_{\rm{sub}}$. A stronger theoretical prior on the host halo mass of typical lenses that accounts for lensing selection effects \citep{Sonnenfeld++23,Tang++25}, or a measurement of the mean host halo mass in our sample, would address this issue. We note that our treatment of the host halo mass as a nuisance parameter for each lens effectively introduces a factor of 2 halo-to-halo variance in the subhalo mass function normalization.  
		
		\subsection{Discussion and outlook}
		
		This work has assembled the largest sample of lensed quasars for a dark matter analysis since \citet{MaoSchneider98} proposed their use as a probe for dark substructure nearly 30 years ago. In addition to the flux ratios, we have demonstrated how modeling of lensed arcs can disentangle small-scale perturbations by halos from uncertainties associated with the mass profile of the main deflector on larger angular scales, leading to the most stringent constraints to date on dark matter free streaming length. Our analysis provides a clear example of how investment of observational resources from multiple space-based observatories delivers transformative dark matter science. 
		
		Extending our analysis to a larger sample of lenses would reduce the statistical uncertainties of the measurement, and push constraints on substructure properties to even lower halo mass scales. This will soon be possible, as cosmological surveys, such as Euclid, the Rubin Observatory, and the Roman Space Telescope discover thousands of new strong lens systems \citep{OguriMarshall10,Shajib++25}. Higher quality imaging data, particularly for systems that lack HST imaging, or where the lensed arcs in HST imaging have low signal to noise, would enable stronger constraints on the mass profile of the main deflector, leading to more robust dark matter inferences. As discussed in Section~\ref{ssec:lmspecific}, in several systems we discovered satellite galaxies of the main deflector by visual inspection of HST data. In other cases, we use high-resolution imaging to identify stellar disks in the main deflector, and omit these systems from the sample. These considerations underscore the importance of having high angular resolution space-based or adaptive optics imaging of the main deflector.
		
		The end-to-end forward modeling of strong gravitational lenses with dark matter substructure will continue to be viable path forward as we discover additional lens systems. The analysis framework we present can be scaled up to incorporate a larger sample size of quadruply imaged quasars, and to model imaging data simultaneously in multiple bands \citep[e.g.][]{Shajib++19,Schmidt++23}. Using lensed arcs to constrain the macromodel also enables doubly-imaged quasars to be included in dark matter inferences \citep{Harvey++20}. 
		
		The techniques we present can be applied to any theory that predicts the abundance and density profiles of halos, including models such as self-interacting dark matter, in which a fraction of halos and subhalos are expected to undergo core collapse \citep{Gilman++21,Gilman++23}, and models that alter the primordial matter spectrum \citep{ZentnerBullock03,Gilman++22,Baldi++24,Esteban++24,Wu++25,Dekker++25,Nadler++25}. In the case of WDM, one additional free parameter can account for the full shape of the transfer function in a variety of models \citep{Stucker++22}, which would enable more robust bounds on the WDM production mechanism. Interpretation of the data used in this paper in the context of self-interacting dark matter will be presented in a forthcoming publication. 
		
		Our lens modeling framework can accommodate other prescriptions for the mass profile of the main deflector, such as separate modeling of stars and dark matter, or additional small-scale degrees of freedom, such as ellipticity gradients and twists \citep[e.g.][]{VandeVyvere++22}. Pixel-based lens models can also introduce additional degrees of small-scale perturbation \citep{Kung++18}. The main requirement for these parameterizations and lens modeling frameworks to be included in our analysis is the fast calculation of deflection angles, which avoids expensive numerical integrals when generating over 100 million lens models, as done in this work. Promising avenues for future research in this regard include multi-Gaussian expansions \citep[e.g.][]{Cappellari02,Shajib++19b,He++24} of the deflector mass profile based on the deflector's surface brightness, lens modeling frameworks that relax global lens model assumptions in favor of a local description of the deflection field \citep[e.g.][]{Paugnat++25}, or the implementation of fast deflection field calculations that can be incorporated alongside parametric lens models, such as the elliptical multipoles \citep{PaugnatGilman25} used in this analysis. 
		
		Lensing enables direct tests of the CDM paradigm, and of any other theory that alters the abundance and structure of low-mass halos, irrespective of whether halos contain enough baryonic mass to form stars and emit light. As we enter a new era of cosmology, one in which large surveys, such as Euclid, Roman, and Rubin, will discover thousands of new lens systems, population-level inferences of dark matter substructure through strong lensing will continue providing groundbreaking insights on the particle nature of dark matter, and of the properties of the early Universe. 
		
		\section*{Acknowledgments}
		We thank Alex Drlica-Wagner, Josh Frieman, Henrik R.~Larsson and Ethan Nadler for helpful discussions. We thank the anonymous referees for constructive feedback. 
		
		DG acknowledges support for this work provided by the Brinson Foundation through a Brinson Prize Fellowship grant. 
		
		AMN, CG, MO and RK acknowledge support support from the National Science Foundation through the grant ``CAREER: An order of magnitude improvement in measurements of the physical properties of dark matter" NSF-AST-2442975.
		
		AMN, TT, XD, HP, CG, MO, RK acknowledge support from the National Science Foundation through the grant ``Collaborative Research: Measuring the physical properties of dark matter with strong gravitational lensing" NSF-AST-2205100, NSF-AST-2206315. 
		
		DW acknowledges support by NSF through grants NSF-AST-1906976 and NSF-AST-1836016, and from the Moore Foundation through grant 8548.
		
		P.M. acknowledges support from the National Science Foundation through grant NSF-AST-2407277. 
		
		SB acknowledges support by the Department of Physics and Astronomy, Stony Brook University
		
		TA acknowledges support from ANID-FONDECYT Regular Project 1240105 and the ANID BASAL project FB210003.
		
		KNA is partially supported by the U.S. National Science Foundation (NSF) Theoretical Physics Program Grant No.\ PHY-2210283. 
		
		V.N.B. acknowledges funding support from STScI grant Nos. HST-GO-17103 and HST-AR-17063. 
		
		SGD acknowledges a generous support from the Ajax Foundation. 
		
		SFH acknowledges support through UK Research and Innovation (UKRI) under the UK government’s Horizon Europe Funding Guarantee (EP/Z533920/1, selected in the 2023 ERC Advanced Grant round) and an STFC Small Award (ST/Y001656/1).
		
		A. K. was supported by the U.S. Department of Energy (DOE) Grant No. DE-SC0009937;  by World Premier International Research Center Initiative (WPI), MEXT, Japan; and by Japan Society for the Promotion of Science (JSPS) KAKENHI Grant No. JP20H05853.
		
		V.M. acknowledges support from ANID FONDECYT Regular grant number 1231418 and Centro de Astrof\'{\i}sica de Valpara\'{\i}so CIDI 21.
		
		The work of LAM and DS was carried out at the Jet Propulsion Laboratory, California Institute of Technology, under a contract with the National Aeronautics and Space Administration (80NM0018D0004).
		
		DS acknowledges the support of the Fonds de la Recherche Scientifique-FNRS, Belgium, under grant No. 4.4503.1 and the Belgian Federal Science Policy Office (BELSPO) for the provision of financial support in the framework of the PRODEX Programme of the European Space Agency (ESA) under contract number 4000142531.
		
		MS acknowledges partial support from NASA grant 80NSSC22K1294. 
		
		K.C.W. is supported by JSPS KAKENHI Grant Numbers JP24K07089, JP24H00221.
		
		This research is based, in part, on data collected at the Subaru Telescope, which is operated by the National Astronomical Observatory of Japan. We are honored and grateful for the opportunity of observing the Universe from Maunakea, which has cultural, historical, and natural significance in Hawaii.
		
		This work is based on observations made with the James Webb Space Telescope through the Cycle 1 program JWST GO-2046 (PI:Nierenberg), and the Hubble Space Telescope through HST-GO-15320, HST-GO-15652, HST-GO-17916 (PI:Treu) and HST-GO-13732 (PI:Nierenberg). Funding from NASA through this programs is gratefully acknowledged.

		Some of the data presented herein were obtained at Keck Observatory, which is a private 501(c)3 non-profit organization operated as a scientific partnership among the California Institute of Technology, the University of California, and the National Aeronautics and Space Administration. The Keck facilities we used were LRIS and OSIRIS. The Observatory was made possible by the generous financial support of the W. M. Keck Foundation.  The authors wish to recognize and acknowledge the very significant cultural role and reverence that the summit of Maunakea has always had within the Native Hawaiian community. We are most fortunate to have the opportunity to conduct observations from this mountain.
		
		This work used computational and storage services provided by the University of Chicago’s Research Computing Center; Caltech's Resnick High Performance Computing Center through Carnegie Science's partnership; the Pinnacles (NSF MRI, $\#$ 2019144) computing cluster at the Cyberinfrastructure and Research Technologies (CIRT) at University of California, Merced; and the Hoffman2 Cluster which is operated by the UCLA Office of Advanced Research Computing’s Research Technology Group.  
		
		\section*{Software} This work made use of {\tt{astropy}}:\footnote{\url{http://www.astropy.org}} a community-developed core Python package and an ecosystem of tools and resources for astronomy \citep{astropy:2013, astropy:2018, astropy:2022};  {\tt{cobyqa}} \citep{rago_thesis,razh_cobyqa}; {\tt{colossus}} \citep{Diemer18};  {\tt{lenstronomy}}\footnote{\url{https://github.com/lenstronomy/lenstronomy}} \citep{lenstronomy18,lenstronomy21}; {\tt{numpy}} \citep{numpy}; {\tt{pyHalo}}\footnote{\url{https://github.com/dangilman/pyHalo}} \citep{Gilman++20}; {\tt{trikde}}\footnote{\url{https://github.com/dangilman/trikde}}; {\tt{samana}}\footnote{\url{https://github.com/dangilman/samana}}; and {\tt{scipy}} \citep{scipy}. 
		
		\section*{Data availability}
		The data used in this article come from HST-GO-15320, HST-GO-15652, HST-GO-17917, HST-GO-13732 and JWST GO-2046\footnote{https://archive.stsci.edu/publishing/doi}. The raw data are publicly available online. Astrometry and and flux ratio measurements are presented by \citet{Nierenberg++14,Nierenberg++20,Keeley++24,Keeley++25}. Reduced imaging data for the systems analyzed in this work are available in the open-source software {\tt{samana}}, which also provides notebooks that perform the lens modeling and scripts to reproduce the dark matter analysis. 
		
		\bibliography{main}
		\bibliographystyle{apsrev4-2}
		
		\appendix
		\section{Convergence testing}
		\label{app:convergence}
		\begin{figure}
			\centering
			\includegraphics[trim=0cm 0.0cm 0cm
			0.cm,width=0.48\textwidth]{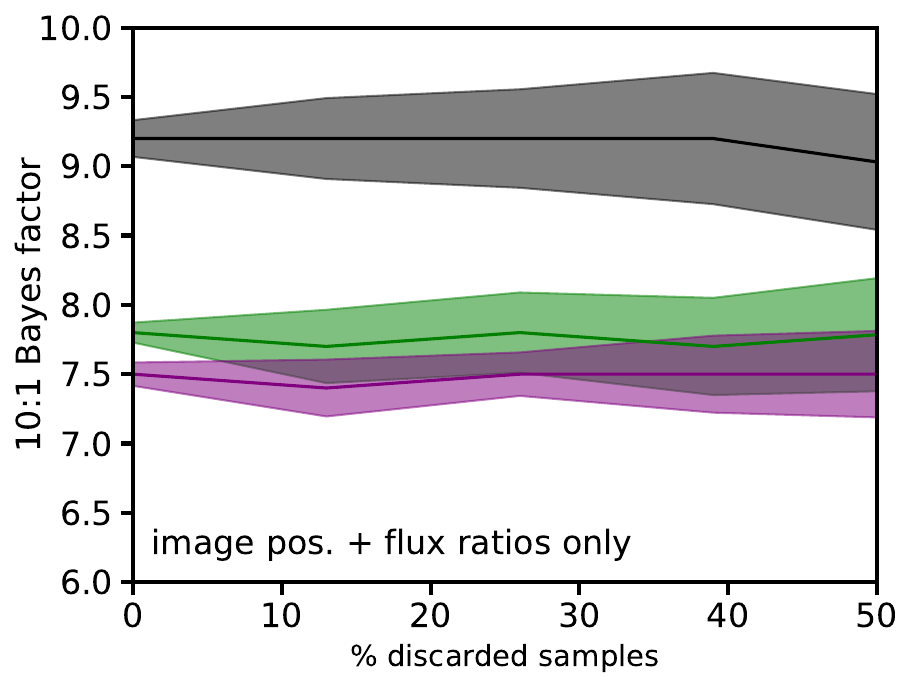}
			\includegraphics[trim=0cm 0.0cm 0cm
			0.cm,width=0.48\textwidth]{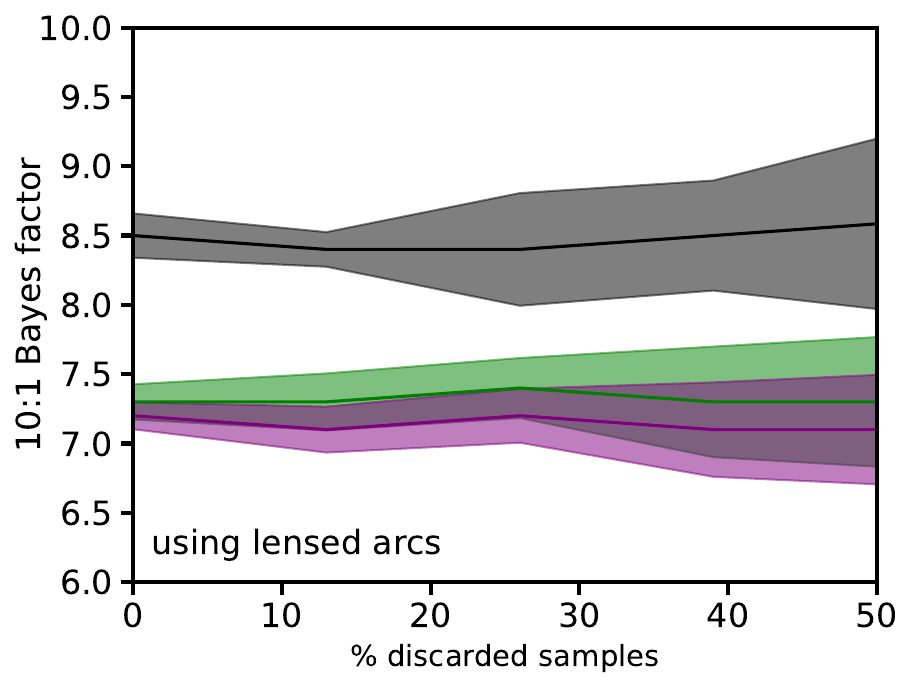}
			\caption{\label{fig:convergencetest} The dependence on the 10:1 Bayes factor constraint on $m_{\rm{hm}}$ on the number of samples used to compute the likelihood functions. The x-axis shows the fraction of discarded samples, and the y-axis shows the resulting constraint on $m_{\rm{hm}}$. The width of the shaded region corresponds to a width of 2 standard deviations. The top panel performs this test with only image positions and flux ratios, and the bottom panel performs the test when incorporating lensed arcs. The colorscale corresponds to a log-uniform prior on $\Sigma_{\rm{sub}}$, and log-normal priors based on {\tt{galacticus}} and $N$-body simulations (green and purple).}
		\end{figure}
		This appendix presents convergence tests that demonstrate we have generated enough samples per lens to compute the likelihood function. We verify that our results are converged by computing the joint posterior distribution shown in Figure \ref{fig:uniformpriors} after discarding a random subset of samples for each lens, and then recomputing the 10:1 Bayes factor constraint on $m_{\rm{hm}}$. The result of this test is shown in Figure \ref{fig:convergencetest}. In each panel, the x-axis shows the fraction of discarded samples when deriving the posterior distribution. The y-axis shows the constraint on $m_{\rm{hm}}$ corresponding to a 10:1 Bayes factor. The color scale is the same as Figure \ref{fig:uniformpriors}, with gray, green and purple representing no prior on $\Sigma_{\rm{sub}}$, a prior based on {\tt{galacticus}}, and based on $N$-body simulations, respectively. 
		
		The width of shaded band corresponds to a $2 \sigma$ fluctuation in the derived constraint, and results from shot-noise in individual posterior distributions. As we discard more samples, the width of the bands increase, because random fluctuations from shot-noise in individual likelihoods becomes more prevalent when we generate fewer realizations. However, we can conclude that we have simulated long enough that our statements regarding the 10:1 Bayes factor constraint are converged at the level $\sim 0.2 $ dex. The uncertainties in the $m_{\rm{hm}}$ bound are larger for differential constraints (such as a 10:1 Bayes factor) compared to integral constraints (such as a confidence interval) because the differential constraints are very sensitive to small changes in the shape of the posterior distribution from statistical fluctuations in the likelihood.  
		
		\section{Kernel density estimation}
		\label{app:kde}
		
		We use a kernel density estimator (KDE) to obtain a continuous approximation of individual likelihood functions. A KDE approximates a target probability distribution, $p\left(\boldsymbol{y}\right)$, using samples, $\boldsymbol{x}$, that we have drawn from the target distribution. The KDE estimator of the target distribution, $\tilde{p}\left(\boldsymbol{y}|\boldsymbol{x}\right)$, is given in terms of these samples 
		\begin{equation}
			\label{eqn:conv}
			\tilde{p}\left(\boldsymbol{y}|\boldsymbol{x}\right) = \frac{1}{N_{\rm{samp}}}\sum_{i=1}^{N_{\rm{samp}}} w_i K\left(\boldsymbol{y} - \boldsymbol{x}_i\right),
		\end{equation}
		where $w_i$ are importance weights for each sample, and $K$ is a kernel function. In the case of this paper, we generate samples using our forward modeling framework, and the importance weights are given by the likelihood function. 
		
		The simplest KDE is a histogram, for which $K\left(\boldsymbol{y}-\boldsymbol{x}_i\right)$ equals 1 or 0. Using histograms to combine multiple independent likelihoods through multiplication faces serious drawbacks for high dimensional parameter spaces, because having zero samples in a bin kills the probability at this coordinate of parameter space for the entire posterior distribution. One can avoid this numerical problem by generating more samples, but in many cases this is not feasible due to computational limitations. In this work, we use a Gaussian KDE, for which $K$ is a multi-variate Gaussian. The covariance matrix is $\boldsymbol{H}$, with entries $H_{ij} = h \hat{H_{ij}}$, where $\hat{H}_{ij}$ the covariance matrix of the samples $\boldsymbol{x}$, and the parameter $h$ is the bandwidth, discussed below. 
		
		KDEs that use extended kernels, such as a Gaussian, systematically down-weight regions of parameter space near the edge of the sampling distribution (they try to smoothly interpolate between some non-zero probability and exactly zero probability outside the sampling distribution). To correct for this known problem, we normalize each kernel by an estimator derived from samples generated uniformly within the sampling distribution, i.e. we take $\tilde{p}\left(\boldsymbol{y} | \boldsymbol{x}\right) \rightarrow  \frac{\tilde{p}\left(\boldsymbol{y} | \boldsymbol{x}\right)}{\tilde{p}\left(\boldsymbol{y}|\boldsymbol{v}\right)}$, where $\tilde{p}\left(\boldsymbol{y}|\boldsymbol{v}\right)$ is derived from samples $\boldsymbol{v}$ drawn from the sampling distribution $\pi_S$. This is a leading-order boundary correction, in the sense that it yields an unbiased estimator when the target distribution has a vanishing gradient at the edge of the sampling distribution. Removing bias in the presence of non-zero gradients requires higher-order corrections \citep[see Section 3A in][]{Lewis19}. The software we use to compute the KDE and to make the triangle plots shown in this paper is publicly available\footnote{\url{https://github.com/dangilman/trikde}}.
		
		In general, a KDE routine will yield the best representation of some target probability distribution when the target distribution is a smoothly-varying function of the parameters, meaning the target distribution is not bimodel and does not have large gradients\footnote{For further discussion of this topic, see \text{https://aakinshin.net/posts/kde-bw/}.}. We can control this aspect of the problem by choosing parameters that should not, based on physical expectations, produce bimodel structure or large gradients in the probability density. 
		
		The choice of bandwidth $h$ is typically  the most important factor when implementing a KDE. While there is no formal definition for the optimum bandwidth unless the target distribution is Gaussian, several ``rule-of-thumb'' estimators provide reasonable starting values. In our analysis, we set the bandwidth along each dimension using Silverman's rule \citep{Silverman86}
		\begin{equation}
			h = \left(n \left(d+2\right) / 4\right)^{\frac{-2}{d+4}}
		\end{equation}
		where $n$ is the number of samples and $d$ is the dimension of the parameter space. We set $n$ equal to the effective sample size, $n_{\rm{eff}}$, which we define as $n_{\rm{eff}}=(1/ /N_{\rm{real}})\sum_{i=1}^{N_{\rm{real}}} p_{i}\left(\datan | \datanprime\right)$, or as the sum of the likelihood weights divided by the number of realizations. In our analysis, this choice results in a plausible representation of the underlying target distribution, in the sense that we obtain correlations between model parameters consistent with physical expectations; for example, the correlation between $\Sigma_{\rm{sub}}$ and $m_{\rm{hm}}$ that is evident in Figure \ref{fig:uniformpriors}. The only modification we make in our implementation is to increase the bandwidth by $10\%$, relative to the baseline Silverman factor. This slightly weakens, by $\sim 0.1$ dex, our constraints on $m_{\rm{hm}}$, but removes ``wiggles'' and islands of probability in the posterior distribution that we believe are unphysical, and which most likely result from shot-noise in individual likelihood functions. 
		
		\section{Globular clusters}
		\label{app:globularclusters}
		\begin{figure*}
			\centering
			\includegraphics[trim=3cm 0.5cm 1cm
			0.cm,width=0.75\textwidth]{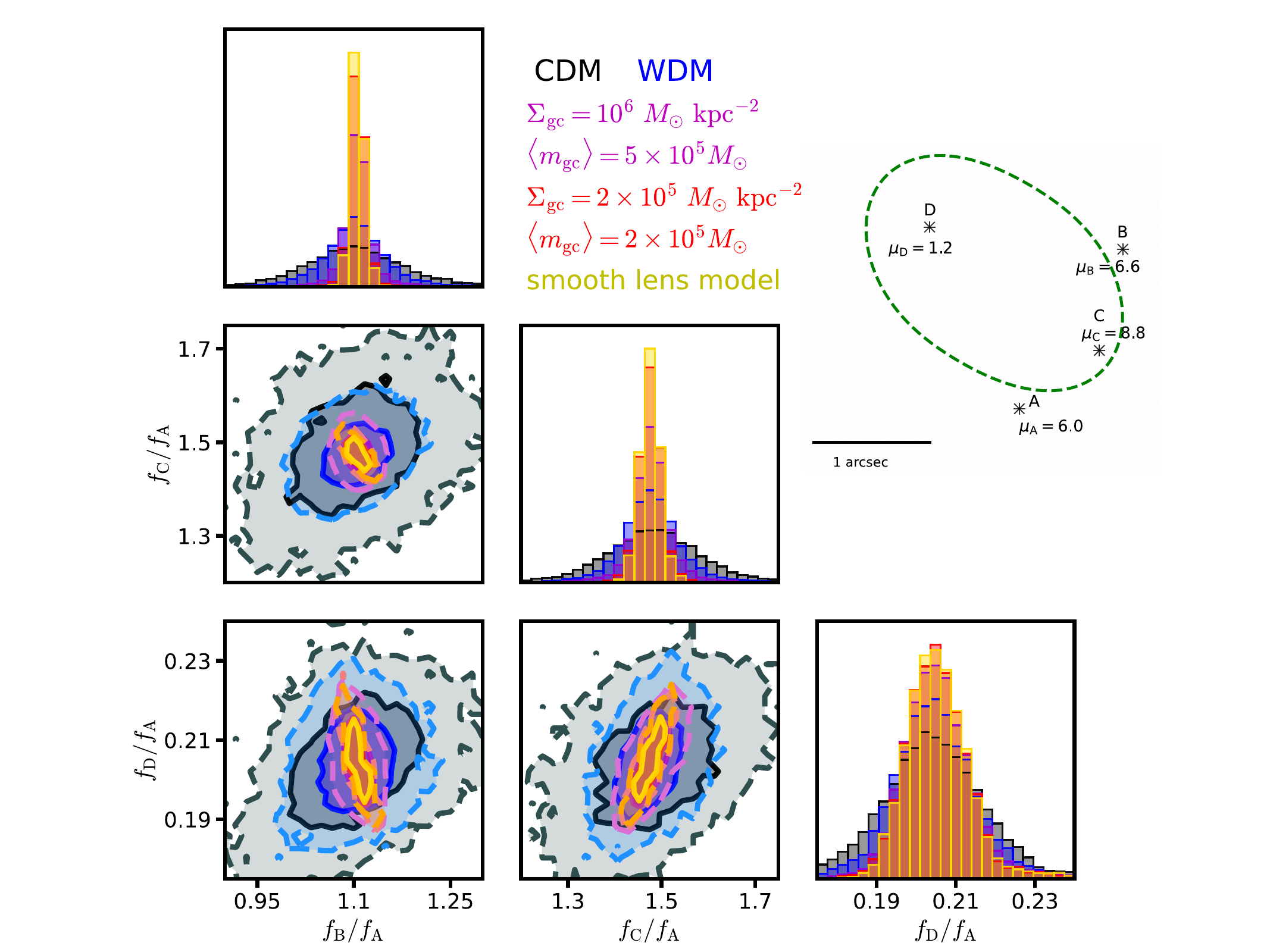}
			\caption{\label{fig:globularclusterfr} The joint distribution of flux ratios for a mock lens with a cusp image configuration that maximizes sensitivity to perturbations by halos and globular clusters. Contours show the flux-ratio distributions in CDM (black), WDM (blue), both including globular clusters with $\Sigma_{\rm{gc}} = 5 \times 10^5 \mathrm{M}_{\odot} \ \rm{kpc^{-2}}$ and $\langle m_{\rm{gc}}\rangle = 2\times 10^{5} \mathrm{M}_{\odot}$.  The magenta and red distributions include perturbations by globular clusters only, and the yellow shows the prediction of a smooth lens model, without halos or GCs. The inset shows a schematic of the mock lens image configuration and magnifications used for the calculation.}
		\end{figure*}
		In our analysis we include globular clusters in the lens model as point masses, as described in Section \ref{ssec:globularclusters}. In this Appendix, we provide additional discussion regarding their lensing effects in the context of flux ratio statistics. 
		
		We can quantify the lensing signal associated with a population of globular clusters by examining how different assumptions regarding the GC mass functions and projected number density affect flux ratio statistics. We make this comparison in relation to the effect of dark matter halos in CDM and WDM to illustrate the relative importance of injecting GCs in the lens model, and changing the properties of dark matter halos. In Figure \ref{fig:globularclusterfr} we show the joint distribution of flux ratios predicted for a mock lens system with a cusp image configuration, which maximizes the sensitivity to small-scale perturbations. The black distribution corresponds to CDM subhalos and field halos only (no GCs) with $\Sigma_{\rm{sub}}=0.1 \ \rm{kpc^{-1}}$. The blue distribution shows the WDM prediction, assuming $m_{\rm{hm}} = 10^8 \mathrm{M}_{\odot}$ and $\Sigma_{\rm{sub}}=0.1 \ \rm{kpc^{-1}}$. The magenta distribution shows perturbations by globular clusters with a surface mass density $\Sigma_{\rm{gc}} =10^6 \mathrm{M}_{\odot} \ \rm{kpc^{-2}}$ and an average GC mass of $\langle m_{\rm{gc}}\rangle =5 \times 10^{5} \mathrm{M}_{\odot}$, and the red distribution has $\Sigma_{\rm{gc}} = 2\times 10^{5} \mathrm{M}_{\odot} \ \rm{kpc^{-2}}$ and $\langle m_{\rm{gc}}\rangle = 2 \times 10^5 \mathrm{M}_{\odot}$. For reference, in our analysis we assumed $\langle m_{\rm{gc}} \rangle = 2 \times 10^5 \mathrm{M}_{\odot}$ and $\Sigma_{\rm{gc}}=10^{5.6} \mathrm{M}_{\odot} \ \rm{kpc^{-2}}$. The width of each log-normal mass function is 0.6 dex. The yellow distribution shows the model-predicted flux ratios with no perturbation from GCs or halos. 
		
		Although GCs evidently cause some degree of perturbation (comparing yellow with red and magenta), the effect of including GCs is much smaller than the difference between CDM and WDM. Making different assumptions for the GC surface mass density and the average GC mass does not change this conclusion (examining red and magenta). Overall, this suggests that including the GCs may slightly weaken constraints on dark matter properties by introducing another source of small-scale perturbation alongside halos, but Figure \ref{fig:globularclusterfr} shows this is a small effect in comparison with altering the halo mass function and concentration--mass relation, as in WDM. For these structures to significantly impact flux ratios studies, we would have to increase their projected number density by over an order of magnitude relative to the measured number density around massive elliptical galaxies.  
		
		\section{Baseline lens models}
		\label{app:baselinelensmodels}
		In this analysis we have performed detailed lens modeling of 24 strong gravitational lenses, including simultaneous reconstruction of the PSF models for systems with HST data. These baseline lens models may serve as a useful starting point for other analyses of these data. Although the current implementation in the open-source software package performs only single-band lens modeling, this can easily be extended to model data in multiple bands.
		
\begin{figure*}
			\includegraphics[trim=5cm 0.5cm 5cm
			2cm,width=0.95\textwidth]{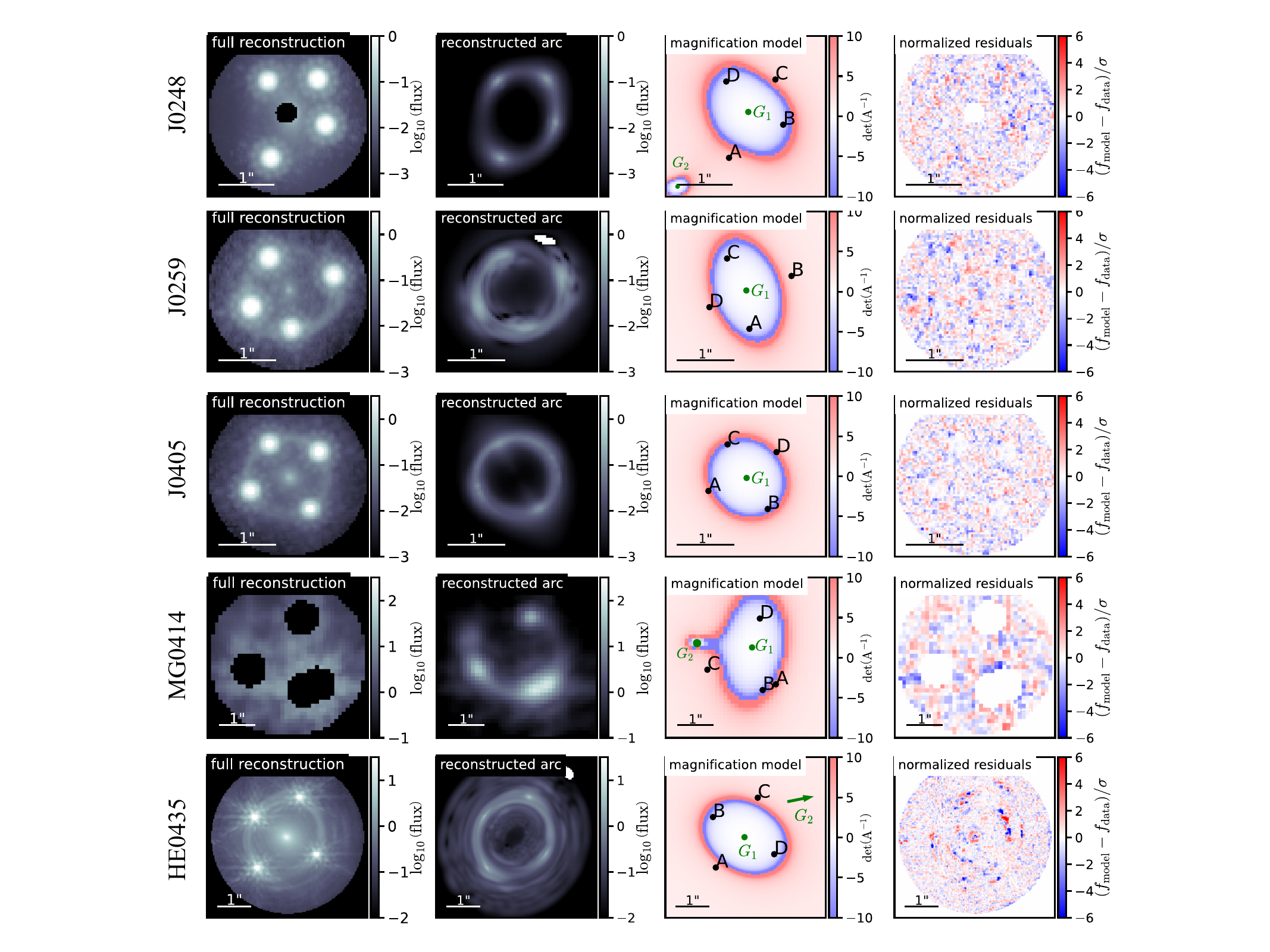}
			\caption{\label{fig:bmodel1} Baseline lens models for J0248, J0259, J0405, and MG0414. From left to right, each column shows the lens mass, lens light, and source light reconstruction; the reconstructed arc; the magnification predicted by the macromodel; and the normalized residuals of all light components. In the magnification plot, image positions are labeled A, B, C, and D, the main deflector centroid is marked as G1, and possible luminous satellites that we include in the lens macromodel are marked as G2/G3. The imaging data masks shown in the left and right columns are those used when computing the image data likelihood, and remove areas where the PSF residuals dominate. The background galaxy ($z = 0.78$) included in the lens model for HE0435 is located off-panel in the direction indicated by the green arrow.}
		\end{figure*}

\begin{figure*}
			\centering
			\includegraphics[trim=5cm 0.5cm 5cm
			2cm,width=0.95\textwidth]{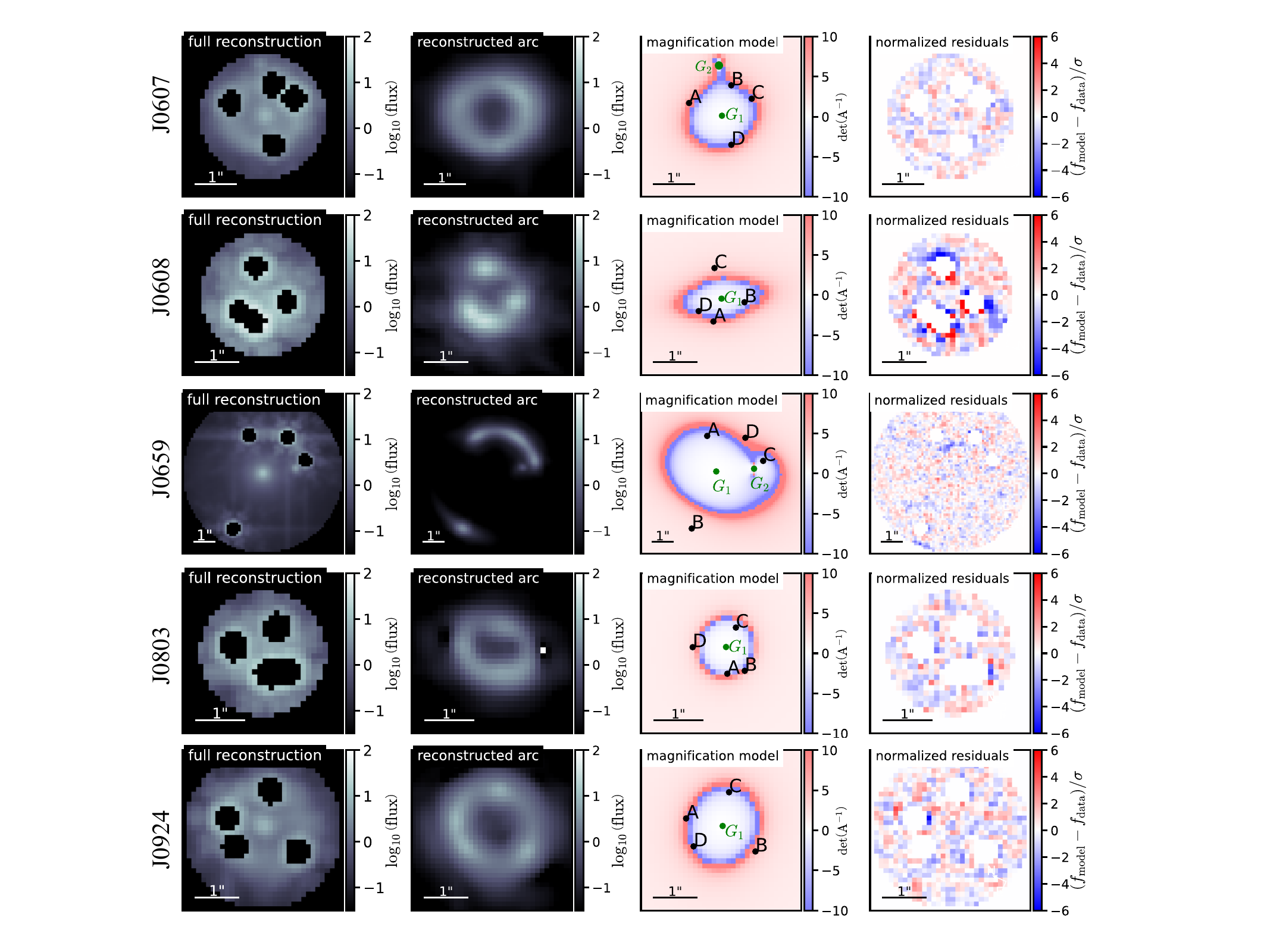}
			\caption{\label{fig:bmodel2} The same as Figure \ref{fig:bmodel1}, but for the systems J0607, J0608, J0659, and J0803, and J0924.}
		\end{figure*}

\begin{figure*}
			\centering
			\includegraphics[trim=5cm 0.5cm 5cm
			2cm,width=0.95\textwidth]{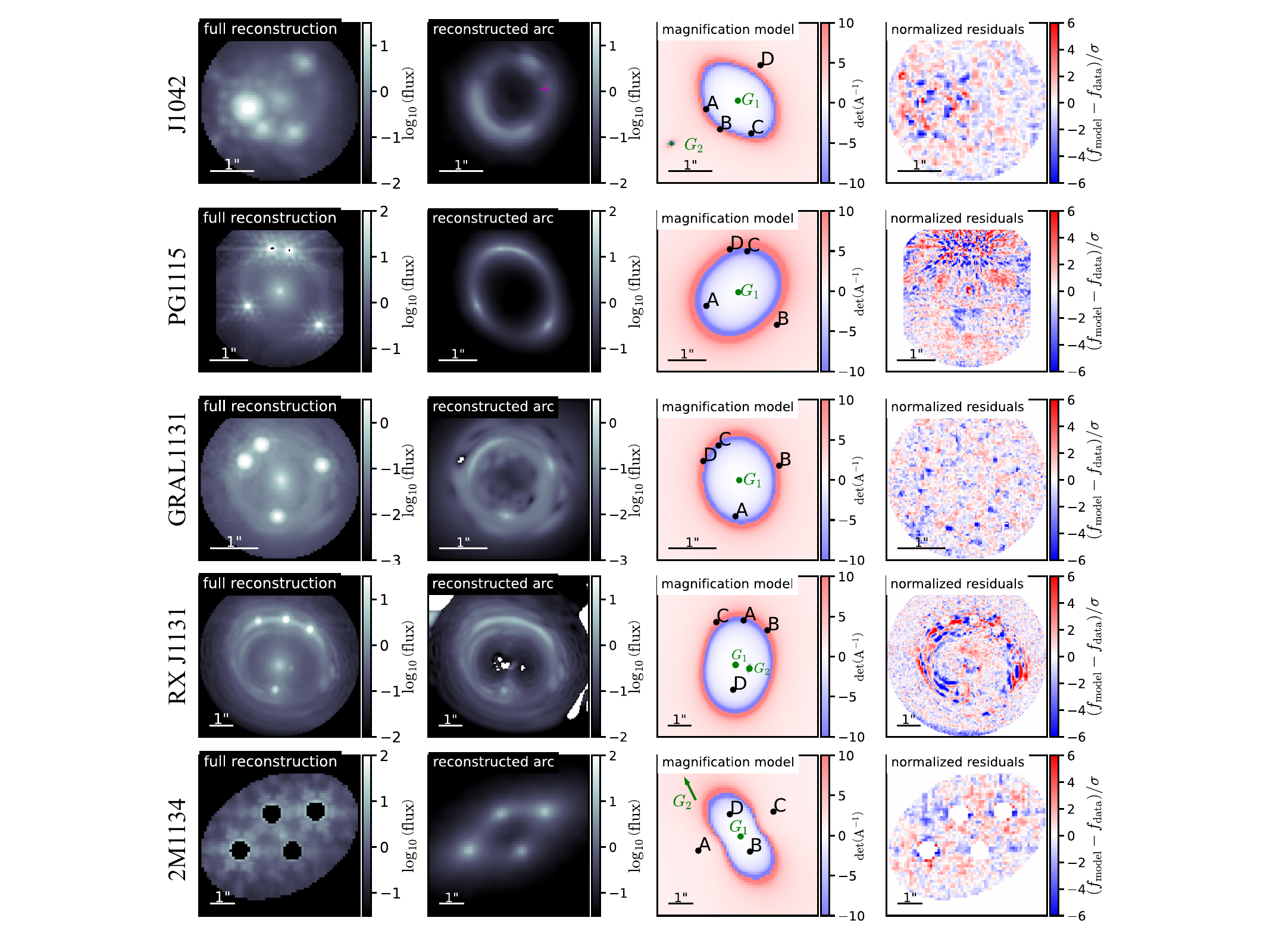}
			\caption{\label{fig:bmodel3} The same as Figure \ref{fig:bmodel1}, but for the systems J1042, PG1115, GRAL1131, RXJ1131, and 2M1134. For PG1115, we mask two saturated pixels at the positions of the merging image pair. For J1042, the location of the feature in the lensed arc associated with the secondary source component (see discussion in Section \ref{sssec:j1042}) is marked in purple. }
		\end{figure*}

\begin{figure*}
			\centering
			\includegraphics[trim=5cm 0.5cm 5cm
			2cm,width=0.95\textwidth]{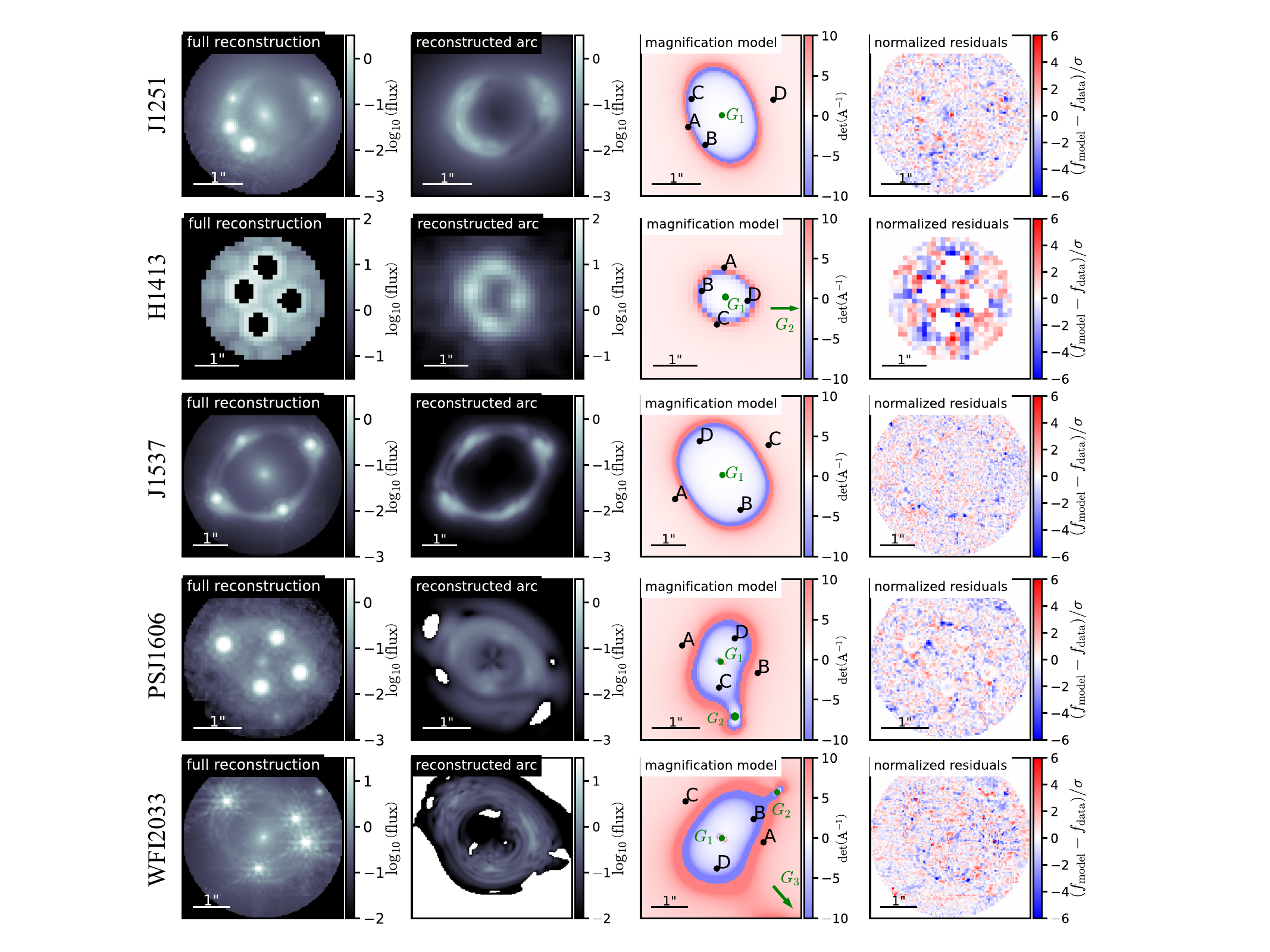}
			\caption{\label{fig:bmodel4}The same as Figure \ref{fig:bmodel1}, but for the systems J1251, H1413, J1537, PSJ1606, and WFI2033.}
		\end{figure*}

\begin{figure*}
			\centering
			\includegraphics[trim=5cm 6cm 5cm
			2cm,width=0.95\textwidth]{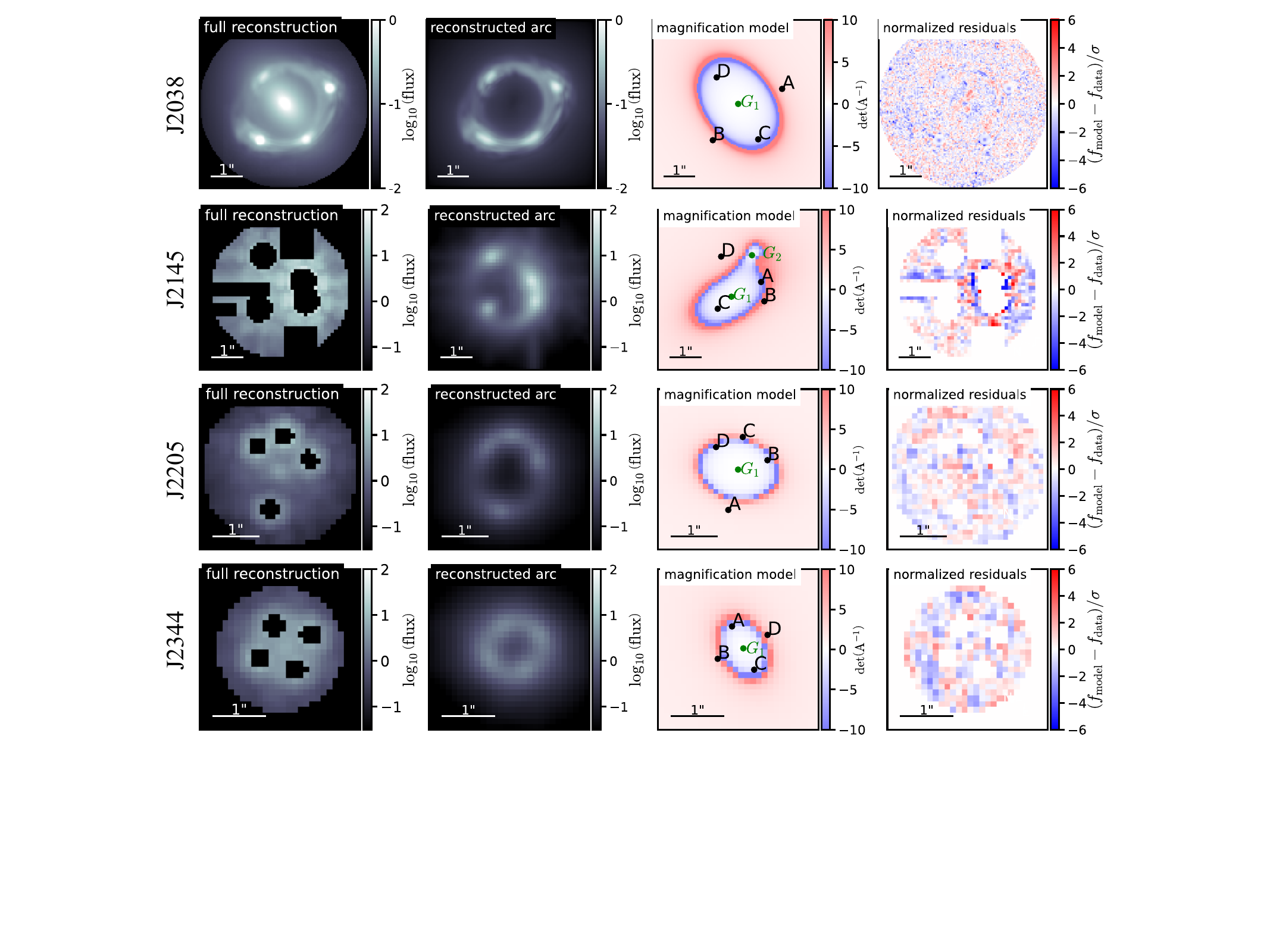}
			\caption{\label{fig:bmodel5} The same as Figure \ref{fig:bmodel1}, but for the systems J2038, J2145, J2205, and J2344.}
		\end{figure*}

		Notebooks that reproduce the baseline lens models, shown in Figures \ref{fig:bmodel1}, \ref{fig:bmodel2}, \ref{fig:bmodel3}, \ref{fig:bmodel4}, and \ref{fig:bmodel5}, are publicly available \footnote{\url{https://github.com/dangilman/samana/tree/main/notebooks}}. The source light model complexity used in the baseline model follows from the calculation of $n_{\rm{max}}$ that minimizes the BIC, or for which the model-predicted flux ratios converges, as discussed in Section \ref{ssec:imglike}. Table \ref{tab:tablemacroparams} lists the macromodel parameters inferred for each lens using the image positions, flux ratios, and lensed arcs. 
		
		\begin{table*}
			\setlength{\tabcolsep}{6pt}
			\caption{\label{tab:tablemacroparams} The effective macromodel parameters inferred for each lens, after marginalizing over substructure in the lens model. We refer to these as effective parameters because they will absorb some features of the (sub)halo population. Therefore, uncertainties will be larger than those typically obtained from strong lens modeling without substructure. We quote results only for the parameters constrained by the imaging data importance weights $w_{\rm{img}}$, full lens model reconstructions can be found in the {\tt{samana}} code repository. For systems without imaging data (PSJ0147, RXJ0911, WFI2026, B1422) we quote the parameters constrained using image positions and flux ratios, for the rest we use image positions, flux ratios, and lensed arcs. From left: Lens system, main deflector mass profile normalization, main deflector axis ratio, external shear strength, logarithmic slope of main deflector mass profile, $m=3$ and $m=4$ multipole strengths (the $m=1$ term is unconstrained for most systems), and the mass of the nearest satellite or companion galaxy (only WFI2033 has more than one satellite/companion). Uncertainties are $68 \%$ confidence intervals. Cases where one of the multipole strengths has a median $\approx 0$ but large uncertainties (e.g. $a_4$ in J0405, J1042) indicate the distribution is bimodal.}
			\setlength\extrarowheight{-2pt}
			\begin{tabular}{cccccccc}
				\hline
				Lens system & $\theta_E$ & $q$ & $\gamma_{\rm{ext}}$ & $\gamma$ & $a_3$ & $a_4$ & $\theta_{E,G2}$ \\
				\hline  \\
				PSJ0147+4630 & $1.85_{-0.03}^{+0.03}$ & $0.82_{-0.09}^{+0.06}$ & $0.16_{-0.04}^{+0.03}$ & $2.08_{-0.1}^{+0.1}$ & $0.0_{-0.005}^{+0.005}$ & $0.0_{-0.01}^{+0.01}$ & -
				\\\\
				J0248+1913 & $0.76_{-0.01}^{+0.01}$ & $0.56_{-0.04}^{+0.04}$ & $0.22_{-0.02}^{+0.02}$ & $2.09_{-0.10}^{+0.10}$ & $0.003_{-0.006}^{+0.005}$ & $-0.008_{-0.009}^{+0.010}$ & $0.05_{-0.02}^{+0.02}$ \\
				\\
				J0259-1635 & $0.74_{-0.01}^{+0.01}$ & $0.77_{-0.06}^{+0.07}$ & $0.05_{-0.02}^{+0.02}$ & $2.13_{-0.09}^{+0.10}$ & $-0.002_{-0.004}^{+0.005}$ & $0.001_{-0.011}^{+0.010}$ & -\\
				\\
				J0405-3308 & $0.702_{-0.005}^{+0.004}$ & $0.84_{-0.05}^{+0.05}$ & $0.04_{-0.03}^{+0.02}$ & $2.13_{-0.10}^{+0.11}$ & $0.002_{-0.003}^{+0.004}$ & $-0.003_{-0.007}^{+0.017}$ & -\\
				\\
				MG0414+0534 & $1.12_{-0.02}^{+0.02}$ & $0.66_{-0.03}^{+0.06}$ & $0.05_{-0.02}^{+0.02}$ & $2.10_{-0.10}^{+0.10}$ & $0.0_{-0.005}^{+0.005}$ & $0.0_{-0.012}^{+0.012}$ & $0.12_{-0.03}^{+0.03}$\\
				\\
				HE0435-1223 & $1.18_{-0.01}^{+0.01}$ & $0.81_{-0.04}^{+0.05}$ & $0.05_{-0.01}^{+0.01}$ & $2.06_{-0.08}^{+0.08}$ & $0.002_{-0.006}^{+0.005}$ & $-0.005_{-0.006}^{+0.007}$ & $0.34_{-0.05}^{+0.05}$\\
				\\
				J0607-2152 & $0.79_{-0.013}^{+0.011}$ & $0.79_{-0.05}^{+0.07}$ & $0.08_{-0.02}^{+0.02}$ & $2.12_{-0.10}^{+0.10}$ & $0.0_{-0.007}^{+0.006}$ & $0.004_{-0.012}^{+0.010}$ & $0.07_{-0.02}^{+0.03}$\\
				\\
				J0608+4229 & $0.65_{-0.01}^{+0.01}$ & $0.46_{-0.05}^{+0.07}$ & $0.09_{-0.03}^{+0.03}$ & $2.12_{-0.10}^{+0.11}$ & $0.0_{-0.005}^{+0.005}$ & $0.005_{-0.013}^{+0.01}$ & -\\
				\\
				J0659+1629 & $2.0_{-0.06}^{+0.07}$ & $0.86_{-0.04}^{+0.07}$ & $0.11_{-0.02}^{+0.02}$ & $2.13_{-0.11}^{+0.10}$ & $0.003_{-0.005}^{+0.005}$ & $-0.002_{-0.010}^{+0.011}$ & $0.39_{-0.08}^{+0.07}$\\
				\\
				J0803+3908 & $0.58_{-0.01}^{+0.01}$ & $0.77_{-0.07}^{+0.07}$ & $0.19_{-0.02}^{+0.02}$ & $2.06_{-0.10}^{+0.10}$ & $-0.002_{-0.005}^{+0.005}$ & $-0.011_{-0.009}^{+0.011}$ & -\\
				\\
				RXJ0911+0551 & $0.85_{-0.08}^{+0.09}$ & $0.84_{-0.06}^{+0.07}$ & $0.28_{-0.03}^{+0.03}$ & $2.07_{-0.09}^{+0.09}$ & $0.0_{-0.005}^{+0.005}$ & $0.0_{-0.011}^{+0.010}$ & $0.35_{-0.11}^{+0.10}$ \\
				\\
				J0924+0219 & $0.89_{-0.01}^{+0.01}$ & $0.67_{-0.04}^{+0.05}$ & $0.08_{-0.02}^{+0.01}$ & $2.12_{-0.09}^{+0.10}$ & $0.003_{-0.005}^{+0.005}$ & $0.001_{-0.016}^{+0.015}$ & -\\
				\\
				J1042+1641 & $0.89_{-0.01}^{+0.01}$ & $0.67_{-0.04}^{+0.05}$ & $0.08_{-0.02}^{+0.05}$ & $2.12_{-0.09}^{+0.10}$ & $0.003_{-0.005}^{+0.005}$ & $0.001_{-0.016}^{+0.015}$ & $0.03_{-0.02}^{+0.02}$\\
				\\
				PG1115+080 & $1.47_{-0.01}^{+0.01}$ & $0.84_{-0.03}^{+0.03}$ & $0.11_{-0.02}^{+0.02}$ & $2.06_{-0.09}^{+0.09}$ & $0.0_{-0.005}^{+0.005}$ & $-0.005_{-0.006}^{+0.008}$ & -\\
				\\
				GRAL1131-4419 & $0.86_{-0.005}^{+0.005}$ & $0.77_{-0.04}^{+0.05}$ & $0.024_{-0.009}^{+0.010}$ & $2.14_{-0.085}^{+0.09}$ & $-0.003_{-0.004}^{+0.004}$ & $-0.005_{-0.007}^{+0.006}$ & -\\
				\\
				RXJ1131-1231 & $1.51_{-0.13}^{+0.14}$ & $0.80_{-0.06}^{+0.05}$ & $0.11_{-0.02}^{+0.02}$ & $2.07_{-0.10}^{+0.10}$ & $0.001_{-0.005}^{+0.005}$ & $0.002_{-0.009}^{+0.009}$ & $0.37_{-0.17}^{+0.14}$\\
				\\
				2M1134-2103 & $1.21_{-0.02}^{+0.01}$ & $0.76_{-0.05}^{+0.04}$ & $0.41_{-0.02}^{+0.02}$ & $2.34_{-0.05}^{+0.06}$ & $0.0_{-0.005}^{+0.005}$ & $0.0_{-0.010}^{+0.010}$ & $0.07_{-0.04}^{+0.06}$\\
				\\
				J1251+2935 & $0.84_{-0.01}^{+0.01}$ & $0.72_{-0.06}^{+0.06}$ & $0.09_{-0.02}^{+0.02}$ & $2.15_{-0.07}^{+0.08}$ & $0.0_{-0.004}^{+0.004}$ & $0.001_{-0.005}^{+0.005}$ & -\\
				\\
				H1413+117 & $0.57_{-0.01}^{+0.01}$ & $0.72_{-0.07}^{+0.06}$ & $0.10_{-0.03}^{+0.03}$ & $2.12_{-0.10}^{+0.10}$ & $0.0_{-0.004}^{+0.004}$ & $-0.004_{-0.008}^{+0.008}$ & $0.54_{-0.09}^{+0.09}$\\
				\\
				B1422+231 & $0.76_{-0.02}^{0.03}$ & $0.57_{-0.06}^{+0.15}$ & $0.22_{-0.05}^{+0.05}$ & $2.14_{-0.10}^{+0.10}$ & $0.0_{-0.005}^{+0.005}$ & $-0.005_{-0.011}^{+0.011}$ & -\\ 
				\\
				J1537-3010 & $1.40_{-0.01}^{+0.01}$ & $0.92_{-0.04}^{+0.04}$ & $0.15_{-0.02}^{+0.02}$ & $2.04_{-0.09}^{+0.09}$ & $0.005_{-0.004}^{+0.004}$ & $-0.002_{-0.007}^{+0.007}$ & -\\
				\\
				PSJ1606-2333 & $0.63_{-0.01}^{+0.01}$ & $0.81_{-0.08}^{+0.08}$ & $0.16_{-0.05}^{+0.04}$ & $2.02_{-0.10}^{+0.11}$ & $0.0_{-0.006}^{+0.005}$ & $-0.001_{-0.009}^{+0.009}$ & $0.17_{-0.05}^{+0.05}$\\
				\\
				WFI2026-4536 & $0.65_{-0.01}^{+0.01}$ & $0.80_{-0.04}^{+0.05}$ & $0.11_{-0.02}^{+0.02}$ & $2.11_{-0.09}^{+0.09}$ & $-0.001_{-0.005}^{+0.005}$ & $-0.001_{-0.008}^{+0.008}$ & -\\
				\\
				WFI2033-4723 & $0.99_{-0.015}^{+0.015}$ & $0.85_{-0.05}^{+0.05}$ & $0.13_{-0.02}^{+0.02}$ & $1.85_{-0.06}^{+0.07}$ & $-0.001_{-0.005}^{+0.005}$ & $-0.013_{-0.008}^{+0.013}$ & $0.08_{-0.02}^{+0.02}$\\
				\\
				J2038-4008 & $1.39_{-0.01}^{+0.01}$ & $0.63_{-0.03}^{+0.03}$ & $0.07_{-0.02}^{+0.02}$ & $2.24_{-0.08}^{+0.08}$ & $0.001_{-0.004}^{+0.004}$ & $0.003_{-0.007}^{+0.008}$ & -\\
				\\
				J2145+6345 & $0.94_{-0.02}^{+0.02}$ & $0.89_{-0.08}^{+0.06}$ & $0.15_{-0.04}^{+0.03}$ & $2.11_{-0.10}^{+0.10}$ & $0.0_{-0.005}^{+0.005}$ & $0.007_{-0.011}^{+0.015}$ & $0.14_{-0.04}^{+0.02}$\\
				\\
				J2205-3727 & $0.76_{-0.01}^{+0.01}$ & $0.82_{-0.09}^{+0.07}$ & $0.06_{-0.02}^{+0.02}$ & $2.09_{-0.10}^{+0.10}$ & $-0.002_{-0.004}^{+0.004}$ & $-0.003_{-0.006}^{+0.007}$ & -\\
				\\
				J2344-3056 & $0.50_{-0.01}^{+0.01}$ & $0.88_{-0.07}^{+0.06}$ & $0.07_{-0.02}^{+0.02}$ & $2.09_{-0.10}^{+0.10}$ & $-0.001_{-0.005}^{+0.005}$ & $0.004_{-0.009}^{+0.013}$ & -\\
				\\
				\hline
			\end{tabular}
		\end{table*}
		

	\end{document}